\pdfoutput=1
\RequirePackage{float}
\let\SNbibliographystyle\bibliographystyle\def\bibliographystyle#1{\SNbibliographystyle{sn-mathphys-num-pub}}
\documentclass[pdflatex,sn-mathphys-num]{sn-jnl}

\usepackage[T1]{fontenc}
\usepackage{lmodern}
\usepackage{graphicx}
\usepackage{amsmath,amssymb}
\usepackage{booktabs,array,tabularx,longtable}
\usepackage{xcolor}
\usepackage{algorithm}
\usepackage{algpseudocode}
\usepackage{placeins}
\usepackage{microtype}
\usepackage{multibib}
\usepackage{tikz}
\newcites{SI}{Supplementary References}
\allowdisplaybreaks[2]
\AtBeginDocument{\setlength{\oddsidemargin}{\dimexpr35.4mm-1in\relax}%
  \setlength{\evensidemargin}{\dimexpr\paperwidth-\textwidth-35.4mm-1in\relax}}

\definecolor{draftblue}{RGB}{28,77,117}
\definecolor{revcolor}{RGB}{160,30,30}
\newif\ifrevisionmarks\revisionmarkstrue

\providecommand{\svec}{}\renewcommand{\svec}{\mathbf{s}}\providecommand{\muvec}{}\renewcommand{\muvec}{\boldsymbol{\mu}}\providecommand{\nuvec}{}\renewcommand{\nuvec}{\boldsymbol{\nu}}
\providecommand{\ind}{}\renewcommand{\ind}{\mathbf{1}}
\providecommand{\old}{}\renewcommand{\old}{\mathrm{old}}
\providecommand{\new}{}\renewcommand{\new}{\mathrm{new}}
\hypersetup{colorlinks=true,linkcolor=draftblue,citecolor=draftblue,urlcolor=draftblue,bookmarksdepth=2,
 pdftitle={Compiling Constraints into Neuromorphic Ising Machines},
 pdfauthor={Faiek Ahsan and Shantanu Chakrabartty}}

\begin{document}

\title[Compiling Constraints into Neuromorphic Ising Machines]{Compiling Constraints into Neuromorphic Ising Machines}
\author[1]{\fnm{Faiek} \sur{Ahsan}}\email{a.faiek@wustl.edu}
\author*[1]{\fnm{Shantanu} \sur{Chakrabartty}}\email{shantanu@wustl.edu}
\affil[1]{\orgdiv{Department of Electrical and Systems Engineering},
\orgname{Washington University in St. Louis},
\orgaddress{\street{One Brookings Drive}, \city{St. Louis}, \state{MO}, \postcode{63130}, \country{USA}}}

\abstract{
Most practical combinatorial optimization problems involve constraints, and an Ising machine that flips one spin at a time can respect them only through energy penalties that raise barriers between feasible solutions. Here we show that a higher-order neuromorphic Ising machine can enforce constraints through programmable transitions rather than penalties. In an autoencoder-like architecture, each inner-layer neuron represents a move, a family of multi-spin transitions that maps every feasible state to another feasible state, receives the energy change of its move through connections that outer-layer neurons multiplicatively switch on or off, and updates the state through the same connections. A compiler turns constraints into this wiring with a number of auxiliary clauses that grows only polynomially for parity, cardinality, one-hot, partition and permutation constraints. Restricting the search to feasible states reduces constraint handling to move design, and we show that the fastest-annealing moves are not those that mix fastest but those that lower the deepest energy barriers, and that moves chosen from the objective lower them further than random ones. On third-order graph matching, the machine needs orders of magnitude fewer iterations than penalty-based neuromorphic machines. Configured as a sampler on a field-programmable gate array, it decodes the surface code with maximum-likelihood accuracy up to code distance 15 within 52\,$\mu$s per shot and, under noisy syndrome measurements, decodes in real time for millisecond measurement rounds up to distance 9 with fewer logical errors than belief-propagation decoders.
}

\maketitle

\section{Introduction}\label{sec:introduction}

Neuromorphic Ising machines harness the collective dynamics of spiking neurons for computation\cite{Chen2025,Ahsan2026}. Each neuron represents a spin, with a spike corresponding to a spin flip; a neuron fires when its input crosses a local, noisy threshold, and annealing this noise allows the network to escape local minima and settle into low-energy states\cite{Chen2025}. Like the preceding optical, quantum, electronic, and probabilistic Ising machines\cite{Mohseni2022,Inagaki2016,McMahon2016,Aadit2022,Bohm2022,Niazi2024}, they map tasks onto energy landscapes explored by physical or emulated dynamical systems\cite{Lucas2014}. Furthermore, they leverage core neuromorphic hardware advantages: asynchronous event-driven updates, highly parallel in-memory computation, and noise-driven dynamics\cite{ChakrabarttyEd2026}. Consequently, these systems have demonstrated improvements in execution time, energy efficiency, and solution quality\cite{Cai2020,Chen2025,Ahsan2026}, while offering scalability on platforms such as SpiNNaker~2, TrueNorth, and Loihi\cite{Mayr2019,Merolla2014,Davies2018,Kudithipudi2025}.

\begin{figure}[!tbp]
\centering
\includegraphics[width=\linewidth]{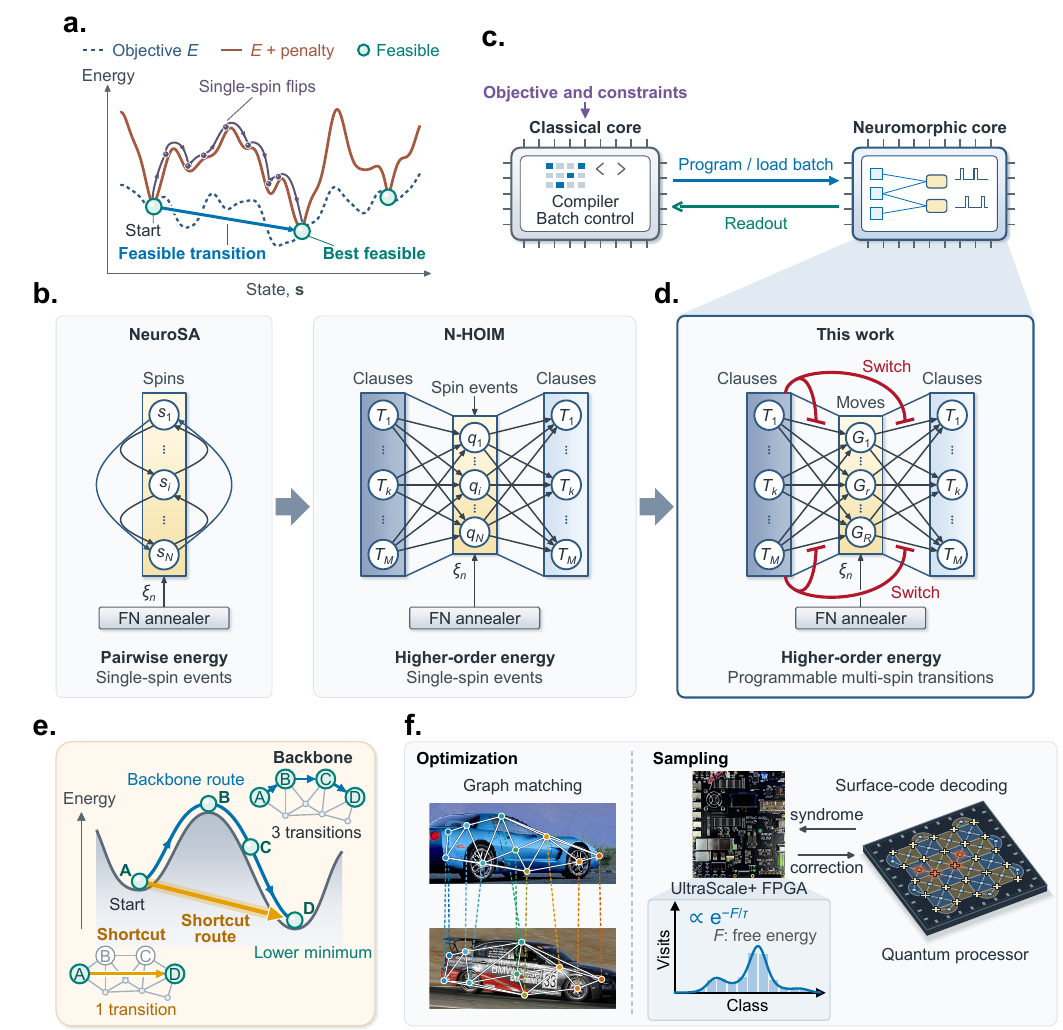}
\caption{\textbf{Transition-programmable higher-order neuromorphic Ising machine.}
\textbf{a}, Energy along a sequence of states of a constrained problem (schematic). Single-spin flips from a feasible state enter infeasible states. When the constraint is added to the objective $E$ as a penalty, these infeasible states raise barriers between feasible states. A spin transition between feasible states reaches the best feasible state without crossing these barriers.
\textbf{b}, Previous neuromorphic Ising machines. NeuroSA represents pairwise interactions between spin neurons. N-HOIM stores the clause outputs $T_1,\ldots,T_M$ of the higher-order clauses, encodes them into the inputs of spin neurons and decodes each accepted spin flip into clause toggles. Both machines use single-spin events. In all three machines (\textbf{b},\textbf{d}), the Fowler--Nordheim (FN) annealer supplies the stochastic threshold $\xi_n$.
\textbf{c}, Programmable system. A classical core compiles the objective and the constraints into a move library and loads it, or successive batches of it, into the neuromorphic core, from which the solution is read out.
\textbf{d}, Transition-programmable architecture of this work. The encoder passes the stored clause outputs through switched connections to the \textit{move} neurons $G_1,\ldots,G_R$, one per move. The input of each \textit{move} neuron is half the energy change of its feasible spin transition from the current state. Red lines: the clause outputs that set the state-dependent switches of the encoder and decoder connections. An accepted transition is decoded, through the same switches, to toggle the corresponding clause outputs.
\textbf{e}, Energy along the feasible states $\mathrm{A}$--$\mathrm{D}$ (schematic) with a backbone move library, which connects all feasible states, and a shortcut move, which joins in one transition two states that the backbone connects only through several transitions. The shortcut bypasses an energy barrier on the backbone route.
\textbf{f}, Optimization and sampling with the same architecture. Optimization (temperature $\tau$ annealed to 0): graph matching between the keypoints of two car photographs of the WILLOW-ObjectClass dataset\cite{Cho2013}. Sampling (temperature $\tau=1$): surface-code decoding on an UltraScale+ FPGA board. Schematic of the intended use: a quantum processor sends the syndrome of its surface code to the decoder, which returns a correction. Here, the syndromes are simulated (Section~\ref{res:surface}). Inset: visits of the logical classes during sampling, proportional to $e^{-F/\tau}$, with $F$ the free energy of the class.}\label{fig:concept}
\end{figure}

Most practical optimization problems, however, impose constraints that couple decision variables and restrict admissible configurations which single-spin-flip Ising machines cannot natively respect. Because a constraint couples several spins, flipping a single spin typically drives a feasible state into an infeasible one. Consequently, constraints are conventionally added to the energy function as penalty terms\cite{Lucas2014}, which introduces three primary drawbacks. First, a penalty strong enough to ensure every feasible state remains lower in energy than its infeasible neighbors elevates energy barriers between feasible states (Fig.~\ref{fig:concept}a). Second, penalty strength requires precise tuning relative to the objective function: if too weak, infeasible states drop below the true optimum; if too strong, the penalty dominates the objective and inflates the dynamic energy range the hardware must represent\cite{Hen2016}. Third, single-spin-flip searches spend a large fraction of their iterations exploring infeasible states\cite{Hadfield2019,Im2026}.

Multi-spin moves mitigate these issues by connecting feasible states directly to one another, thereby bypassing high-energy barriers (Fig.~\ref{fig:concept}a). Although such constraint-preserving moves are standard in classical simulated annealing\cite{Johnson1991}, Ising-type solvers adopt them only within narrow domains. Existing methods rely on specific constraint forms or variable representations: constraint-preserving driver Hamiltonians and mixing operators in quantum annealing and the quantum alternating operator ansatz\cite{Hen2016,Hadfield2019}; one-hot variables enforced by construction in Potts annealers\cite{Whitehead2023} or via binary encodings\cite{Garg2025} (which only ensure single-variable validity while still requiring penalties for inter-variable constraints like permutations); digital encoding circuits that derive problem variables from probabilistic bits\cite{Iyer2025} at the cost of routing all energy updates through the circuit; and heuristics restricted to fixed move sets, such as permutation-preserving exchanges\cite{Bagherbeik2020,Im2026,Kong2026}. Crucially, none of these architectures can program arbitrary moves for general constrained problems directly into hardware, nor can they evaluate high-order multi-spin moves locally through local interconnections as an Ising machine evaluates a single spin flip.

Previous neuromorphic Ising machines share this limitation (Fig.~\ref{fig:concept}b). In NeuroSA, every spin is a neuron and pairwise synapses carry the interactions\cite{Chen2025}, so objectives with interactions among three or more spins must be reduced to pairwise form, which adds auxiliary spins and widens the range of coupling strengths\cite{Boros2002}. The higher-order neuromorphic Ising machine (N-HOIM) avoids this reduction with an autoencoder-like architecture\cite{Ahsan2026}. It expresses the energy of $N$ spins $\svec=(s_1,\ldots,s_N)$, $s_i\in\{-1,+1\}$, as
\begin{equation}
E(\svec)=E_\mathrm{const}-\sum_{k=1}^{M}J_kT_k(\svec),\qquad T_k(\svec)=\prod_{i=1}^{N}s_i^{H_{ki}},
\label{eq:intro-energy}
\end{equation}
where $H_{ki}\in\{0,1\}$ marks whether spin $i$ belongs to the $k$th spin product, $J_k$ is the weight of that product and $E_\mathrm{const}$ is a constant. Following N-HOIM, we call each spin product a clause and its sign $T_k\in\{-1,+1\}$ the clause output; a clause of $\ell$ spins encodes an interaction of order $\ell$, so Eq.~\eqref{eq:intro-energy} represents interactions of any order. N-HOIM encodes the $M$ clause outputs using outer-layer neurons (Fig.~\ref{fig:concept}b), and inner-layer neurons encode the flip of one spin. Each inner-layer neuron therefore sums the weighted outputs of the clauses that contain its spin, which is proportional to the energy change of the flip. The neuron fires when this input crosses a noisy threshold (determined by an annealer), and the firing selectively toggles the clause outputs recursively similar to an autoencoder (Fig.~\ref{fig:concept}b). For a constrained problem, however, the energy is minimized or sampled only over the feasible set $\Omega\subseteq\{-1,+1\}^{N}$, and N-HOIM, which changes one spin per event (or several only when they do not interact), can enforce constraints only through penalties.

Here, we propose a neuromorphic Ising machine that programs spin transitions as well as the energy landscape, ensuring that its state search operates exclusively within the feasible space of a higher-order energy landscape. This construction is based on the observation that a spin transition from a feasible state $\boldsymbol{\mu} \in \Omega$ to another feasible state $\boldsymbol{\nu} \in \Omega$ ($\boldsymbol{\mu}\to\boldsymbol{\nu}$) changes a clause $T_k$ exactly when an odd number of its component spins change (see Methods Section~\ref{meth:model}). The switch variable $\alpha_{k,\boldsymbol{\mu}\leftrightarrow\boldsymbol{\nu}} \in \{0,1\}$ equals 1 if this transition changes $T_k$ and 0 otherwise. Consequently, the net energy change resulting from the transition $\boldsymbol{\mu}\to\boldsymbol{\nu}$ is given exactly in terms of the switch variables, clauses, and weights as
\begin{equation}
\Delta E_{\boldsymbol{\mu}\to\boldsymbol{\nu}}=2\sum_{k=1}^{M}J_k\,\alpha_{k,\boldsymbol{\mu}\leftrightarrow\boldsymbol{\nu}}\,T_k(\boldsymbol{\mu})
\label{eq:intro-deltae}
\end{equation}
(see Methods Section~\ref{meth:model}). Note that the bidirectional arrow in the index $\alpha_{k,\boldsymbol{\mu}\leftrightarrow\boldsymbol{\nu}}$ indicates that the switch variable is symmetric for both transition directions. While Eq.~\eqref{eq:intro-deltae} suggests an N-HOIM autoencoder architecture~\cite{Ahsan2026} where each feasible spin transition is mapped to an individual inner-layer neuron, the combinatorial explosion of feasible transitions renders this direct mapping intractable for practical, large-scale problems. The proposed transition-programmable architecture (Fig.~\ref{fig:concept}d) resolves this bottleneck by abstracting primitive spin transitions into sets of feasible actions, or ``moves.'' This strategy parallels motor control in biological nervous systems, where cortical and spinal circuits encode low-dimensional motor primitives and muscle synergies or coordinated actions that are physically feasible by construction—rather than controlling individual joint angles independently\cite{dAvella2005}. Within this framework, the move library constitutes a repertoire of primitives, anchored by a minimal set of backbone moves that spans the feasible state space. We detail this abstraction for a representative toy problem in Results Section~\ref{res:toy}, Methods Section~\ref{meth:model}, and Fig.~\ref{fig:toy}, and present its generalization to complex constraints in Supplementary Section~\ref{supp:closure}. In the transition-programmable architecture (Fig.~\ref{fig:concept}d), each inner-layer neuron represents a distinct move $G_r$ ($r=1,\dots,R$), and the switch variable $\alpha_{k,G_r} \in \{0,1\}$ determines whether clause $T_k$ can trigger move $G_r$. The energy change per move, $\Delta E_{G_r}$, is expressed as
\begin{equation}
\Delta E_{G_r} = 2\sum_{k=1}^{M}J_k\,\alpha_{k,G_r}\,T_k=2\,v_{G_r},
\label{eq:intro-input}
\end{equation}
which is directly proportional to the total synaptic input $v_{G_r}$ received by \textit{move} neuron $r$. A \textit{move} neuron fires when its input falls below a stochastic threshold $\xi_n$ generated by a Fowler--Nordheim annealer\cite{Chen2025}, which enforces Metropolis acceptance criteria at the current annealing temperature (Methods Section~\ref{meth:benchmarks}). Consistent with the autoencoder topology, the output of a firing \textit{move} neuron toggles the clause outputs $T_k$, which forms a feedback path that likewise passes through the switches $\alpha_{k,G_r}$ (Fig.~\ref{fig:concept}d). Neuromorphic architectures exploit activity sparsity: only neurons whose state changes generate events~\cite{ChakrabarttyEd2026, Kudithipudi2025}. The state-dependent switching adds a second form of dynamic sparsity, connectivity sparsity, which is familiar from sparse artificial neural networks~\cite{Mocanu2018}: the machine's own stored state switches on and off the connectivity, so that only the connections relevant to the current state carry the signal. Finally, we demonstrate in Results Section~\ref{res:surface} that implementing this neuromorphic core on a field-programmable gate array (FPGA) significantly accelerates computational performance. To tackle full-scale practical problems, this neuromorphic core is co-designed with a classical processing unit (comprising CPUs or GPUs), as illustrated in Fig.~\ref{fig:concept}c. The classical core compiles the problem objective and constraints into the target move library and subsequently programs these configuration parameters into the neuromorphic hardware.

For parity, cardinality, one-hot, partition and permutation constraints with objectives of bounded order, we show that each move adds only polynomially many clauses and connections to the machine (Supplementary Table~\ref{tab:supp-closure}). The machine therefore grows only polynomially with problem size, because it is built incrementally from a minimal set of backbone moves, polynomially many for these constraints, chosen to guarantee connectivity across all feasible states. Shortcut moves are then added to the backbone to shorten the average path length between distant states (Fig.~\ref{fig:concept}e), constructing a topology analogous to a small-world network~\cite{Watts1998}. Such shortcuts allow the search to bypass energy barriers rather than cross them, a classical analogue of quantum tunneling in quantum annealing\cite{Denchev2016}, though transition acceptance remains governed by the stochastic dynamics of the machine rather than a tunneling amplitude. Using exact energy landscapes, we demonstrate that objective-informed shortcuts lower the deepest barriers more effectively than random shortcuts and enable faster annealing, despite producing slower mixing. Furthermore, when hardware capacity is limited, shortcuts can be loaded and processed in sequential batches. As a neuromorphic architecture, each move (or neuron) is updated asynchronously through local interconnections, enabling moves that do not share overlapping clauses to execute in parallel.

We demonstrate this architecture across two primary modalities (Fig.~\ref{fig:concept}f): (i) optimization via annealing to identify low-energy states, and (ii) probabilistic sampling for inference. As an optimizer, the solver is evaluated on graph matching problems with third-order similarity terms. Emulation results show that the proposed machine requires orders of magnitude fewer iterations than conventional neuromorphic architectures that enforce constraints using penalty terms. As a sampler, we apply the machine to decode surface codes—a prominent class of quantum error-correcting codes characterized by parity constraints. Implemented on a field-programmable gate array (FPGA), the decoder samples the logical classes of corrections given a single fault-free syndrome measurement per shot. It achieves logical error rates approaching maximum-likelihood decoding within microseconds per shot, matching the accuracy of a tensor-network decoder up to code distance 15. Under noisy syndrome measurements in real time, the FPGA implementation completes decoding well within the millisecond measurement windows of trapped-ion and neutral-atom processors, outperforming baseline CPU decoders by a margin that increases with code distance below the fault-tolerance threshold.

\section{Results}\label{sec:results}

\subsection{Visualizing operating principle using a toy problem}\label{res:toy}

\begin{figure}[!tbp]
\centering
\includegraphics[width=\linewidth]{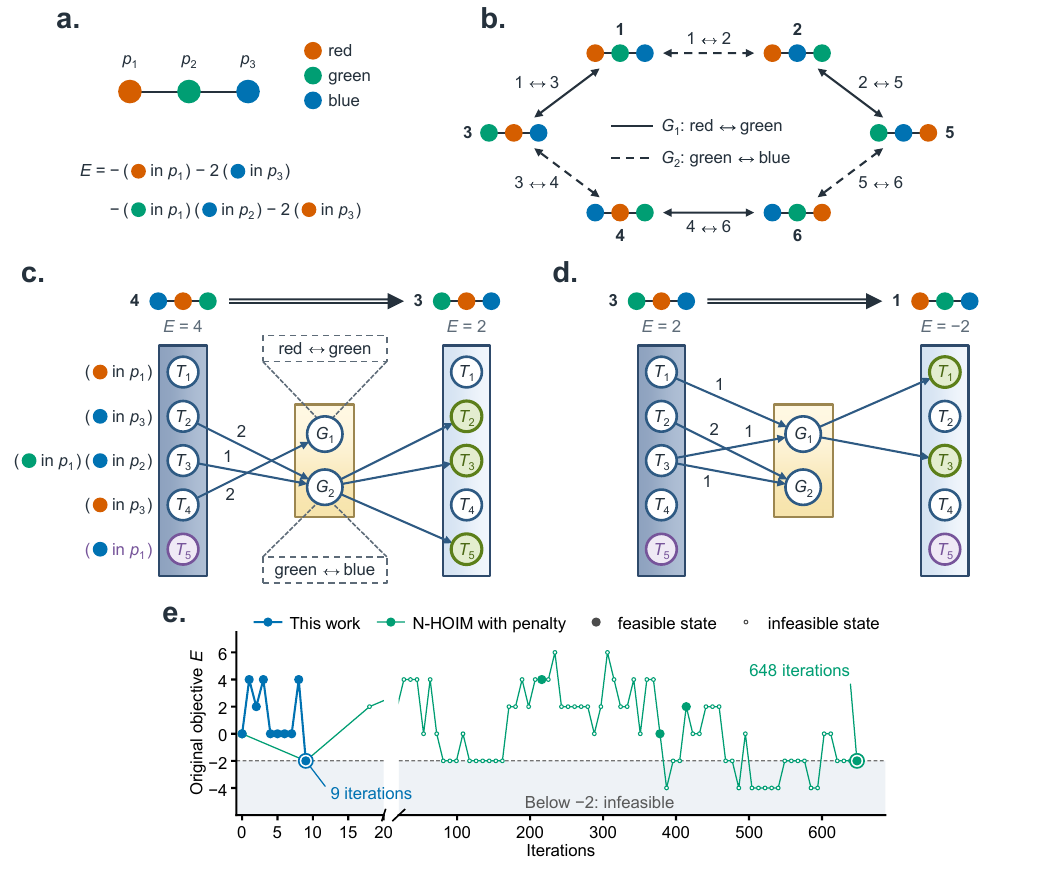}
\caption{\textbf{Operating the architecture on a small assignment problem: three colors in three positions.}
\textbf{a}, The problem: three colors, red, green and blue, are to be arranged in positions $p_1$, $p_2$ and $p_3$. The cost to be minimized is $E$, where each bracket is $+1$ when its statement is true and $-1$ otherwise.
\textbf{b}, The six feasible states and the spin transitions between them. Solid arrows: transitions of the move $G_1$ (red $\leftrightarrow$ green). Dashed arrows: transitions of $G_2$ (green $\leftrightarrow$ blue). Together, the two moves connect all six states.
\textbf{c}, \textbf{d}, Operation of the compiled architecture over two iterations. Methods Section~\ref{meth:model} shows how the compiler obtains this architecture from the objective and the constraints. Only active connections are shown: in the encoder, every connection whose switch is on, and in the decoder, the connections of the firing \textit{move} neuron whose switch is on. The connections that are on differ between \textbf{c} and \textbf{d} (connectivity sparsity), and only the firing \textit{move} neuron and the clause outputs that it toggles generate events (activity sparsity).
\textbf{c}, The accepted transition from state 4 to state 3. Left: the term of the objective that corresponds to each clause. Purple: the auxiliary clause $T_5$. Olive: the clause outputs that the decoder toggles.
\textbf{d}, The next transition, from state 3 to the optimum, state 1, drawn as in \textbf{c}.
\textbf{e}, Original objective $E$ versus iterations up to the first visit of the feasible optimum (state 1, $E=-2$), for this work and for N-HOIM. N-HOIM minimizes $E+4P_\mathrm{pen}$, where $P_\mathrm{pen}$ is the penalty of the assignment constraints (Methods Section~\ref{meth:benchmarks}). One run of each machine, whose first visit of the optimum (after 9 and 648 iterations) equals that machine's median over 200 runs. Both machines use the same annealing schedule and the same starting assignment. One iteration is one proposed spin transition for this work and one single-spin update test for N-HOIM.}\label{fig:toy}
\end{figure}

We first demonstrate the architecture's operation on an assignment problem small enough to enumerate all of its feasible states (Fig.~\ref{fig:toy}). Three colors red, green, and blue are to be assigned to three distinct positions ($p_1$, $p_2$, and $p_3$), with exactly one color per position. The cost function $E$ to be minimized is shown in Fig.~\ref{fig:toy}a. It comprises four weighted terms, each defined as a single bracketed indicator or the product of two bracketed indicators; each bracket evaluates to $+1$ if its contained statement is true and $-1$ otherwise. Exactly six configurations are feasible, denoted as states 1 through 6 (Fig.~\ref{fig:toy}b). Six spin transitions interlink these states, each resulting from one of two elementary operations: swapping the positions of red and green ($\text{red} \leftrightarrow \text{green}$) or swapping the positions of green and blue ($\text{green} \leftrightarrow \text{blue}$).

Figures~\ref{fig:toy}c and \ref{fig:toy}d illustrate two iterations of the compiled architecture, which evaluates and executes each spin transition exclusively through clause outputs. Methods Section~\ref{meth:model} details how the compiler constructs this architecture and subsequently generalizes the procedure. The architecture’s clauses correspond to the terms of the cost function, defined as $T_1=(\text{red in }p_1)$, $T_2=(\text{blue in }p_3)$, $T_3=(\text{green in }p_1)(\text{blue in }p_2)$, and $T_4=(\text{red in }p_3)$, yielding the total energy $E=-T_1-2T_2-T_3-2T_4$. The two elementary operations map to two distinct moves, each represented by a \textit{move} neuron: $G_1$ for swapping red and green ($\text{red}\leftrightarrow\text{green}$) and $G_2$ for swapping green and blue ($\text{green}\leftrightarrow\text{blue}$). Which clauses a given move toggles depends on the current system state. For example, swapping red and green toggles $T_1$ ($\text{red in }p_1$) only when blue is not in $p_1$. If blue occupies $p_1$, neither red nor green can enter or leave $p_1$, leaving $T_1$ unchanged. Consequently, the connection from $T_1$ to $G_1$ requires a switch that remains closed (active) precisely when blue is not in $p_1$. Because no original cost term explicitly tracks whether blue is in $p_1$, the compiler inserts an auxiliary clause $T_5=(\text{blue in }p_1)$ with zero weight. No additional auxiliary clauses are required (Methods Section~\ref{meth:model}). In general, the compiler sets each switch from one stored clause output and adds that clause with zero weight (an auxiliary clause) when the objective does not contain it. Supplementary Section~\ref{supp:closure} and Supplementary Table~\ref{tab:supp-closure} establish upper bounds on the number of auxiliary clauses introduced during compilation, while Supplementary Algorithm~\ref{alg:compile} outlines the step-by-step compilation workflow. For visual clarity, Figs.~\ref{fig:toy}c and \ref{fig:toy}d depict only the active connections for the current state. Figure~\ref{fig:toy}c initiates the search from state 4 ($E=4$). The input to a \textit{move} neuron represents half the net energy change of the spin transition generated by its corresponding move from the current state. In state 4, red resides in $p_2$ and green in $p_3$. Swapping red and green places red in $p_3$, which toggles $T_4$ from $-1$ to $+1$ while leaving all other clauses invariant. Given that $T_4$ has a weight coefficient of 2 in the energy cost, this move yields an energy difference of $\Delta E = -4$, corresponding to a \textit{move} neuron input $v_{G_1} = -2$. Analogously, $v_{G_2} = -1$. The architecture calculates both inputs concurrently via the active switched connections in the encoder. Because both inputs are negative—indicating that both candidate spin transitions reduce total energy—either neuron may fire, with the firing sequence determined stochastically. In Fig.~\ref{fig:toy}c, $G_2$ fires first. Green and blue swap positions, and the decoder toggles three clause outputs ($T_2$, $T_3$, and $T_5$), transitioning the system to state 3 ($E=2$). Because the clause outputs have updated, a new set of switches activates during the subsequent iteration, altering the active network topology (Fig.~\ref{fig:toy}d). Guided by these updated connections, $G_1$ fires, toggles $T_1$ and $T_3$, and drives the system directly into the optimal feasible state 1 ($E=-2$). The active connections thus change with every accepted transition. This connectivity sparsity follows from the compression of spin transitions into moves: each \textit{move} neuron represents all three spin transitions of its move (Fig.~\ref{fig:toy}b), and in each state its switches turn on only the connections of the transition that the move makes from that state. Only the clause outputs that the firing neuron toggles generate events, and Section~\ref{res:event} quantifies the operations that this activity sparsity saves. On this benchmark toy problem, our architecture reaches the feasible optimum in far fewer iterations than standard N-HOIM, which relies on single-spin flips while attempting to minimize the objective augmented by penalty terms for violated constraints (Fig.~\ref{fig:toy}e). Across 200 independent trials initialized from state 6, our machine exhibits a median iteration count of 9 to reach the feasible optimum ($E=-2$), compared to 648 iterations for N-HOIM. Both solvers utilize an identical annealing schedule optimized for N-HOIM (Methods Section~\ref{meth:benchmarks}). Here, an iteration is defined as one proposed multi-spin move for our architecture versus one single-spin update evaluation for N-HOIM. Figure~\ref{fig:toy}e traces a representative execution trajectory for each solver reaching the optimum at its respective median iteration count. Critically, N-HOIM traverses numerous infeasible intermediate states, several of which display unpenalized objective values $E$ lower than the true feasible optimum. This comparative analysis highlights the foundational mechanics of our constraint-preserving framework, setting the stage for evaluations on larger-scale optimization problems.

\subsection{Graph matching results with third-order similarity terms}\label{res:matching}

\begin{figure}[!tbp]
\centering
\includegraphics[width=\linewidth]{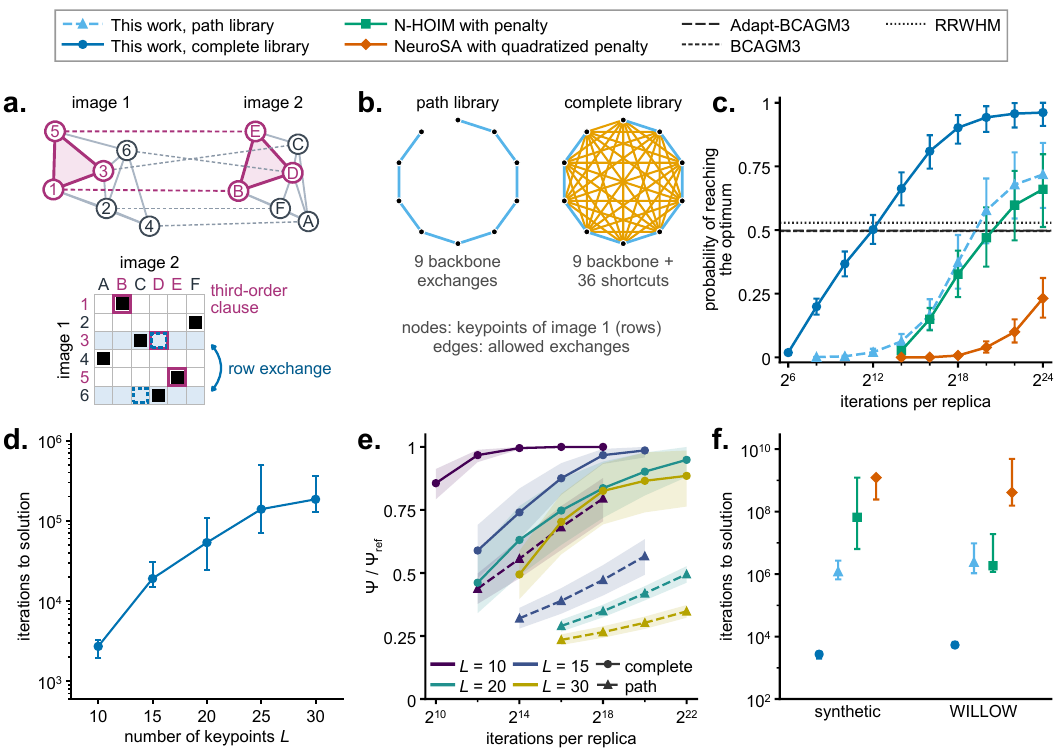}
\caption{\textbf{Graph matching with third-order (triangle) similarity terms.}
\textbf{a}, Schematic instance with $L=6$ keypoints per image, 1--6 in image 1 and A--F in image 2, each image with its own graph (grey lines). Dashed lines: the current assignment. Reddish purple: the triangles 1--3--5 and B--D--E, whose term is one third-order clause that contributes to the objective only when 1, 3 and 5 are assigned to B, D and E. Assignment matrix (rows: image 1, columns: image 2): black cells mark the current assignment, one per row and column. Outlined: the three cells of the clause, which is off because cell (3, D) is empty. Blue: an exchange of rows 3 and 6, which keeps a permutation and turns the clause on.
\textbf{b}, Path library of the $L-1$ exchanges between consecutive rows (backbone, light blue) and complete library of all $L(L-1)/2$ exchanges (backbone with all shortcuts, orange).
\textbf{c}, Probability that a replica reaches the certified optimum versus iterations per replica on 40 WILLOW-ObjectClass\cite{Cho2013} pairs ($L=10$), for this work with the backbone alone (path library) and with all shortcuts (complete library), N-HOIM with a penalty and NeuroSA with a quadratized penalty. Each method uses the settings selected on separate training pairs, the same at every budget. One iteration is one proposed spin transition for this work and one single-spin update test for the penalty machines. Black lines: probability that one restart of each domain solver reaches the certified optimum (line styles in the key). Method symbols: key at top (also for \textbf{f}).
\textbf{d}, Iterations to solution (Methods Section~\ref{meth:problems}) of this work with all shortcuts versus the number of keypoints $L$ on synthetic instances, with the schedule selected for each size on separate training instances. Target: $\Psi\geq0.98\,\Psi_\mathrm{ref}$.
\textbf{e}, Best objective found, $\Psi/\Psi_\mathrm{ref}$ (mean over replicas and over 20 synthetic instances), versus iterations per replica for $L=10$, 15, 20 and 30 (colors). Solid lines: complete library. Dashed lines: path library. Schedule: automatically calibrated (Methods Section~\ref{meth:benchmarks}). Each $L$ is shown up to the largest budget run with the complete library.
\textbf{f}, Iterations to solution at $L=10$ on 20 synthetic instances and 40 WILLOW pairs. Target: the certified optimum. Settings as in \textbf{c}, selected on the training instances of each problem set.
Replicas: 256 (64 for NeuroSA on WILLOW). $\Psi_\mathrm{ref}$: best value found by three domain solvers (certified optimum at $L=10$). Error bars and bands: 95\% bootstrap confidence intervals over instances.}\label{fig:matching}
\end{figure}

For this specific application, we demonstrate the architecture acting as an optimizer, annealing it to a low temperature to identify a low-energy feasible state. Graph matching maps $L$ keypoints from a source image onto $L$ keypoints of a target image. Incorporating third-order similarity terms maximizes the geometric resemblance between triangles of keypoints in the first image and their corresponding assigned triangles in the second\cite{Duchenne2011,Lee2011}. Unlike pairwise distances, the interior angles of a triangle remain invariant under translation, rotation, and uniform scaling. Consequently, triangular structures provide crucial geometric constraints when local keypoint appearance is ambiguous, such as in regions with repetitive patterns, uniform textures, or low feature distinctiveness\cite{Duchenne2011}. Establishing such correspondences between image feature sets represents a foundational step in feature tracking, image classification, visual retrieval, object detection, shape matching, and wide-baseline stereo vision\cite{Duchenne2011}. The objective function sums the angle-similarity scores $\psi_{ijl;abc}$ across pairs of triangles (Supplementary Section~\ref{supp:matching-formulation}):
\begin{equation}
\begin{aligned}
\max_{\mathbf{x}\in\{0,1\}^{L\times L}}\quad&\Psi(\mathbf{x})=\sum_{(i,j,l;\,a,b,c)}\psi_{ijl;abc}\,x_{ia}\,x_{jb}\,x_{lc},\qquad x_{ia}=\frac{1+s_{ia}}{2},\\
\text{subject to}\quad&\sum_{a=1}^{L}x_{ia}=1\ \ \forall i,\qquad\sum_{i=1}^{L}x_{ia}=1\ \ \forall a,
\end{aligned}
\label{eq:res-matching-objective}
\end{equation}
where the binary variable $x_{ia}=1$ indicates that keypoint $i$ in Image 1 is assigned to keypoint $a$ in Image 2. The spins $s_{ia}$ form an assignment matrix in which rows correspond to keypoints in Image 1 and columns correspond to keypoints in Image 2. Because each objective term multiplies three assignment variables, the resulting energy $E=-\Psi$ comprises clauses of up to third order. The constraints give the assignment matrix a one-to-one permutation structure (Fig.~\ref{fig:matching}a). Maximizing this third-order objective subject to permutation constraints is NP-hard\cite{Nguyen2017}. A row-exchange operation swaps the assigned targets of two keypoints, thereby preserving the permutation constraint by construction. Whether a row exchange alters a specific clause depends on the current assignments of those two keypoints; hence, every exchange requires state-dependent switches and its own set of auxiliary clauses (Supplementary Section~\ref{supp:closure}). Here, we compare two move libraries (Fig.~\ref{fig:matching}b). The path library, consisting of the $L-1$ exchanges between adjacent rows, serves as the minimal backbone that guarantees connectivity across all $L!$ feasible permutations, as adjacent swaps can compose into any arbitrary permutation. Any non-adjacent exchange acts as a shortcut move; the complete library, combining the backbone with all available shortcuts, encompasses all $L(L-1)/2$ possible row exchanges.

We first investigate whether restricting the search space to the feasible set offers a performance advantage over neuromorphic architectures that enforce constraints via penalty terms. All evaluated architectures, including ours, are compared on an iteration basis rather than by execution time. Here, an iteration is one proposed update: one proposed move for our machine and one single-spin update test for N-HOIM and NeuroSA. This metric evaluates the underlying search algorithms independently of hardware implementations, emulation platforms, or execution speeds. We benchmark our machine against two penalty-enforcing architectures—N-HOIM and NeuroSA—on 40 photograph pairs from the WILLOW-ObjectClass dataset\cite{Cho2013} with $L=10$ keypoints, where ground-truth global optima are certified by enumerating all $10!$ feasible permutations. NeuroSA first reduces third-order terms to pairwise form prior to execution (Supplementary Section~\ref{supp:matching-formulation}). Utilizing all shortcuts (the complete library), a single replica of our machine reaches the certified optimum within the maximum budget of $2^{24}$ iterations with a probability of 0.96, compared to 0.72 for the backbone alone (path library), 0.66 for N-HOIM, and 0.23 for NeuroSA (Fig.~\ref{fig:matching}c). At a reduced budget of $2^{12}$ iterations, the complete library achieves the optimum with a probability of 0.50, matching single-restart performance of domain-specific hypergraph-matching solvers BCAGM3 and Adapt-BCAGM3\cite{Nguyen2017} (0.50) and closely tracking RRWHM\cite{Lee2011} (0.53). Each algorithm employs schedule parameters—and penalty strengths, where applicable—tuned to minimize median iterations to solution on independent training instances (Methods Sections~\ref{meth:benchmarks} and~\ref{meth:problems}; Supplementary Section~\ref{supp:settings} and Supplementary Table~\ref{tab:supp-settings}). While hyperparameter selection determines an operating point, Fig.~\ref{fig:matching}c displays the full success probability distribution across all iteration budgets. To evaluate scaling behavior on larger problem instances, we generated synthetic benchmarks. Each instance distributes $L$ keypoints uniformly at random within a unit square and maps them to a target image via random rotation, uniform scaling, and additive Gaussian noise with standard deviation $\sigma=0.05$ (Methods Section~\ref{meth:problems}; Supplementary Section~\ref{supp:matching-formulation}). Beyond $L=10$, global optima can no longer be certified by brute-force enumeration; we therefore define a reference objective value $\Psi_\mathrm{ref}$ as the best objective score identified across three baseline hypergraph-matching solvers. The complete library continues to reach the target objective threshold ($\Psi \geq 0.98\,\Psi_\mathrm{ref}$) up to $L=30$ (Fig.~\ref{fig:matching}d). With size-calibrated annealing schedules, its median iterations to solution scale from $2.7\times10^{3}$ at $L=10$ to $1.9\times10^{5}$ at $L=30$.

Shortcut moves become increasingly critical as problem scale $L$ expands (Fig.~\ref{fig:matching}e). Comparing both move libraries under an automatically calibrated annealing schedule (Methods Section~\ref{meth:benchmarks}) at the maximum budget evaluated for the complete library shows clear performance divergence (Fig.~\ref{fig:matching}e). While the path library was evaluated for at least as many iterations, the normalized mean best objective achieved by the complete library drops only from 0.99 to 0.89 of $\Psi_\mathrm{ref}$ between $L=15$ and $L=30$, whereas the path library degrades sharply from 0.57 to 0.35. Because this performance gap widens with $L$, we subsequently analyze the precise topological and energetic mechanisms by which shortcut moves reduce iterations to solution. At $L=10$ (Fig.~\ref{fig:matching}f), penalty-enforcing architectures require significantly more iterations to solution than our machine operating with the complete library. This overhead ranges from a factor of 340 (N-HOIM on WILLOW) to $4.5\times10^{5}$ (NeuroSA on synthetic instances)—representing an efficiency gain of two to more than five orders of magnitude. On WILLOW, the complete library requires a median of $5.4\times10^{3}$ iterations to reach the solution. This performance disparity stems from search space geometry: because any single-spin flip from a valid permutation transitions into an infeasible state, penalty-based architectures can traverse between valid assignments only via high-energy, infeasible paths. Within our framework, the restricted path library requires 443 times more iterations to solution than the complete library on synthetic benchmarks. A higher objective value $\Psi$ is meaningful only if it directly translates into superior matching accuracy. 

On the WILLOW dataset, the top-performing replica of the complete library discovers the certified optimum on every image pair within an iteration budget of $2^{8}$ or greater (Supplementary Fig.~\ref{fig:supp-matching}a). At this solution quality, keypoint correspondences match human expert annotations on 59.5\% of keypoints, demonstrating that matching accuracy is constrained by the formulation of the objective function itself rather than search convergence. We verified these empirical findings through two control experiments. First, doubling keypoint spatial deformation ($\sigma=0.10$) decreases the target hit rate at $L=30$, while performance remains stable at $L=20$ and $L=25$ (Supplementary Fig.~\ref{fig:supp-matching}b, c). At $L=30$, the fraction of replicas attaining $0.98\,\Psi_\mathrm{ref}$ within $2^{22}$ iterations decreases from 0.80 to 0.57. Second, validation across six synthetic instances confirms that optimization trajectories driven by the complete library result genuinely from thermal annealing dynamics rather than uniform state space exploration (Supplementary Section~\ref{supp:control} and Supplementary Fig.~\ref{fig:supp-control}). Under identical candidate move sequences, iteration budgets, and random seeds, no replicas of a random walk (accepting all proposed transitions) achieved more than 0.27 of $\Psi_\mathrm{ref}$ on any instance. Similarly, isothermal runs maintained at the initial high temperature averaged less than half of $\Psi_\mathrm{ref}$ across all test cases.

\subsection{Optimizing shortcuts to limit the number of \textit{move} neurons}\label{res:library}

In graph matching, incorporating all shortcut moves into the path backbone reduces the required iteration count by more than two orders of magnitude. However, the complete library scales rapidly: each of its $L(L-1)/2$ row exchanges demands its own auxiliary clauses, which outnumber the original objective clauses for all keypoint counts $L \geq 40$ (Supplementary Section~\ref{supp:resources} and Supplementary Fig.~\ref{fig:supp-size}). Consequently, we investigate two key questions: which subset of shortcuts should be prioritized when the hardware allocation of \textit{move} neurons is constrained, and whether loading shortcuts dynamically in sequential batches can substitute for storing the complete library simultaneously. Under uniform proposal probabilities, expanding the move library would progressively dilute the backbone share—the fraction of proposals allocated to backbone moves—reducing it to just $2/L$ for the complete library. To isolate the intrinsic topological impact of the move library from this proposal frequency bias, we fix the overall backbone share by assigning each move a selection probability proportional to an allocated weight (Methods Eq.~\eqref{eq:proposal}). Specifically, when $K$ shortcuts are introduced alongside the $L-1$ backbone moves, each shortcut is assigned a weight of 1, while each backbone move receives a scaled weight of $w_\mathrm{B}=\max\left(1, \mathrm{round}[K/(L-1)]\right)$. Under this weighting scheme, the backbone retains a total proposal fraction of $(L-1)w_\mathrm{B}/\left[(L-1)w_\mathrm{B}+K\right]$, maintaining a stable selection share of approximately one half whenever $K \geq L$ shortcuts are added (Supplementary Section~\ref{supp:librarysupp}).

\begin{figure}[!tbp]
\centering
\includegraphics[width=\linewidth]{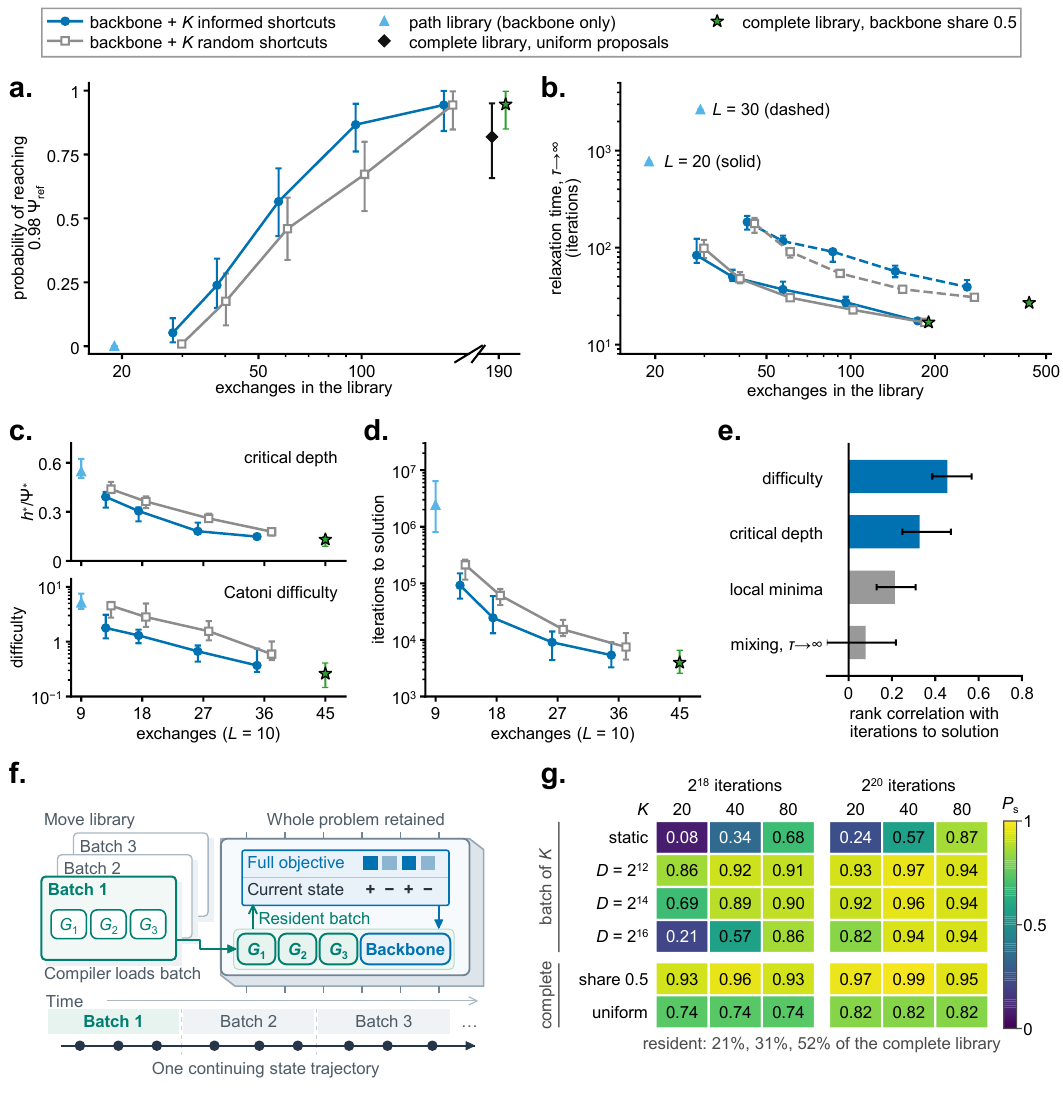}
\caption{\textbf{Choosing and batching shortcuts in graph matching.}
\textbf{a}, Probability that a replica reaches $0.98\,\Psi_\mathrm{ref}$ within $2^{20}$ iterations (mean over instances) versus the number of exchanges in the move library ($L=20$). Libraries: the backbone of 19 path exchanges plus $K=10$--160 informed or random shortcuts. References: the backbone alone (path library) and the backbone with all 171 shortcuts (complete library of 190 exchanges, after the axis break) with uniform proposals or with backbone share 0.5.
\textbf{b}, Relaxation time at infinite temperature of the libraries of \textbf{a} ($L=20$, solid lines) and of the same construction at $L=30$ (dashed lines).
\textbf{c}--\textbf{e}, Exact energy landscapes at $L=10$. Libraries: the backbone of 9 path exchanges plus $K=4$, 9, 18 or 27 informed or random shortcuts, and the backbone with all 36 shortcuts (complete library). Random shortcuts: median over ten draws per instance.
\textbf{c}, Critical depth $h^\ast$, in units of the optimum $\Psi^\ast$, and Catoni difficulty.
\textbf{d}, Iterations to solution (Methods Section~\ref{meth:problems}). Target: $\Psi^\ast$. Budgets: $2^6$ to $2^{16}$ iterations ($2^{20}$ for the backbone alone).
\textbf{e}, Spearman rank correlation between each property of the landscape and the iterations needed to reach $\Psi^\ast$ with 99\% probability at one common budget (Supplementary Section~\ref{supp:shortcut-theory}), among the 11 move libraries with $K$ shortcuts of one instance. Mixing: relaxation time at infinite temperature. Bars: median over the 80 combinations of instance and $K$.
\textbf{f}, Batching (schematic).
\textbf{g}, Probability $P_\mathrm{s}$ that a replica reaches $0.98\,\Psi_\mathrm{ref}$ within $2^{18}$ (left) and $2^{20}$ (right) iterations (mean over instances, $L=20$). Columns: number $K$ of informed shortcuts loaded beside the backbone. Row static: the first batch only. Rows $D$: the batch is replaced every $D$ iterations. Last two rows: complete library with backbone share 0.5 and with uniform proposals. Batched minus static, with 95\% confidence intervals: Supplementary Table~\ref{tab:supp-dwell}.
All panels: 20 synthetic test instances. Annealing: 256 replicas per instance, automatically calibrated schedule (Methods Section~\ref{meth:benchmarks}). Error bars: 95\% bootstrap confidence intervals over instances (of the median in \textbf{b}--\textbf{e}). $\Psi_\mathrm{ref}$: best value found by three domain solvers.}\label{fig:library}
\end{figure}

We evaluate two distinct strategies for selecting the $K$ non-backbone shortcut exchanges: selecting shortcuts based on the objective function alone prior to searching (informed shortcuts), or drawing them at random (random shortcuts). Two keypoints whose similarity terms concentrate on identical candidate targets are likely to exchange those targets within high-objective configurations. For a given keypoint $i$, let $\psi_{ia}$ denote the sum of all similarity weights in Eq.~\eqref{eq:res-matching-objective} containing $x_{ia}$, and let $\phi_i(a)=\psi_{ia}/\sum_b\psi_{ib}$ represent the resulting target probability distribution. Informed shortcuts select the $K$ keypoint pairs whose distributions exhibit the greatest overlap, measured by $\sum_a\min[\phi_i(a),\phi_j(a)]$ (Supplementary Section~\ref{supp:matching-formulation}). Conversely, random shortcuts select $K$ pairs drawn uniformly at random per instance (median over ten independent draws at $L=10$). At $L=20$, the backbone augmented with $K$ informed shortcuts attains $0.98\,\Psi_\mathrm{ref}$ within $2^{20}$ iterations more frequently than when augmented with $K$ random shortcuts across all tested sizes $K \in [10, 80]$ (Fig.~\ref{fig:library}a). At $K=80$, the respective success probabilities are 0.87 for informed shortcuts and 0.67 for random shortcuts, whereas no trial using the backbone alone achieves this threshold. With 160 shortcuts, both informed and random selection match the success probability of the complete library (backbone with all 171 shortcuts) under a fixed backbone share of 0.5. At $L=30$, using an automatically calibrated annealing schedule, informed selection yields higher success rates once the move library incorporates at least one-third of the complete library's exchanges (Supplementary Fig.~\ref{fig:supp-library}a).

We next analyze why shortcuts accelerate convergence, and why informed selection offers a distinct advantage. Introducing shortcuts shortens state-space path lengths; however, structural path reduction alone does not guarantee reduced annealing times (Supplementary Section~\ref{supp:shortcut-theory}). High-temperature annealing first mixes across the feasible state space, after which the system must escape local minima as the temperature decreases. These two distinct regimes depend on different properties of the move library. At high temperatures, where nearly all proposed transitions are accepted, an exchange library shuffles targets among keypoints via a random walk on its underlying exchange graph, where vertices correspond to keypoints and edges denote exchange operations. The characteristic relaxation time of this process measures the iteration count over which the slowest-decaying correlation with the initial configuration decays by a factor of $\mathrm{e}$. Mathematically, this relaxation time equals the inverse spectral gap of the exchange-graph Laplacian, weighted by proposal probabilities\cite{Caputo2010}. The path backbone alone shifts a target by at most one position per accepted spin transition, causing its relaxation time to scale as $O(L^3)$. In contrast, a shortcut between distant keypoints transfers targets across the network in a single move step. Consequently, the complete library (with a backbone share of 0.5) reduces relaxation times 46-fold at $L=20$ (from 772 to 17 iterations) and 98-fold at $L=30$ (from 2,647 to 27 iterations; Fig.~\ref{fig:library}b). Even ten random shortcuts decrease relaxation time eightfold at $L=20$. While informed shortcuts enable significantly faster relaxation than the path backbone alone, their mixing rates are up to 1.7 times slower than those of an equivalent number of random shortcuts (Supplementary Section~\ref{supp:shortcut-theory}). Rapid mixing thus accounts for the general benefit of shortcut insertion, but it does not explain the specific performance advantage of informed shortcuts. We therefore examine low-temperature dynamics. At a low temperature $\tau$, the characteristic escape time from a local minimum of depth $h$ scales as $\mathrm{e}^{h/\tau}$, where depth represents the minimal energy barrier that any escape trajectory to a lower-energy state must overcome. The critical depth $h^\ast$—defined as the maximum depth across all non-optimal local minima—governs the maximum cooling rate allowed for asymptotic convergence to the global optimum\cite{Hajek1988}. Furthermore, the Catoni difficulty factor—the maximum ratio between a non-optimal local minimum's depth and its energy excess above the optimum—determines the maximum convergence rate achievable within a finite iteration budget\cite{Catoni1992}. A shortcut reduces a local minimum's depth only if one of its constituent spin transitions bypasses the highest-energy state along the optimal escape path. Consequently, barrier suppression depends explicitly on the objective energy landscape rather than on the move graph topology alone (Fig.~\ref{fig:concept}e; Supplementary Section~\ref{supp:shortcut-theory}). For instance, a configuration can become trapped in a local minimum when two keypoints with overlapping similarity distributions are each assigned the target that yields a higher objective value for the other. Exchanging their targets directly bypasses the high-energy intermediate states required by single-step path moves. By enumerating all $10!$ configurations for $L=10$ benchmark instances, we extracted every local minimum and evaluated its corresponding depth across move libraries (Supplementary Section~\ref{supp:shortcut-theory}). 

While informed and random shortcuts eliminate comparable numbers of local minima, informed shortcuts suppress the critical depth $h^\ast$ more effectively and reduce the Catoni difficulty by a larger margin (Fig.~\ref{fig:library}c). For a fixed shortcut count $K$, the median Catoni difficulty under informed selection ranges between 0.39 and 0.62 times that of random selection. Consistent with their lower energy barriers, informed shortcuts require substantially fewer iterations to achieve solution convergence (Fig.~\ref{fig:library}d). At $L=10$, under an automatically calibrated schedule targeting certified global optima, informed shortcuts require 0.4 to 0.7 times as many iterations to solution as random shortcuts across all tested move-library sizes. Likewise, when employing annealing schedules optimized independently for each move library on separate training sets, informed shortcuts consistently outperform random shortcuts across all shortcut capacities (Supplementary Table~\ref{tab:supp-its10}). 

We compare move libraries of equal size to identify which structural property correlates most strongly with the iteration count required to reach a solution. For each $L=10$ problem instance and each shortcut capacity $K$, we evaluated 11 distinct move libraries (one informed and ten random). Across these configurations, the iteration count required by a given library on a specific instance (Fig.~\ref{fig:library}e) increases with Catoni difficulty (median Spearman rank correlation $\rho_s = 0.46$) and critical depth ($\rho_s = 0.33$). It correlates less strongly with the total number of local minima ($\rho_s = 0.21$) and exhibits only a weak correlation with the relaxation time at infinite temperature ($\rho_s = 0.08$). These empirical findings highlight two primary mechanisms by which shortcut moves accelerate solution search. First, any shortcut move bridges distant state-space assignments, thereby accelerating high-temperature mixing. Second, among move libraries of identical size, those whose shortcuts selectively lower the deepest energy barriers crossed during cooling require significantly fewer iterations. Informed shortcuts—selected directly from objective function couplings—suppress these energy barriers more effectively than random shortcuts, despite exhibiting slightly slower high-temperature mixing rates (Supplementary Section~\ref{supp:shortcut-theory}).

We next address the second question: whether dynamically loading shortcut moves in sequential batches can substitute for maintaining a fully resident complete library. When hardware constraints prevent storing all shortcuts simultaneously, the host classical core can stream shortcuts sequentially in batches alongside the resident path backbone. Because objective clauses and current state vectors persist across reconfigurations, a single optimization trajectory proceeds continuously across batch transitions (Fig.~\ref{fig:library}f). For $L=20$ instances, we ranked all 171 candidate shortcuts by overlap score and partitioned them into consecutive batches of size $K$. A static library retains the top batch—comprising the $K$ informed shortcuts with highest overlap scores—loaded for the entire duration of the run. Conversely, a batched library rotates through these batches sequentially, maintaining each for a fixed replacement interval of $D$ iterations. Each loading phase reinitializes the auxiliary clauses, switches, and input states of the newly introduced moves from the current system configuration. Static and batched libraries sharing the same concurrent shortcut capacity (as well as the complete library operating with a backbone share of half) employ identical annealing schedules calibrated on the backbone augmented by the initial batch. Only the complete library operating under uniform proposal probabilities is calibrated independently (Supplementary Section~\ref{supp:librarysupp}). Dynamic batching consistently achieves the target objective threshold ($0.98\,\Psi_\mathrm{ref}$) more frequently than the static baseline across all evaluated parameter combinations, with improvements exceeding sampling uncertainty margins (Fig.~\ref{fig:library}g and Supplementary Table~\ref{tab:supp-dwell}). Evaluated parameter sets spanned all combinations of concurrent shortcut capacity ($K \in \{20, 40, 80\}$), iteration budget ($2^{18}$ or $2^{20}$ iterations), and replacement interval ($D \in \{2^{12}, 2^{14}, 2^{16}\}$). Provided the total budget permits executing multiple batch rotations, dynamic batching recovers most of the performance gain that the complete library (with backbone share 0.5) achieves over static designs (Supplementary Section~\ref{supp:librarysupp}). The performance benefit of batching is most pronounced at $K=20$ loaded shortcuts, which—combined with the backbone—represent only 39 of the 190 total possible exchanges. At this capacity, the probability of reaching $0.98\,\Psi_\mathrm{ref}$ within $2^{20}$ iterations increases from 0.24 under a static library to 0.93 when rotating batches every $D=2^{12}$ iterations, and to 0.82 at $D=2^{16}$ iterations. Note that the hardware overhead associated with batch reloading depends on specific platform implementation and is omitted from iteration count comparisons.

\FloatBarrier
\subsection{Surface-code decoding on a field-programmable gate array}\label{res:surface}

For the second application, we demonstrate the same architecture operating as a probabilistic sampler. At a fixed temperature, the fraction of iterations spent within a given set of feasible states estimates the Boltzmann probability of that set which is useful for decoding quantum error-correcting codes. The rotated surface code of distance $d$ encodes one logical qubit into $d^2$ data qubits. Each check operation measures a stabilizer, defined as the product of $X$ or $Z$ Pauli operators across two or four neighboring qubits. The resulting measurement syndrome identifies which checks are violated, and the decoder must determine an appropriate error correction to restore the system\cite{Bombin2007,Fowler2012}. Corrections that differ only by stabilizer operators belong to the same logical class and exert an identical effect on the underlying encoded information; however, only one logical class accurately neutralizes the true physical error (Supplementary Section~\ref{supp:surface-formulation}). Maximum-likelihood (ML) decoding selects the logical class possessing the highest aggregate probability\cite{Dennis2002}, which can differ from the class containing the single most probable error string. For general stabilizer codes, identifying the most likely individual error is NP-hard, while calculating the most likely logical class is \#P-complete\cite{Iyer2015}. Consequently, most practical decoders search for a single representative error configuration—such as minimum-weight perfect matching\cite{Dennis2002}, union-find\cite{Delfosse2021}, and belief propagation with ordered-statistics decoding (BP-OSD)\cite{Panteleev2021,Roffe2020}. Alternatively, tensor networks approximate logical class probabilities\cite{Bravyi2014}, though their computational cost scales steeply with both code distance and approximation accuracy. While thermal sampling at a fixed temperature $\tau=1$ directly estimates class probabilities based on state visitation frequencies\cite{Wootton2012}, software implementations are inherently bottlenecked by sequential move evaluations. To overcome this latency constraint, we implemented our parallel neuromorphic architecture directly on the field-programmable gate array (FPGA) of an RFSoC 4x2 board.

\begin{figure}[!tbp]
\centering
\includegraphics[width=\linewidth]{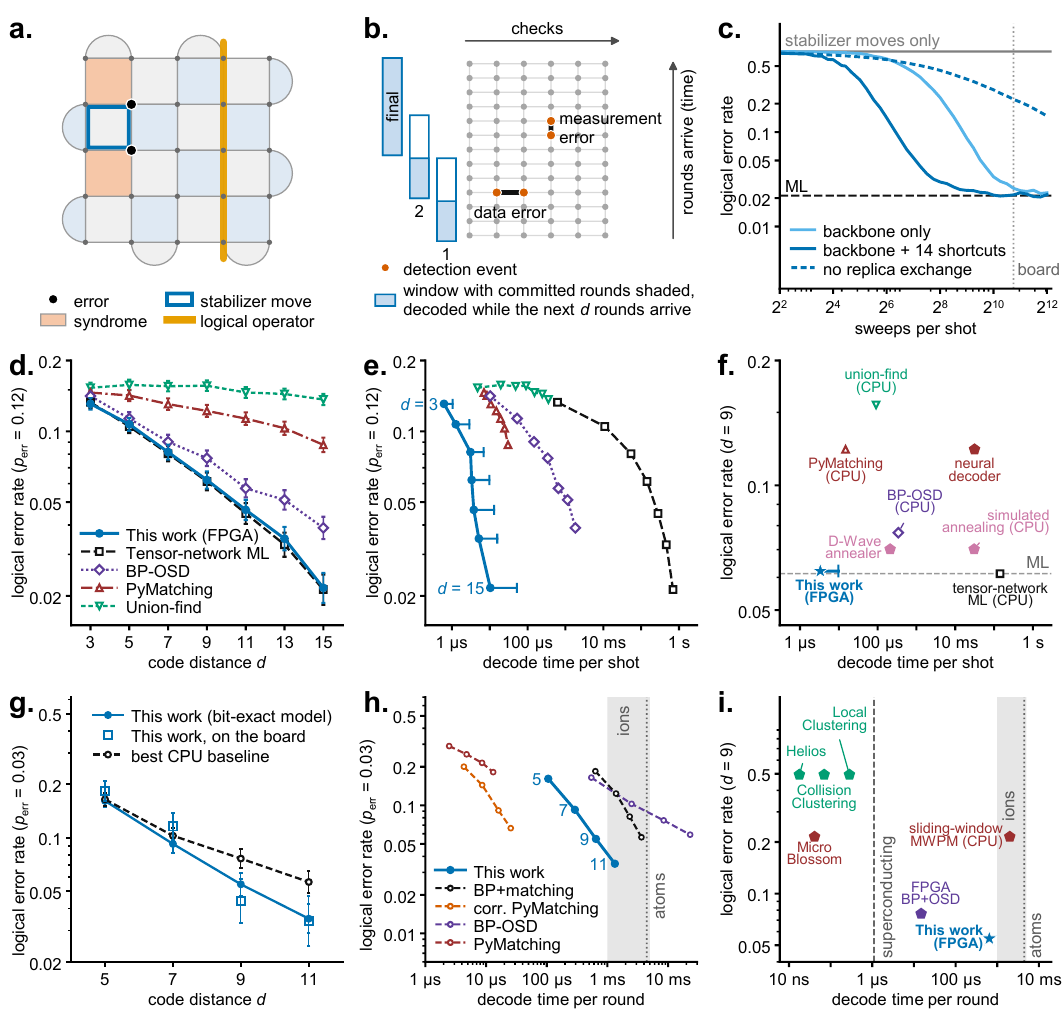}
\caption{\textbf{Surface-code decoding on the FPGA, one syndrome per shot and in real time.}
Rotated surface code on the field-programmable gate array (FPGA) of an RFSoC 4x2 board, one design per code distance $d$. Logical error rates (LERs) of this work: bit-exact software model of each design. All decoders that we ran use the same shots. CPU decoders: one core of an Intel Xeon Gold 6548Y+. Error bars: 95\% Wilson confidence intervals. \textbf{c}--\textbf{f}: one syndrome per shot. \textbf{g}--\textbf{i}: real time.
\textbf{a}, One syndrome per shot (schematic, $d=5$, depolarizing noise, measurement without error). A logical operator keeps the syndrome and changes the class.
\textbf{b}, Real time (schematic, $d=3$, phenomenological noise): checks of one error type along one axis, rounds along the other. $4d$ noisy rounds and one round without error. Windows of $2d$ rounds commit their first $d$, the final window all.
\textbf{c}, LER versus sweeps per shot, $d=15$, $p_\mathrm{err}=0.12$. Backbone: stabilizer moves and one logical operator per error type. Shortcuts: the other $d-1=14$ per error type. No replica exchange: all shortcuts, one replica at $\tau=1$. Black dashed line: maximum-likelihood (ML) LER. Dotted line: board budget.
\textbf{d}, LER versus $d$, $p_\mathrm{err}=0.12$. BP-OSD: belief propagation with ordered-statistics decoding.
\textbf{e}, LER versus decode time per shot, $d=3$--15, $p_\mathrm{err}=0.12$. This work: mean with early stopping (dots) to the fixed budget (bar ends). CPU markers in \textbf{e}, \textbf{f}: as in \textbf{d}.
\textbf{f}, Decoders of one syndrome per shot, $d=9$, $p_\mathrm{err}=0.12$. Pentagons: published times\cite{Varsamopoulos2020,Sakashita2026,Guo2025}, plotted at the LER of their kind of algorithm on our shots (pink: lowest-energy correction, red: matching).
\textbf{g}, LER per shot (shots as in \textbf{b}) versus $d$, $p_\mathrm{err}=0.03$. Squares: the board on 1,000 other shots. Best CPU baseline: lowest LER of correlated PyMatching, BP+matching (belief propagation with matching) and BP-OSD, in the same windows.
\textbf{h}, LER versus decode time per round (window time divided by $d$), $p_\mathrm{err}=0.03$. Labels: $d$. Shaded: trapped-ion rounds of 1--5\,ms\cite{Ye2026}. Dotted line: the 4.45\,ms neutral-atom round\cite{Bluvstein2026}.
\textbf{i}, Real-time decoders, $d=9$, $p_\mathrm{err}=0.03$. Pentagons: published times per round\cite{Liyanage2024,Barber2025,Ziad2025,Wu2025,Bascones2025,Skoric2023}, at the LER of their algorithm on our shots (green: union-find, red: matching, purple: BP-OSD). Times measured at lower $p_\mathrm{err}$ favor them. BP+OSD: worst-case time. MWPM: minimum-weight perfect matching. Shading, dotted line: as in \textbf{h}. Dashed line: the 1.1\,$\mu$s superconducting round\cite{GoogleQAI2025}.}\label{fig:surface}
\end{figure}

We evaluate decoding performance across two distinct experimental settings. A shot is defined as a single sampled error pattern paired with its corresponding measurement syndrome. In the first setting, each shot contains a single syndrome measured without readout error (code-capacity noise model; Fig.~\ref{fig:surface}a and Supplementary Section~\ref{supp:oneshot}). Under this model, each data qubit undergoes an $X$, $Y$, or $Z$ error with equal probability $p_\mathrm{err}/3$ (depolarizing noise), where $p_\mathrm{err}$ represents the physical error rate. In the second setting, syndrome checks are measured continuously over repeated measurement rounds prior to logical qubit readout (quantum memory experiment; Fig.~\ref{fig:surface}b). Here, each measurement round introduces additional depolarizing errors, and each check outcome is itself corrupted with probability $p_\mathrm{err}$ (a depolarizing variant of phenomenological noise\cite{Dennis2002}; Supplementary Section~\ref{supp:realtime}). In this operating regime, the decoder must process syndromes continuously at or above the physical measurement rate (real-time decoding). A single syndrome round requires approximately one microsecond on superconducting qubit architectures\cite{GoogleQAI2025}, whereas neutral-atom and trapped-ion processors operate or are projected to operate on millisecond timescales\cite{Bluvstein2026,Ye2026}.

We represent an error correction using two spins per qubit $i$, $s^X_i$ and $s^Z_i$, which take the value $-1$ when the correction applies an $X$ or $Z$ component to the qubit, respectively. The energy function corresponds to the negative log-probability of the correction (Methods Section~\ref{meth:problems}; Supplementary Section~\ref{supp:surface-formulation}). Under a depolarizing noise model, this energy equals $\omega_3$ times the total number of physical errors across the qubits, up to an additive constant, ensuring that a $Y$ error (comprising both $X$ and $Z$ errors) is counted only once, and the measured syndrome restricts the corrections to the feasible set $\Omega$:
\begin{equation}
\begin{aligned}
E&=E_\mathrm{const}-\frac{\omega_3}{4}\sum_{i=1}^{d^2}\left(s^X_i+s^Z_i+s^X_is^Z_i\right),\qquad \omega_3=\ln\frac{3(1-p_\mathrm{err})}{p_\mathrm{err}},\\
\Omega&=\Bigl\{\svec:\ \textstyle\prod_{i}(s^X_i)^{(H_Z)_{ji}}=(-1)^{(\mathrm{syn}_Z)_j},\ \ \prod_{i}(s^Z_i)^{(H_X)_{ji}}=(-1)^{(\mathrm{syn}_X)_j}\ \ \forall j\Bigr\}.
\end{aligned}
\label{eq:surface-energy}
\end{equation}
Here, $H_Z$ denotes the check matrix for $Z$ checks, where $(H_Z)_{ji}=1$ if check $j$ acts on qubit $i$, and $(\mathrm{syn}_Z)_j=1$ if check $j$ is violated; $H_X$ and $\mathrm{syn}_X$ are defined likewise for the $X$ checks. Whereas a conventional single-spin-flip Ising machine would incorporate each check into the energy function as a fourth-order penalty term (or second-order on code boundaries)\cite{Ahsan2026}, our move architecture enforces syndrome constraints natively by design (Fig.~\ref{fig:surface}a). Because every candidate move reverses a fixed set of spins, all switch variables remain constant and compilation introduces zero auxiliary clauses (Supplementary Section~\ref{supp:closure}). A stabilizer move flips the $X$ components of the qubits involved in one $X$ check, or the $Z$ components of those in one $Z$ check, thereby preserving both the measurement syndrome and the logical class. A logical operator move spanning a linear chain of $d$ qubits across the code lattice, likewise preserves the syndrome but transitions the system between logical classes in a single {\it move} step. Stabilizer moves combined with one representative logical operator per error type connect all feasible corrections to form the {\it move} backbone. The remaining $d-1$ logical operators per error type, positioned along alternative spatial paths, act as shortcut moves. Unless noted otherwise, the complete {\it move} library consists of the backbone augmented with all shortcuts. For a single syndrome measurement per shot, the decoder samples state configurations at temperature $\tau=1$, where the Boltzmann weight $\mathrm{e}^{-E(\boldsymbol{\mu})}$ of a valid correction $\boldsymbol{\mu}$ is directly proportional to its conditional probability given the observed syndrome (Supplementary Section~\ref{supp:oneshot}). The aggregate probability of a logical class $\gamma$, containing the set of valid corrections $\Omega_\gamma \subseteq \Omega$, is given by
\begin{equation}
P(\gamma)=\frac{\sum_{\boldsymbol{\mu}\in\Omega_\gamma}\mathrm{e}^{-E(\boldsymbol{\mu})}}{\sum_{\boldsymbol{\mu}\in\Omega}\mathrm{e}^{-E(\boldsymbol{\mu})}}.
\label{eq:class-probability}
\end{equation}
The numerator of Eq.~\eqref{eq:class-probability} is equivalent to $\mathrm{e}^{-F_\gamma}$, where $F_\gamma$ represents the free energy of class $\gamma$; thus, maximum-likelihood (ML) decoding returns the logical class exhibiting the lowest free energy. Because no individual correction explicitly evaluates this partition sum, Metropolis dynamics executed at $\tau=1$ exploit the Boltzmann distribution on $\Omega$ as their stationary distribution (Methods Section~\ref{meth:benchmarks}). Consequently, the proportion of iterations spent within logical class $\gamma$ directly estimates $P(\gamma)$, allowing the decoder to output the most frequently visited logical class. When few qubits carry errors, thermal sampling at $\tau=1$ rarely induces logical class transitions, as applying a logical operator requires overcoming an energy barrier of up to $d\,\omega_3$. To overcome this barrier, the decoder operates multiple parallel instances of the machine (termed replicas) across a temperature spectrum ranging from $\tau=1$ to $\tau=10$. After every second Monte Carlo sweep (during which every candidate move is proposed once), adjacent replicas at temperatures $\tau < \tau'$ with energy states $E$ and $E'$ exchange operating temperatures with probability
\begin{equation}
\min\Bigl\{1,\exp\Bigl[\Bigl(\frac{1}{\tau}-\frac{1}{\tau'}\Bigr)(E-E')\Bigr]\Bigr\}
\label{eq:replica-exchange}
\end{equation}
via parallel tempering\cite{Hukushima1996}. This swap rule preserves the joint stationary Boltzmann distribution across all replicas, ensuring that the target replica at $\tau=1$ still samples logical classes according to the probability distribution in Eq.~\eqref{eq:class-probability}. High-temperature transitions between logical classes subsequently propagate down to the $\tau=1$ sampling thread through successive replica exchanges.

We employ graph coloring techniques\cite{Aadit2022} to evaluate and execute candidate moves concurrently on the FPGA architecture. Moves that share no common clauses and do not alter one another's switch variables fall into a single color class. All candidate moves within a color class are processed in parallel within a single clock cycle (Supplementary Section~\ref{supp:fpga} and Supplementary Figs.~\ref{fig:supp-arch1} and~\ref{fig:supp-arch2}). Under a single-syndrome-per-shot model, the complete move set partitions into six color classes across all code distances; thus, a complete Monte Carlo sweep requires six clock cycles, or eight clock cycles on average when accounting for replica exchange overhead. The FPGA implementation matches its high-level software model bit-for-bit; consequently, the logical error rates (LERs) reported for our decoder are derived directly from the software model (Methods Section~\ref{meth:fpga}). The backbone move library determines whether the decoder can access the correct logical class, whereas shortcut moves and replica exchange dictate the convergence rate (Fig.~\ref{fig:surface}c). At distance $d=15$ and physical error rate $p_\mathrm{err}=0.12$, using stabilizer moves alone fails to induce logical class transitions, leaving the LER locked at the baseline value of the initial error correction across all sweep budgets. Incorporating all shortcut moves allows the LER to reach the maximum-likelihood (ML) decoding limit of 0.021—as estimated by a benchmark tensor-network decoder—within approximately $2^{10}$ sweeps per shot, compared to roughly $2^{12}$ sweeps for the backbone library alone. 

Conversely, omitting replica exchange (operating a single sampling thread at $\tau=1$ with all shortcuts) yields a significantly degraded LER of 0.15 after $2^{12}$ sweeps. Because every candidate move is proposed once per sweep, introducing shortcut moves increases the frequency of class-change proposals; this comparison reflects both the structural path reduction and the increased proposal rate. For single-syndrome decoding at $p_\mathrm{err}=0.12$, our decoder achieves the ML-optimal LER across all tested distances from $d=3$ to $d=15$ (Fig.~\ref{fig:surface}d). Evaluated on identical syndrome shots at $d=15$, alternative decoders—BP-OSD, minimum-weight perfect matching (PyMatching\cite{Higgott2022,Higgott2025}), and union-find—incur 1.8, 4.1, and 6.3 times as many logical errors as our decoder, respectively (Methods Section~\ref{meth:problems}). Across the entire evaluated parameter space ($p_\mathrm{err}=0.08$--0.19 and $d=3$--15), our decoder produces only 0.7\% more total logical errors than ML decoding (Supplementary Fig.~\ref{fig:supp-dec}a and Section~\ref{supp:qec-results}). The fault-tolerance threshold represents the physical error rate below which scaling to larger code distances monotonically suppresses logical errors. Near this threshold ($p_\mathrm{err} \geq 0.14$), the fixed execution budget of each FPGA design does not fully reach the asymptotic ML limit; there, the decoder instead executes a larger budget of $2^{12}$ sweeps per shot for $d \geq 5$, requiring 0.09--0.13\,ms per shot (Supplementary Section~\ref{supp:oneshot}). Under this budget, the threshold crossing between $d=11$ and $d=15$ occurs at $p_\mathrm{err}=0.182$ (95\% confidence interval [CI]: 0.177--0.185) for our decoder, compared to 0.184 (95\% CI: 0.180--0.186) for ML decoding (Supplementary Section~\ref{supp:qec-results}). Within statistical sampling uncertainty, our decoder matches the ML threshold estimate. Both values lie slightly below published theoretical thresholds for this noise model, such as 0.189 for optimal decoding\cite{Bombin2012} and 0.187 for tensor-network decoding fitted across $d=9$--21\cite{Tuckett2018}, noting that finite-distance crossings provide an estimate of the asymptotic threshold.

The FPGA board decodes individual shots on microsecond timescales (Fig.~\ref{fig:surface}e): 1.1\,$\mu$s at $d=3$ and 52\,$\mu$s at $d=15$ under fixed-budget execution. Implementing an optional early-stopping criterion terminates sampling once a specific logical class count reaches a predefined threshold, reducing the mean decoding latency to 0.6--10\,$\mu$s. Early-stopping parameters were tuned on independent validation shots and shift the resulting LER in Fig.~\ref{fig:surface}d by at most 0.0014 (Supplementary Section~\ref{supp:oneshot}). Furthermore, the proportion of total samples allocated to the selected class provides a confidence metric for each decoding decision. Discarding the 10\% least confident decisions reduces the logical error rate 8-fold at $d=15$ (Supplementary Section~\ref{supp:board} and Supplementary Fig.~\ref{fig:supp-board}b). Among decoders operating in the single-syndrome regime, our architecture achieves the ML-optimal error rate thousands of times faster than tensor-network decoders and achieves lower latency than all alternative benchmarked decoders at $d=9$ (Fig.~\ref{fig:surface}f). At $d=9$ and $p_\mathrm{err}=0.12$, our decoder averages 3.3\,$\mu$s per shot with a worst-case latency of 9.8\,$\mu$s. By comparison, the tensor-network decoder in qecsim (executed single-threaded on a CPU) requires 0.14\,s per shot—a 15,000 to 43,000-fold latency overhead. The neural decoder\cite{Varsamopoulos2020} (evaluated on our noise model) and CPU-based simulated annealing\cite{Sakashita2026} require 31\,ms per shot at this distance. A D-Wave quantum annealer\cite{Guo2025} requires 0.21\,ms per shot (counting only interface communication and annealing duration). While hardware benchmarks were conducted under varying system conditions, these baseline differences are negligible compared to the orders-of-magnitude latency gap observed (Supplementary Section~\ref{supp:qec-results}). Each published decoder is plotted at the LER that its algorithmic class reaches on our shots. For quantum and classical annealers, this performance corresponds to finding the lowest-energy error configuration, which yields higher error rates than sampling the true maximum-likelihood logical class (Methods Section~\ref{meth:problems}; Supplementary Section~\ref{supp:qec-results}).

In real-time decoding, the spin variables explicitly index the measurement round $t$ (equivalent to time). The spins $s^X_{i,t}$ and $s^Z_{i,t}$ take the value $-1$ when the correction contains an $X$ or a $Z$ error on data qubit $i$ between rounds $t-1$ and $t$, which persists in all later rounds. This persistence is reflected in the same measurement outcome in round $t$ and all later rounds and any measurement error will produce a disagreement. The measurement error could also be corrected by introducing additional spins referred to as measurement-error spins $s^\mathrm{m}_{j,t}$.
Each measurement-error spin introduces a first-order energy clause, and the corrections must explain every detection event (Supplementary Section~\ref{supp:realtime}) as specified by modified constraints in $\Omega$:
\begin{equation}
\begin{aligned}
E&=E_\mathrm{const}-\frac{\omega_3}{4}\sum_{i,t}\left(s^X_{i,t}+s^Z_{i,t}+s^X_{i,t}s^Z_{i,t}\right)-\frac{\omega_\mathrm{m}}{2}\sum_{j,t}s^\mathrm{m}_{j,t},\qquad \omega_\mathrm{m}=\ln\frac{1-p_\mathrm{err}}{p_\mathrm{err}},\\
\Omega&=\Bigl\{\svec:\ s^\mathrm{m}_{j,t-1}\,s^\mathrm{m}_{j,t}\textstyle\prod_{i}(s^X_{i,t})^{(H_Z)_{ji}}=(-1)^{(\mathrm{syn}_Z)_{j,t}+(\mathrm{syn}_Z)_{j,t-1}},\\
&\hphantom{{}=\Bigl\{\svec:\ }s^\mathrm{m}_{j,t-1}\,s^\mathrm{m}_{j,t}\textstyle\prod_{i}(s^Z_{i,t})^{(H_X)_{ji}}=(-1)^{(\mathrm{syn}_X)_{j,t}+(\mathrm{syn}_X)_{j,t-1}}\ \ \forall j,t\Bigr\}.
\end{aligned}
\label{eq:realtime-energy}
\end{equation}
Here, $(\mathrm{syn}_Z)_{j,t}$ and $(\mathrm{syn}_X)_{j,t}$ equal 1 if check $j$ is violated in round $t$, so that the right-hand side is $-1$ at a detection event, a change of this outcome between rounds $t-1$ and $t$. Before the first round, all outcomes are 0 and $s^\mathrm{m}_{j,-1}=+1$, and the final round is measured without error ($s^\mathrm{m}_{j,t}=+1$). Stabilizer moves and logical operators operate within a single measurement round. Real-time decoding adds a third type of move, the time-like move, which links consecutive rounds. A time-like move reverses $s^X_{i,t}$ and $s^X_{i,t+1}$. The reversal changes the accumulated $X$ error on qubit $i$, $\prod_{t'\leq t}s^X_{i,t'}$, only in round $t$, for instance, by shifting an $X$ error from between rounds $t-1$ and $t$ to between rounds $t$ and $t+1$. This operation simultaneously flips $s^\mathrm{m}_{j,t}$ for the one or two $Z$ checks $j$ acting on qubit $i$, so that the correction still explains every detection event. The corresponding switch variables remain constant, so compilation adds no auxiliary clauses. Time-like moves for $Z$ errors are defined symmetrically by interchanging $X$ and $Z$ indices. The real-time decoder operates across sliding time windows of $2d$ rounds (Fig.~\ref{fig:surface}b), accepting detection events as input. Each window is decoded concurrently as the subsequent $d$ rounds are acquired, after which its initial $d$ rounds are committed to fix their final corrections. The succeeding window advances by $d$ rounds, initializing from the corrected syndrome of the newly committed correction at the boundary round\cite{Dennis2002,Skoric2023}. The terminal window commits all remaining rounds, culminating in a final error-free measurement round that concludes the memory experiment. While an active window remains open at its upper temporal boundary, time-like moves can continuously cancel any logical operator, making distinct logical classes indistinguishable within the open window (Supplementary Section~\ref{supp:realtime}). Consequently, each non-terminal window commits the lowest-energy state configuration identified across all parallel replicas, whereas only the final window commits the most visited logical class at $\tau=1$. A trial shot fails if the cumulative committed correction across all sliding windows differs from the true physical error string by a non-trivial logical operator; the logical error rate (LER) represents the proportion of failed shots. For direct benchmarking, four CPU decoders process identical shots across equivalent sliding windows: standard PyMatching, correlated PyMatching\cite{Fowler2013,Paler2023} (which accounts for correlations between $X$ and $Z$ errors), belief propagation combined with matching (BP+matching)\cite{Higgott2023}, and BP-OSD. At each distance $d$ and physical error rate $p_\mathrm{err}$, the best CPU baseline is defined as the best-performing decoder among correlated PyMatching, BP+matching, and BP-OSD.

Our real-time decoder produces fewer logical errors than the best CPU baseline across all evaluated code distances and physical error rates, with its comparative performance advantage widening as code distance increases (Fig.~\ref{fig:surface}g). Averaged across $p_\mathrm{err}=0.025$--0.04 (the physical error rate range evaluated at every distance), the best CPU baseline generates 1.03 times as many logical errors as our decoder at $d=5$, a relative error multiplier that increases monotonically at each step in distance up to 1.51 at $d=11$. All averaged relative ratios exhibit a 95\% confidence interval strictly greater than 1. Empirical LERs measured directly on the FPGA board across 1,000 independent evaluation shots per distance (represented by squares) match software simulation results within expected statistical sampling margins. On shots that played no part in choosing the number of sweeps, threshold crossing curves for $d=7$ and $d=9$ intersect at $p_\mathrm{err}=0.047$ for our decoder and at 0.046 for BP+matching (the highest crossing threshold among CPU baselines)—a difference well within statistical sampling uncertainty. At $d=11$, hardware logic resource constraints limit the FPGA design to 24 parallel replicas, whereas software simulations indicate that 32 replicas yield approximately a 10\% further reduction in logical errors. Operating with 24 replicas, the threshold crossing between $d=9$ and $d=11$ occurs at a slightly lower error rate for our decoder than for BP+matching. Consequently, at hardware-constrained replica capacities, our decoder suppresses the LER near the fault-tolerance threshold relative to CPU baselines without shifting the absolute threshold value upward.

The physical FPGA board maintains processing throughput matching 1\,ms syndrome measurement rounds up to $d=9$, and keeps pace with the 4.45\,ms round duration of neutral-atom processors at $d=11$ (Fig.~\ref{fig:surface}h). Throughput is evaluated by the average processing time per round, defined as total window decoding latency divided by the $d$ committed rounds per window. This metric scales from 0.11\,ms at $d=5$ to 0.65\,ms at $d=9$ and 1.33\,ms at $d=11$ (Supplementary Table~\ref{tab:supp-designs}). The finalized correction for any given measurement round is rendered exactly one window decoding interval after the final round of its window is received (5.8\,ms at $d=9$). Each window must complete commitment before the final round of the subsequent window arrives to ensure the new window initializes from its corrected syndrome. Operating with 1\,ms measurement rounds for $d=3$--9 and 4.45\,ms rounds at $d=11$—matching neutral-atom experimental regimes\cite{Bluvstein2026}—the board registered zero deadline overruns across 1,000 test shots per distance (Supplementary Fig.~\ref{fig:supp-board}c). Note that syndrome data were pre-stored on-chip and released via hardware timers; quantum-processor-to-FPGA link latency is excluded from this test. At $d=11$, single-core CPU implementations of sliding-window BP+matching and BP-OSD require 3.6\,ms and 23\,ms per round, respectively, while exhibiting higher logical error rates than our decoder. Conversely, PyMatching (with and without error correlations) requires only 26\,$\mu$s and 13\,$\mu$s per round, respectively, but incurs higher LERs than BP+matching.

While most published real-time decoders exhibit lower execution latency than our architecture, their underlying algorithms yield substantially higher logical error rates under our evaluated noise conditions (Fig.~\ref{fig:surface}i). FPGA-based decoders leveraging union-find\cite{Liyanage2024}, clustering\cite{Barber2025,Ziad2025}, and matching\cite{Wu2025} algorithms achieve sub-microsecond per-round processing times at $d=9$, tailored for the microsecond measurement cycles of superconducting qubits. However, when evaluated on our test shots at $p_\mathrm{err}=0.03$, these alternative algorithms generate 3.9 to 9.0 times more logical errors than our decoder. Because these hardware architectures were sized and benchmarked at low physical error rates ($p_\mathrm{err} = 0.1\%$) with execution times that scale dynamically with detection event density, their operational latencies would expand under our higher noise rates. This comparative analysis demonstrates algorithmic behavior near the fault-tolerance threshold—where union-find fails on approximately half of all shots—rather than under low-noise regimes for which those decoders were engineered (Supplementary Section~\ref{supp:qec-results}). An FPGA implementation of BP+OSD\cite{Bascones2025} demonstrates a deterministic worst-case latency of 15\,$\mu$s per round independent of physical error rate, yet its algorithm yields 1.4 times as many logical errors as ours. A CPU-based sliding-window matching decoder\cite{Skoric2023} requires 2\,ms per round when evaluated at $p_\mathrm{err} = 2\%$. Consequently, at $d=9$ and $p_\mathrm{err}=0.03$, our decoder achieves superior logical error suppression compared to all benchmarked algorithms while operating comfortably within the millisecond round duration characteristic of neutral-atom and trapped-ion processors.

\FloatBarrier

\subsection{Event-driven updates vs recomputation}\label{res:event}

In graph matching, the computational cost of a state search was evaluated in iterations, whereas in quantum error-correcting decoding, it was measured in sweeps. However, the exact cost of a single iteration depends fundamentally on how the input states of the \textit{move} neurons are maintained. Our clocked digital emulation evaluates one \textit{move} neuron per iteration while retaining three defining characteristics of neuromorphic operation: state acceptance governed by a stochastic threshold, accepted transitions functioning as discrete event spikes that toggle clause outputs through interconnections, and continuously maintained neuron inputs (analogous to membrane potentials). Following an accepted state transition, event-driven updating applies localized updates governed by Supplementary Eqs.~\eqref{eq:event-clause-local} and~\eqref{eq:event-gate-local} (Supplementary Section~\ref{supp:local-updates}) exclusively through connections whose associated clause outputs or switch states have changed, and thereby exploits the activity sparsity of the architecture. Conversely, many digital Ising implementations, including architectures based on probabilistic bits\cite{Aadit2022,Nikhar2024} and simulated bifurcation algorithms\cite{Goto2019}, recompute a spin's effective field from the current states of its neighbors at every update step. For pairwise interactions, Isakov et al.\ noted that an accepted spin flip updates approximately as many stored energy differences as a full recomputation reads, whereas a rejected proposal updates none. Storing energy differences is therefore computationally superior when proposal acceptance rates remain low\cite{Isakov2015}.

For higher-order, multi-spin moves, an accepted transition can update significantly more or far fewer inputs than a single recomputation reads. We therefore extend this operation-counting framework to higher-order architectures. The recomputation strategy maintains no internal state inputs, instead evaluating the input of the proposed \textit{move} neuron directly from current clause outputs at each iteration. Given identical pseudorandom number sequences, both strategies compute identical transition inputs, accept and reject the exact same candidate proposals, and traverse identical trajectory paths through the state space. We can therefore quantify the operation counts of both strategies precisely along a single simulated trajectory per replica. Here, one operation is defined as either a single local input update across one connection or a single memory read of one weighted connection (where reading a switched connection includes evaluating the clause output that controls its switch). Clause toggles resulting from an accepted transition are common to both strategies and are omitted from this operation count. Consider an execution window of $N_\mathrm{w}$ iterations containing $N_\mathrm{acc}$ accepted state transitions (excluding non-updating self-loops that map a state to itself). The recomputation approach evaluates all input connections of the proposed move at every iteration—averaging $f_\mathrm{in}$ connections per proposal requiring $N_\mathrm{w}f_\mathrm{in}$ total read operations. In contrast, event-driven updating executes zero operations following a rejected proposal, and applies an average of $f_\mathrm{out}$ local updates after each accepted state transition, performing $N_\mathrm{acc}f_\mathrm{out}$ total updates. The ratio of event-driven operations to recomputation operations simplifies directly to the product of the acceptance rate $a$ and the structural sharing ratio $\rho$:
\begin{equation}
\frac{N_\mathrm{acc}f_\mathrm{out}}{N_\mathrm{w}f_\mathrm{in}}=a\rho,\qquad a=\frac{N_\mathrm{acc}}{N_\mathrm{w}},\qquad\rho=\frac{f_\mathrm{out}}{f_\mathrm{in}}.
\label{eq:arho}
\end{equation}
Because both operational counts are evaluated along the exact same trajectory, Eq.~\eqref{eq:arho} holds strictly across every execution window (Supplementary Section~\ref{supp:event} and Supplementary Fig.~\ref{fig:supp-event}a). Under uniform proposal distributions, recomputing inputs across all $R$ candidate \textit{move} neurons at each iteration requires reading all $Rf_\mathrm{in}$ connections; relative to that baseline, the operation ratio scales as $a\rho/R$.

\begin{figure}[!tbp]
\centering
\includegraphics[width=\linewidth]{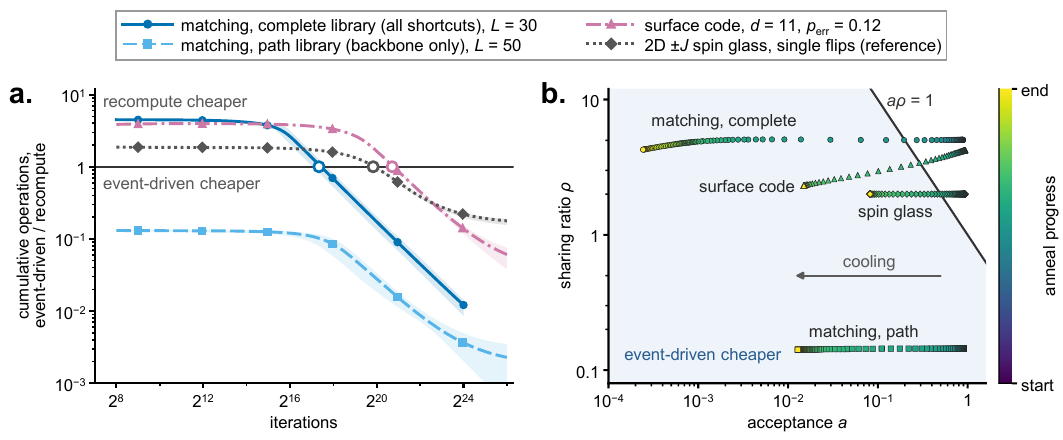}
\caption{\textbf{When event-driven updates need fewer operations than recomputation.}
\textbf{a}, Cumulative operations of event-driven updating divided by those of recomputation, versus
iterations along one slow anneal per workload. Recomputation: the input of the proposed \textit{move} neuron is recomputed at every iteration. In event-driven updating, an accepted spin transition updates the maintained inputs of the \textit{move} neurons it affects. Lines: pooled over replicas and instances. Bands: interquartile range of the
per-replica ratio. Open circles: break-even point (the ratio stays below 1 afterwards). Both
strategies are counted exactly on the same runs, because both strategies produce identical trajectories. \textbf{b}, Acceptance $a$ (accepted state changes per iteration) versus
sharing ratio $\rho$ (local updates per accepted state change divided by connection reads per iteration) for log-spaced windows of the same
anneals. Symbols: one per window. Shape: workload (key). Color: progress through the anneal. Per iteration, the
event-driven / recomputation operation ratio equals $a\rho$ exactly. Line: $a\rho=1$. Shading: event-driven updating needs fewer operations. Protocol: one Fowler--Nordheim (FN) annealing schedule (Methods Eq.~\eqref{eq:fn-schedule}) from the calibrated start temperature to 1/1,000 of it. Matching workloads: complete library (backbone with all shortcuts) at $L=30$ and path library (backbone alone) at $L=50$. Surface-code workload: $d=11$, depolarizing $p_\mathrm{err}=0.12$. Spin-glass workload: 2D $\pm J$ spin glass on a $16\times16$ torus, single spin flips. Schedule lengths, instances and replicas: Supplementary Section~\ref{supp:event}. One operation: one local update of one input (event-driven) or one connection read (recomputation).}\label{fig:event}
\end{figure}

We track both operation counts along a single slow annealing schedule per workload (see protocol in Supplementary Section~\ref{supp:event}): graph matching using all shortcuts (the complete library) versus using the backbone alone (the path library), decoding the surface code, and, as a pairwise benchmark, a two-dimensional $\pm J$ spin glass under single-spin flips\cite{Edwards1975}. Across all evaluated workloads, the cumulative operation count of event-driven updating falls below that of recomputation early in the annealing trajectory (Fig.~\ref{fig:event}a). Specifically, this crossover occurs within the initial 1--3\% of execution for the complete matching library, the surface code, and the spin glass, and holds from the very start of annealing for the path library. By the conclusion of the anneal, event-driven updating requires only 0.2\% (path library) to 18\% (spin glass) of the operations demanded by single-move recomputation. When compared against the alternative baseline of recomputing every \textit{move}-neuron input at every iteration, event-driven updating utilizes less than 3\% of total operations throughout the entire run. Figure~\ref{fig:event}b illustrates the underlying mechanics of this efficiency gain. The structural sharing ratio $\rho$ is governed primarily by the move library topology. It ranges between 2 and 5 for the complete matching library, the surface code, and the spin glass, but drops to 0.14 for the path library. Conversely, the acceptance rate $a$ decreases by several orders of magnitude as temperature declines. As a result, every workload with $\rho > 1$ eventually crosses the break-even threshold $a\rho = 1$, whereas the path library (with $\rho < 1$) remains strictly below $a\rho = 1$ from the outset. Crucially, the acceptance rate does not decay to zero because iso-energetic spin transitions ($\Delta E = 0$) are accepted unconditionally. At the end of the anneal, this acceptance floor reaches about 0.08 per iteration (spin glass) and remains strictly below $1/\rho$ across all workloads; thus, the event-driven advantage persists through the end of execution, bounded only by the inherent degeneracy of the energy landscape. Event-driven updating therefore yields substantial operation savings throughout the low-temperature regime, where the Fowler--Nordheim annealing schedule concentrates the majority of its iterations to refine solution quality\cite{Chen2025,Ahsan2026}. To the extent that energy consumption in a physical neuromorphic architecture scales with the volume of input update operations, the condition $a\rho < 1$ also defines the regime where an operational energy advantage begins. Note that we report exact algorithmic operation counts rather than wall-clock execution time or measured physical power.

\section{Discussion}\label{sec:discussion}

Programming spin transitions shifts the handling of constraints from a problem of energy design into one of \textit{move} design. Consider a set of reversible moves that generate the exact inverse of each spin transition, operating under symmetric proposal probabilities and Metropolis acceptance criteria implemented via a stochastic threshold. Under these conditions, the transition-programmable architecture is functionally isomorphic to classical simulated annealing restricted to the feasible state space (Methods Section~\ref{meth:benchmarks}; Supplementary Section~\ref{supp:barriers}). Because its temperature evolution follows a Fowler--Nordheim schedule, the machine can theoretically guarantee convergence to the global ground state of a constrained optimization problem—provided that its backbone move library maintains global connectivity across all feasible states, the annealing rate enters the logarithmic regime ($\tau_n \approx \tau_\mathrm{FN}/\ln n$), and the initial scale parameter $\tau_\mathrm{FN}$ is bounded below by the critical depth $h^\ast$ of the landscape\cite{Hajek1988}. In practice, executing hardware annealing under this schedule enables neuromorphic architectures to reliably hit low-energy ground states across independent runs\cite{Chen2025,Ahsan2026}.

The transition-programmable architecture applies to a significantly broader class of problems than the two benchmark applications demonstrated in this work. In our architecture, each state-dependent switch is controlled by a single clause output—originating from either an objective clause or an auxiliary clause. A single clause output is sufficient whenever every target spin of the destination state equals a product of current-state spins, up to a sign factor. Because a clause is itself a spin product and spin values satisfy $s_j^2=1$, the product of any two spin products reduces to another spin product. Consequently, whether a clause undergoes a sign inversion following a candidate move is determined by a single spin-product function of the current state configuration. Expressed in binary variables, such transitions correspond to affine transformations over the Galois field GF(2). This mathematical class encompasses every move operation that permutes or flips spins, including pairwise spin swaps, row exchanges, categorical variable recolorings, and codeword additions across standard combinatorial neighborhoods. As detailed in Supplementary Table~\ref{tab:supp-closure}, this class alone covers a vast range of practically relevant constrained optimization problems. These problem families utilize fixed spin reversals under parity constraints, pairwise swaps under cardinality and one-hot constraints, and row exchanges under partition and permutation constraints. The number of auxiliary clauses that they require is bounded per move: none for parity constraints, whose switches are all constant, at most one for cardinality and one-hot constraints, and polynomially many for partition and permutation constraints with objectives of bounded order (Supplementary Section~\ref{supp:closure}). Furthermore, candidate moves within this framework need not be reversible. For instance, a cyclic three-row exchange represents a valid permutation of spins; its inverse transformation (the counter-rotation) constitutes a distinct move that interfaces with the same clause outputs but operates through separate switches, input structures, and destination mappings (Methods Section~\ref{meth:model}). Affine moves also encompass state-dependent transformations that are not strict permutations. 
For example, a conditional reversal flips a designated subset of spins if and only if the product of a disjoint subset of spins equals $-1$. Such a reversal requires only a single spin product per switch control signal. The underlying architectural derivation imposes no specific structural assumptions on the underlying optimization problem (Methods Section~\ref{meth:model}). 

For specific problem formulations, candidate moves falling outside the affine class can still be implemented with a single spin product per switch whenever algebraic identities governing the feasible state space reduce each switch function to an existing stored product. More generally, because any arbitrary Boolean function of spin variables can be expressed as a weighted linear combination of spin products, the architecture can in principle represent any candidate transition move. However, the required number of spin-product terms per switch, and consequently the auxiliary-clause overhead, may grow exponentially with problem dimension for general non-affine moves. To prevent this exponential scaling for moves whose switch functions depend on a localized subset of state variables, a natural structural extension introduces switches controlled by multi-input Boolean functions. Controlling switches via low-order Boolean functions of a few clause outputs bounds the auxiliary-clause overhead to polynomial growth, extending the architecture to more complex state-dependent operations—such as executing a row exchange conditionally upon an odd parity constraint. While the core mathematical derivation remains identical under this extension, its physical hardware implementation lies beyond the scope of the present demonstration.

Making state transitions explicit transforms the move library into a key design parameter possessing distinct dynamical properties. The backbone move library dictates reachability across the feasible state space, whereas shortcut moves govern convergence acceleration—not primarily through path-length reduction, but through energy barrier suppression. While any shortcut move accelerates high-temperature mixing, the iteration count among move libraries of identical capacity correlates directly with the highest energy barriers encountered during cooling. Selecting shortcuts guided by objective function couplings lowers these critical energy barriers more effectively than random selection. Crucially, introducing shortcuts cannot substantially increase the relaxation time: when the backbone retains a fixed proposal probability share $\beta$, the relaxation time at any constant temperature is bounded above by $1/\beta$ times that of the backbone alone, regardless of the shortcut configuration (Supplementary Section~\ref{supp:shortcut-theory}). The design challenge thus shifts from whether to incorporate shortcuts to identifying which specific shortcuts suppress energy barriers sufficiently to justify their dedicated allocation of \textit{move} neurons, interconnections, and auxiliary clauses. Informed shortcuts selected prior to execution substantially lower these energy barriers, while dynamic batching enables a single optimization trajectory to leverage a larger effective shortcut space than hardware capacity allows simultaneously. Furthermore, move selection could be learned dynamically. By tracking move acceptance rates and state visitation frequencies online, an adaptive learning rule could identify persistent local energy minima, dynamically introducing shortcuts to lower the surrounding barriers while pruning underutilized moves during runtime execution.

Demonstrating both graph matching and surface-code decoding establishes that a single transition-programmable architecture can natively execute both optimization and probabilistic sampling tasks. The physical hardware components, including candidate moves, clause circuits, and \textit{move} neurons, remain invariant across both modes; only the operational temperature schedule and readout protocol change. Crucially, the two operational paradigms target fundamentally distinct quantities. When configured as an annealer for decoding, the system returns the single lowest-energy correction string (representing the most likely individual error vector). Conversely, maximum-likelihood (ML) decoding requires identifying the logical class possessing the lowest free energy $F_\gamma$, a global quantity defined over exponentially many valid error configurations that cannot be stored directly within fixed clause weights. Thermal sampling at $\tau=1$ directly evaluates these partition sums via state visitation frequencies, providing an explicit confidence metric for every decoding decision without incurring extra computational overhead. The physical decoding performance reported here is bound primarily by the assumed noise model and hardware chip capacity. For real-time decoding, our experiments employ a phenomenological noise model, which omits fault mechanisms originating within the measurement circuit gates themselves. Under a full circuit-level noise model, a single physical fault can generate multiple correlated detection events across space and time. Because our architecture supports clauses of arbitrary order, such multi-qubit fault correlations can be incorporated natively into the objective function—analogous to how $X$–$Z$ error correlations within $Y$ errors are currently captured via second-order clause interactions. Extending the framework to circuit-level noise models represents a natural immediate progression. Furthermore, chip-level logic resources currently constrain real-time decoding throughput to millisecond syndrome rounds (rather than the microsecond regimes of superconducting qubits) and limit maximum code distances to $d=15$ for single-syndrome shots and $d=11$ for continuous real-time operation. Scaling to larger application-specific integrated circuits (ASICs) or adopting multi-chip distributed decoding topologies will naturally extend both processing bandwidth and code-distance capacity.

The transition-programmable architecture also enables graph coloring techniques to drive massively parallel update schemes\cite{Aadit2022}—in this context, executed across \textit{move} neurons (Section~\ref{res:surface} and Supplementary Section~\ref{supp:fpga}). Graph coloring achieves maximum efficiency when move operations are sparsely connected to energy clauses. Neuromorphic hardware, in which large arrays of \textit{move} neurons evaluate their input states concurrently, could leverage color classes in an identical manner. A related extension building on neuromorphic parallelism involves partitioning large problem instances across distributed multi-machine clusters. In this approach, each machine searches a designated sub-domain while synchronizing periodically or sharing state boundary conditions with adjacent units. Unlike dynamic move batching—which rotates loaded move libraries while maintaining a single unified state vector—such spatial decomposition directly splits the state space itself, making it particularly well suited for weakly coupled physical and combinatorial systems.

We have demonstrated that event-driven processing, a defining operational feature of neuromorphic architecture~\cite{ChakrabarttyEd2026}, maintains a distinct computational advantage within this architecture over the recomputation strategies employed by conventional digital Ising machines\cite{Aadit2022,Nikhar2024,Goto2019}, even when evaluated within a clocked digital emulation environment. For pairwise interactions, maintaining updated input states requires fewer total operations whenever proposal acceptance rates remain low\cite{Isakov2015}. For higher-order multi-spin moves, the efficiency ratio is governed by both the move library topology and the operating temperature; this neuromorphic advantage holds strictly throughout the low-temperature regime of the anneal, where the cooling schedule concentrates the vast majority of its iterations. In contrast, our current FPGA decoder implementation recomputes its low-bit-width integer inputs every clock cycle to enable the fully parallel application of complete color classes. The full operational efficiency of event-driven updating would be realized in spiking hardware implementations, where event updates arrive asynchronously. Crucially, the clause outputs and their corresponding local updates map directly onto the complementary ON--OFF neuron pairs of the NeuroSA architecture\cite{Chen2025} (Supplementary Section~\ref{supp:onoff}), establishing a concrete structural pathway toward a physical spiking neuromorphic implementation.

From a neuromorphic point of view, the most distinctive feature of the proposed architecture is that the neurons here represent actions rather than states. In NeuroSA, a neuron represents a spin, and in N-HOIM, a neuron represents the flip of one spin. Here, a neuron represents a move. Even though the architecture was derived from the algebra of constraints rather than from a neural circuit, three of its features have counterparts in nervous systems. First, its neurons represent feasible actions rather than individual variables, as muscle synergies do. Cortical and spinal circuits do not represent joint angles one at a time but muscle synergies and motor primitives, low-dimensional coordinated actions that are feasible by construction with respect to the body’s constraints\cite{dAvella2005}. The move library is a repertoire of primitives and the backbone is the minimal repertoire that spans the feasible space. The penalty approach corresponds instead to controlling every muscle independently, with the constraints imposed by penalties. Second, switching a connection on and off by the state of a third unit is a form of dynamic sparsity\cite{Mocanu2018,Kudithipudi2025} and has precedents in higher-order Boltzmann machines\cite{Sejnowski1986,Memisevic2010} and a cellular counterpart in presynaptic inhibition, which silences one synapse without affecting the postsynaptic neuron’s other inputs. Here the switching is not learned but derived from the constraints, and it lets a fixed set of weights serve every feasible state. Third, a shortcut compresses a sequence of backbone transitions into a single transition, as habit formation chunks action sequences. Whether these correspondences are more than functional is beyond this work, but the last analogy suggests how a move library could be learned.

\FloatBarrier
\section{Methods}\label{sec:methods}

\subsection{Derivation of the transition-programmable architecture}\label{meth:model}

For $N$ Ising spins $\svec=(s_1,\ldots,s_N)$, let $\Omega\subseteq\{-1,+1\}^{N}$ denote the set of feasible states. The energy is
\begin{equation}
E(\svec)=E_{\mathrm{const}}-\sum_{k=1}^{M}J_kT_k(\svec),\qquad \svec\in\Omega,
\label{eq:energy-clauses}
\end{equation}
with
\begin{equation}
T_k(\svec)=\prod_{i=1}^{N}s_i^{H_{ki}},\qquad H_{ki}\in\{0,1\}.
\label{eq:hamiltonian}
\end{equation}
Following N-HOIM\cite{Ahsan2026}, we call each spin product $T_k$ a clause and its stored sign, $T_k\in\{-1,+1\}$, a clause output. The $k$th row of $H$ identifies the spins in clause $k$, $J_k$ is the weight of clause $k$, and $M$ is the number of objective clauses, the clauses weighted in the objective. The solver can act as an optimizer that finds $\arg\min_{\svec\in\Omega}E(\svec)$, or as a sampler that draws $\svec\in\Omega$ from the Boltzmann distribution $\propto\mathrm{e}^{-E(\svec)/\tau}$ at a temperature $\tau$.

Now consider a spin transition $\muvec\to\nuvec$ between two feasible states $\muvec,\nuvec\in\Omega$. Since $T_k$ can only remain the same or reverse sign, write
\begin{equation}
T_k(\nuvec)=\bigl[1-2\alpha_{k,\muvec\leftrightarrow\nuvec}\bigr]T_k(\muvec).
\label{eq:clause-toggle-form}
\end{equation}
It follows that
\begin{equation}
\alpha_{k,\muvec\leftrightarrow\nuvec}=\frac{1-T_k(\nuvec)T_k(\muvec)}{2}=\frac{1-\prod_{i=1}^{N}(\mu_i\nu_i)^{H_{ki}}}{2}.
\label{eq:alpha-definition}
\end{equation}
The product $\mu_i\nu_i$ is $-1$ exactly when spin $i$ changes. So $\alpha_{k,\muvec\leftrightarrow\nuvec}=1$ when an odd number of the spins of clause $k$ change, and zero otherwise. Substituting Eq.~\eqref{eq:clause-toggle-form} into Eq.~\eqref{eq:energy-clauses} gives
\begin{equation}
\Delta E_{\muvec\to\nuvec}=E(\nuvec)-E(\muvec)=2\sum_{k=1}^{M}J_k\,\alpha_{k,\muvec\leftrightarrow\nuvec}\,T_k(\muvec).
\label{eq:deltae-derived}
\end{equation}
Because the same clauses change sign in either direction, $\alpha_{k,\muvec\leftrightarrow\nuvec}$ is symmetric in the two states. The energy change, in contrast, is evaluated with the clause outputs of the current state and changes sign with the direction, $\Delta E_{\nuvec\to\muvec}=-\Delta E_{\muvec\to\nuvec}$.

Equation~\eqref{eq:deltae-derived} holds for any spin transition. The solver starts from a feasible state $\muvec_0\in\Omega$, obtained in a problem-dependent way. A neuron in the solver hardware then proposes a spin transition from the current state, evaluates its $\Delta E$ by Eq.~\eqref{eq:deltae-derived}, and accepts or rejects it stochastically. The instantiated spin transitions must therefore make every feasible state reachable. Instantiating a neuron for every spin transition, however, is not practical, because there are $|\Omega|(|\Omega|-1)/2$ pairs of feasible states and $|\Omega|$ typically grows exponentially with $N$. Instead, we use the symmetries of the problem to group spin transitions, and instantiate one neuron for each group. We show this on the following example.

\paragraph{An example toy problem}\label{meth:assignment-example} Beads of three colors, red, green and blue, are linearly arranged in a row of three positions with the constraint that only one color can occupy one position at a time (Fig.~\ref{fig:toy}a). Each configuration of the beads has an energy associated with it and is given by
\begin{equation}
E=-(\text{red in }p_1)-2\,(\text{blue in }p_3)-(\text{green in }p_1)(\text{blue in }p_2)-2\,(\text{red in }p_3),
\label{eq:ex-problem}
\end{equation}
where the positions are labeled from left to right as $p_1$, $p_2$ and $p_3$ (Fig.~\ref{fig:toy}a). The problem can be converted to an equivalent higher-order Ising problem by associating a spin variable with each bracket, which is $+1$ when the color is in the position and $-1$ otherwise. If the colors red, green and blue are denoted as $x=1,2,3$ and if we define one spin for each color and position according to
\begin{equation*}
s_{i=3(x-1)+a}=\begin{cases}+1 & \text{if color }x\text{ is in position }p_a,\\ -1 & \text{otherwise,}\end{cases}\qquad x,a\in\{1,2,3\},
\end{equation*}
the higher-order Ising problem corresponding to minimizing the energy in Eq.~\eqref{eq:ex-problem} is
\begin{equation}
\begin{aligned}
\min_{\svec}\ &E(\svec)=-s_1-2s_9-s_4s_8-2s_3\\
\text{subject to}\ &s_1+s_2+s_3=s_4+s_5+s_6=s_7+s_8+s_9=-1,\\
&s_1+s_4+s_7=s_2+s_5+s_8=s_3+s_6+s_9=-1.
\end{aligned}
\label{eq:ex-spins}
\end{equation}
Comparing with Eqs.~\eqref{eq:energy-clauses} and \eqref{eq:hamiltonian} gives $E_\mathrm{const}=0$ and $M=4$ objective clauses,
\begin{equation}
\begin{gathered}
T_1=s_1,\qquad T_2=s_9,\qquad T_3=s_4s_8,\qquad T_4=s_3,\\[2pt]
H=\begin{pmatrix}1&0&0&0&0&0&0&0&0\\0&0&0&0&0&0&0&0&1\\0&0&0&1&0&0&0&1&0\\0&0&1&0&0&0&0&0&0\end{pmatrix},\qquad J=\begin{pmatrix}1\\2\\1\\2\end{pmatrix}.
\end{gathered}
\label{eq:ex-HJ}
\end{equation}

\paragraph{Feasible states} 
Of the $2^9$ possible spin states, six feasible states satisfy the constraints in Eq.~\eqref{eq:ex-spins} which we enumerate here:
\begin{center}\small
\begin{tabular}{@{}c|ccccccccc@{}}
\toprule
State & $s_1$ & $s_2$ & $s_3$ & $s_4$ & $s_5$ & $s_6$ & $s_7$ & $s_8$ & $s_9$\\
\midrule
1 & $+1$ & $-1$ & $-1$ & $-1$ & $+1$ & $-1$ & $-1$ & $-1$ & $+1$\\
2 & $+1$ & $-1$ & $-1$ & $-1$ & $-1$ & $+1$ & $-1$ & $+1$ & $-1$\\
3 & $-1$ & $+1$ & $-1$ & $+1$ & $-1$ & $-1$ & $-1$ & $-1$ & $+1$\\
4 & $-1$ & $+1$ & $-1$ & $-1$ & $-1$ & $+1$ & $+1$ & $-1$ & $-1$\\
5 & $-1$ & $-1$ & $+1$ & $+1$ & $-1$ & $-1$ & $-1$ & $+1$ & $-1$\\
6 & $-1$ & $-1$ & $+1$ & $-1$ & $+1$ & $-1$ & $+1$ & $-1$ & $-1$\\
\bottomrule
\end{tabular}
\end{center}

\paragraph{Spin transitions that connect $\Omega$} 
Consider the graph whose vertices are the feasible states, with an edge between two states whenever the solver can make a spin transition between them. We call $\Omega$ connected when every feasible state is reachable. Six states admit 15 bidirectional edges, and a spanning tree, with $|\Omega|-1=5$ edges, is the smallest connected choice. Two spanning trees are specified here:
\begin{center}
\begin{tikzpicture}[baseline=(current bounding box.center),scale=0.8,
  st/.style={circle,fill=black,inner sep=1.4pt},
  lab/.style={font=\scriptsize,fill=white,inner sep=0.8pt}]
\node[font=\small] at (0,2.0) {Tree 1};
\node[st] (v1) at (120:1.15) {};
\node[font=\footnotesize] at (120:1.5) {1};
\node[st] (v2) at (60:1.15) {};
\node[font=\footnotesize] at (60:1.5) {2};
\node[st] (v3) at (0:1.15) {};
\node[font=\footnotesize] at (0:1.5) {3};
\node[st] (v4) at (-60:1.15) {};
\node[font=\footnotesize] at (-60:1.5) {4};
\node[st] (v5) at (-120:1.15) {};
\node[font=\footnotesize] at (-120:1.5) {5};
\node[st] (v6) at (180:1.15) {};
\node[font=\footnotesize] at (180:1.5) {6};
\draw (v1) -- node[lab] {4} (v2);
\draw (v2) -- node[lab] {6} (v3);
\draw (v3) -- node[lab] {4} (v4);
\draw (v4) -- node[lab] {6} (v5);
\draw (v5) -- node[lab] {4} (v6);
\end{tikzpicture}\hfill
{\small\setlength{\tabcolsep}{4pt}\setlength{\aboverulesep}{0pt}\setlength{\belowrulesep}{0pt}\renewcommand{\arraystretch}{1.2}%
\begin{tabular}{@{}cc@{\enspace}|@{\enspace}cc@{}}
\toprule
\multicolumn{2}{c@{\enspace}|@{\enspace}}{Tree 1} & \multicolumn{2}{c}{Tree 2}\\
\midrule
Transition & \begin{tabular}{@{}c@{}}\# changed\\[-2pt] spins\end{tabular} & Transition & \begin{tabular}{@{}c@{}}\# changed\\[-2pt] spins\end{tabular}\\
\midrule
$1\leftrightarrow2$ & 4 & $1\leftrightarrow2$ & 4\\
$2\leftrightarrow3$ & 6 & $2\leftrightarrow5$ & 4\\
$3\leftrightarrow4$ & 4 & $5\leftrightarrow6$ & 4\\
$4\leftrightarrow5$ & 6 & $6\leftrightarrow4$ & 4\\
$5\leftrightarrow6$ & 4 & $4\leftrightarrow3$ & 4\\
\bottomrule
\end{tabular}}\hfill
\begin{tikzpicture}[baseline=(current bounding box.center),scale=0.8,
  st/.style={circle,fill=black,inner sep=1.4pt},
  lab/.style={font=\scriptsize,fill=white,inner sep=0.8pt}]
\node[font=\small] at (0,2.0) {Tree 2};
\node[st] (v1) at (120:1.15) {};
\node[font=\footnotesize] at (120:1.5) {1};
\node[st] (v2) at (60:1.15) {};
\node[font=\footnotesize] at (60:1.5) {2};
\node[st] (v3) at (0:1.15) {};
\node[font=\footnotesize] at (0:1.5) {3};
\node[st] (v4) at (-60:1.15) {};
\node[font=\footnotesize] at (-60:1.5) {4};
\node[st] (v5) at (-120:1.15) {};
\node[font=\footnotesize] at (-120:1.5) {5};
\node[st] (v6) at (180:1.15) {};
\node[font=\footnotesize] at (180:1.5) {6};
\draw (v1) -- node[lab] {4} (v2);
\draw (v2) -- node[lab,pos=0.28] {4} (v5);
\draw (v5) -- node[lab] {4} (v6);
\draw (v6) -- node[lab,pos=0.75] {4} (v4);
\draw (v4) -- node[lab] {4} (v3);
\end{tikzpicture}
\end{center}

Both trees connect $\Omega$. We choose tree 2, which uses only four-spin transitions, the smallest change between two feasible states. These transitions suffice to connect $\Omega$, and every longer transition is a sequence of them, for example $1\to5=1\to2\to5$. Whether a longer transition accelerates the search depends on the energies of the intermediate states that it bypasses. We therefore first connect $\Omega$ with the smallest changes, and add further transitions afterwards.

\paragraph{Grouping spin transitions} 
Tree 2 still has five spin transitions, which would require one neuron each. One neuron per spin transition can be redundant, because several spin transitions can be grouped by a common operation on the spins, such as a swap or a permutation of spin values. We therefore reduce the spin transitions symbolically to such operations. Each symbolic operation then needs only one neuron. Consider $1\leftrightarrow2$. From state 1 to state 2, $s_5$ and $s_9$ change from $+1$ to $-1$, and $s_6$ and $s_8$ from $-1$ to $+1$. Swapping the values of $s_5$ and $s_9$ with those of $s_6$ and $s_8$ therefore turns state 1 into state 2. One way to do so is the pair of swaps $s_5\leftrightarrow s_8$ and $s_6\leftrightarrow s_9$. However, these two swaps alone do not preserve the constraints of Eq.~\eqref{eq:ex-spins}, because under
$s_5\leftrightarrow s_8,\ s_6\leftrightarrow s_9$ the constraint $s_4+s_5+s_6$ becomes $s_4+s_8+s_9$, which is not a constraint. Adding the swap $s_4\leftrightarrow s_7$ restores them: under
$s_4\leftrightarrow s_7,\ s_5\leftrightarrow s_8,\ s_6\leftrightarrow s_9$, the constraint $s_4+s_5+s_6$ becomes $s_7+s_8+s_9$, which is again a constraint.

The same construction for other transitions in tree 2 gives
\begin{center}\small
\begin{tabular}{@{}l@{\qquad}c@{\qquad}c@{}}
\toprule
Transition & Swaps of changed spins & Swap added for constraints\\
\midrule
$1\leftrightarrow2$ & $s_5\leftrightarrow s_8$, $s_6\leftrightarrow s_9$ & $s_4\leftrightarrow s_7$\\
$2\leftrightarrow5$ & $s_1\leftrightarrow s_4$, $s_3\leftrightarrow s_6$ & $s_2\leftrightarrow s_5$\\
$5\leftrightarrow6$ & $s_4\leftrightarrow s_7$, $s_5\leftrightarrow s_8$ & $s_6\leftrightarrow s_9$\\
$6\leftrightarrow4$ & $s_2\leftrightarrow s_5$, $s_3\leftrightarrow s_6$ & $s_1\leftrightarrow s_4$\\
$4\leftrightarrow3$ & $s_4\leftrightarrow s_7$, $s_6\leftrightarrow s_9$ & $s_5\leftrightarrow s_8$\\
\bottomrule
\end{tabular}
\end{center}
The swaps $(s_4\leftrightarrow s_7,\ s_5\leftrightarrow s_8,\ s_6\leftrightarrow s_9)$ appear for $1\leftrightarrow2$, $5\leftrightarrow6$ and $4\leftrightarrow3$, and the swaps $(s_1\leftrightarrow s_4,\ s_2\leftrightarrow s_5,\ s_3\leftrightarrow s_6)$ for $2\leftrightarrow5$ and $6\leftrightarrow4$. These two operations reduce the five transitions to two symbolic mappings, $G_1$ and $G_2$,
\begin{equation}
\begin{aligned}
G_1:&\quad (s_1,s_2,s_3,s_4,s_5,s_6,s_7,s_8,s_9)\ \longrightarrow\ (s_4,s_5,s_6,s_1,s_2,s_3,s_7,s_8,s_9),\\
G_2:&\quad (s_1,s_2,s_3,s_4,s_5,s_6,s_7,s_8,s_9)\ \longrightarrow\ (s_1,s_2,s_3,s_7,s_8,s_9,s_4,s_5,s_6).
\end{aligned}
\label{eq:ex-g1}
\end{equation}

Applied to all six states, $G_1$ and $G_2$ produce the spin transitions
\begin{equation}
G_1:\ 1\leftrightarrow3,\ 2\leftrightarrow5,\ 4\leftrightarrow6,\qquad\qquad G_2:\ 1\leftrightarrow2,\ 3\leftrightarrow4,\ 5\leftrightarrow6.
\label{eq:grouped-transition}
\end{equation}
Since every feasible state is an arrangement of the three colors, we can now identify $G_1$ as swapping the positions of red and green (red\,$\leftrightarrow$\,green), and $G_2$ as swapping the positions of green and blue (green\,$\leftrightarrow$\,blue) (Fig.~\ref{fig:toy}b). Thus, for a large problem whose structure is known beforehand, we can identify such symbolic mappings without enumerating $\Omega$ (Sections~\ref{res:matching} and~\ref{res:surface}). In general, we call such a symbolic mapping $G$ a move. 

Let $\{G_r: r=1,\ldots,R\}$ be a set of moves. We call it a backbone when it connects $\Omega$, as $\{G_1,G_2\}$ does (Fig.~\ref{fig:toy}b). Similarly, the remaining four-spin transitions $1\leftrightarrow6$, $2\leftrightarrow4$ and $3\leftrightarrow5$ swap the positions of red and blue (red\,$\leftrightarrow$\,blue). We call such further moves, added to a backbone, shortcuts (Section~\ref{res:library}). In hardware, each move $G_r$ is instantiated as one \textit{move} neuron, denoted by the same symbol.

\paragraph{Transition switches} 
For each spin transition $\muvec\leftrightarrow\nuvec$, Eq.~\eqref{eq:alpha-definition} gives one binary value per clause,
\begin{equation*}
\alpha_{k,\muvec\leftrightarrow\nuvec}=\frac{1-\prod_{i=1}^{N}(\mu_i\nu_i)^{H_{ki}}}{2}.
\end{equation*}
The values $\alpha_{k,\muvec\leftrightarrow\nuvec}$ for the transitions of $G_1$ and $G_2$ are listed below, with the spins $s_7$ and $s_9$, which $G_1$ leaves unchanged, and $s_3$, which $G_2$ leaves unchanged,
\begin{center}\small\setlength{\tabcolsep}{3pt}
\begin{tabular}[t]{@{}c|cccc|cc@{}}
\toprule
 & \multicolumn{4}{c|}{$\alpha_{k,\muvec\leftrightarrow\nuvec}$ for $G_1$} & & \\
\cmidrule(lr){2-5}
$\muvec\leftrightarrow\nuvec$ & $k=1$ & $k=2$ & $k=3$ & $k=4$ & $s_7$ & $s_9$\\
\midrule
$1\leftrightarrow3$ & 1 & 0 & 1 & 0 & $-1$ & $+1$\\
$2\leftrightarrow5$ & 1 & 0 & 1 & 1 & $-1$ & $-1$\\
$4\leftrightarrow6$ & 0 & 0 & 0 & 1 & $+1$ & $-1$\\
\bottomrule
\end{tabular}\hfill
\begin{tabular}[t]{@{}c|cccc|c@{}}
\toprule
 & \multicolumn{4}{c|}{$\alpha_{k,\muvec\leftrightarrow\nuvec}$ for $G_2$} & \\
\cmidrule(lr){2-5}
$\muvec\leftrightarrow\nuvec$ & $k=1$ & $k=2$ & $k=3$ & $k=4$ & $s_3$\\
\midrule
$1\leftrightarrow2$ & 0 & 1 & 1 & 0 & $-1$\\
$3\leftrightarrow4$ & 0 & 1 & 1 & 0 & $-1$\\
$5\leftrightarrow6$ & 0 & 0 & 0 & 0 & $+1$\\
\bottomrule
\end{tabular}
\end{center}

The tables show that these values depend on the current state, through $s_7$ and $s_9$ for $G_1$ and through $s_3$ for $G_2$. The switch $\alpha_{k,G_r}$ from clause $k$ to the move $G_r$ is
\begin{center}\small
\begin{tabular}{@{}c|cc@{}}
\toprule
$k$ & $\alpha_{k,G_1}$ & $\alpha_{k,G_2}$\\
\midrule
1 & $(1-s_7)/2$ & $0$\\
2 & $0$ & $(1-s_3)/2$\\
3 & $(1-s_7)/2$ & $(1-s_3)/2$\\
4 & $(1-s_9)/2$ & $0$\\
\bottomrule
\end{tabular}
\end{center}

\paragraph{Auxiliary clauses} 
The goal for this synthesis is to formulate the architecture only based on the clause outputs $T_1,\ldots,T_4$ and the \textit{move} neurons $G_r$. The spins $s_1,\ldots,s_9$ in the above table constitute hidden variables. We therefore rewrite the switches above in terms of clause outputs. The switches depend on $s_7$, $s_9$ and $s_3$, and the objective clauses are
\begin{equation*}
T_1=s_1,\qquad T_2=s_9,\qquad T_3=s_4s_8,\qquad T_4=s_3.
\end{equation*}
The spins $s_9$ and $s_3$ are already the outputs of $T_2$ and $T_4$, but no objective clause has the output $s_7$. To eliminate this hidden spin, we append the clause
\begin{equation}
T_5=s_7,\qquad J_5=0.
\label{eq:ex-gate}
\end{equation}
A zero-weight clause appended to supply a switch is an auxiliary clause. An auxiliary clause also needs switches for its connections to the moves. Repeating for $T_5$ the computation of the switches above gives the left table below. All switches can then be written in terms of clause outputs (right), and no further clause is needed,
\begin{center}\small
\begin{tabular}{@{}c|cc@{}}
\toprule
$k$ & $\alpha_{k,G_1}$ & $\alpha_{k,G_2}$\\
\midrule
1 & $(1-s_7)/2$ & $0$\\
2 & $0$ & $(1-s_3)/2$\\
3 & $(1-s_7)/2$ & $(1-s_3)/2$\\
4 & $(1-s_9)/2$ & $0$\\
\midrule
5 & $0$ & $(1-s_1)/2$\\
\bottomrule
\end{tabular}
\quad$\Longrightarrow$\quad
\begin{tabular}{@{}c|cc@{}}
\toprule
$k$ & $\alpha_{k,G_1}$ & $\alpha_{k,G_2}$\\
\midrule
1 & $(1-T_5)/2$ & $0$\\
2 & $0$ & $(1-T_4)/2$\\
3 & $(1-T_5)/2$ & $(1-T_4)/2$\\
4 & $(1-T_2)/2$ & $0$\\
\midrule
5 & $0$ & $(1-T_1)/2$\\
\bottomrule
\end{tabular}
\end{center}
A move $G$ is affine when each spin after the move equals a product of spins before the move, up to sign (Section~\ref{sec:discussion}). For affine moves, each state-dependent switch is set by the output $T_j$ of only one clause, a spin product, as
\begin{equation}
\alpha_{k,G_r}=\frac{1\pm T_j}{2}.
\label{eq:gate-switch}
\end{equation}
If $T_j$ is an objective clause, its output is reused. Otherwise, $T_j$ is appended as an auxiliary clause. In general, auxiliary clauses are appended until the clause outputs supply every switch. Let $\widetilde M$ denote the total number of objective and auxiliary clauses, with $J_k=0$ for $k>M$. Supplementary Section~\ref{supp:closure} bounds the number of auxiliary clauses appended in this way, and Supplementary Algorithm~\ref{alg:compile} gives the complete procedure.

\paragraph{Encoder architecture} 
Substituting the switches into Eq.~\eqref{eq:deltae-derived} gives $\Delta E_{G_r}=2v_{G_r}$ with
\begin{equation}
v_{G_r}=\sum_{k=1}^{M}J_k\,\alpha_{k,G_r}\,T_k.
\label{eq:encoder}
\end{equation}
We call $v_{G_r}$ the input of \textit{move} neuron $G_r$. This switched, weighted projection from the clause outputs to the \textit{move} neurons is the encoder of the proposed autoencoder-like architecture (Fig.~\ref{fig:concept}d). With the switches in terms of clause outputs, the inputs of the two \textit{move} neurons are
\begin{equation}
\begin{aligned}
v_{G_1}&=1\cdot\left(\frac{1-T_5}{2}\right)T_1+1\cdot\left(\frac{1-T_5}{2}\right)T_3+2\cdot\left(\frac{1-T_2}{2}\right)T_4,\\
v_{G_2}&=2\cdot\left(\frac{1-T_4}{2}\right)T_2+1\cdot\left(\frac{1-T_4}{2}\right)T_3.
\end{aligned}
\label{eq:assignment-fields}
\end{equation}
In state 4, $(T_1,\ldots,T_5)=(-1,-1,+1,-1,+1)$ gives $v_{G_1}=-2$ and $v_{G_2}=-1$ (Fig.~\ref{fig:toy}c).

\paragraph{Sampler, annealer and decoder architecture} 
An accepted transition must also update the clause outputs, for example $T_2$, $T_3$ and $T_5$ when $4\to3$ is accepted (Fig.~\ref{fig:toy}c). Let $q_r\in\{0,1\}$ be the output of \textit{move} neuron $G_r$. The neuron fires, and its transition is accepted, when its input lies below a stochastic threshold $\xi_r$,
\begin{equation}
q_r=\begin{cases}1 & \text{if } v_{G_r}\leq\xi_r,\\ 0 & \text{otherwise.}\end{cases}
\label{eq:firing}
\end{equation}
A later section gives the distribution of $\xi_r$ and its annealing schedule. As in N-HOIM\cite{Ahsan2026}, asynchronous operation requires that at most one transition be accepted at a time,
\begin{equation}
\sum_{r=1}^{R}q_r\leq1.
\label{eq:single-commit}
\end{equation}
The switches of the state before the transition determine the toggle delivered to each clause,
\begin{equation}
\chi_k=\sum_{r=1}^{R}\alpha_{k,G_r}\,q_r\in\{0,1\},
\label{eq:decoder-sigma-async}
\end{equation}
and the clause output is updated as
\begin{equation}
T_k\leftarrow\left(1-2\chi_k\right)T_k,\qquad k=1,\ldots,\widetilde M.
\label{eq:decoder-clause}
\end{equation}
The accepted event thus reaches the clause outputs through the same switches as in the encoder, but without the weights $J_k$, because it determines only whether a clause output toggles. This unweighted projection from the \textit{move} neurons to the clause outputs is the decoder (Fig.~\ref{fig:concept}d). In the example, accepting $4\to3$ in $G_2$ gives $(\chi_{1},\ldots,\chi_{5})=(0,1,1,0,1)$, so $T_2$, $T_3$ and the auxiliary clause $T_5$ toggle (Fig.~\ref{fig:toy}c).

\subsection{Stochastic transition dynamics and FN annealing}\label{meth:benchmarks}

\paragraph{Asynchronous updates and their digital emulation} In an asynchronous neuromorphic realization, the condition of Eq.~\eqref{eq:single-commit} follows from the timing of events. The digital experiments emulate this condition by proposing one \textit{move} neuron per iteration. At iteration $n$, a \textit{move} neuron $G_{r_n}$ is drawn from the moves loaded at that iteration, $\mathcal L_n\subseteq\{1,\ldots,R\}$, with fixed positive weights $w_r$,
\begin{equation}
\Pr(r_n=r)=\frac{w_r\,\ind\{r\in\mathcal L_n\}}{\sum_{r'\in\mathcal L_n}w_{r'}}.
\label{eq:proposal}
\end{equation}
Except in the library experiments, every move is loaded and $w_r=1$, so the neuron is drawn uniformly. In the library experiments, the backbone moves carry a weight $w_\mathrm{B}\geq1$, which fixes the fraction of proposals that the backbone receives, and every shortcut carries the weight 1. In batched runs, $\mathcal L_n$ contains the backbone and the batch $b(n)=\lfloor n/D\rfloor\bmod N_\mathrm{b}$ of the $N_\mathrm{b}$ batches of shortcuts, each loaded for $D$ iterations (Section~\ref{res:library}). The stochastic threshold of Eq.~\eqref{eq:firing} at iteration $n$ is
\begin{equation}
\xi_n=-\frac{\tau_n}{2}\ln u_n,\qquad u_n\sim\mathrm{Uniform}(0,1),
\label{eq:threshold}
\end{equation}
which is exponentially distributed with mean $\tau_n/2$ at temperature $\tau_n$.

\paragraph{Metropolis acceptance} Because $\Delta E_{G_{r_n}}=2v_{G_{r_n}}$, Eq.~\eqref{eq:firing} accepts the selected transition with probability $\Pr(\xi_n\geq v_{G_{r_n}})=\min(1,\mathrm{e}^{-\Delta E_{G_{r_n}}/\tau_n})$, the Metropolis rule\cite{Kirkpatrick1983}. With reversible moves, state-independent weights and a self-loop counted as an iteration, like a rejected proposal, the proposal of Eq.~\eqref{eq:proposal} is symmetric (Supplementary Section~\ref{supp:barriers}). At a fixed temperature, the dynamics on a connected library therefore has the Boltzmann distribution $\propto\mathrm{e}^{-E(\svec)/\tau}$ on $\Omega$ as its stationary distribution.

\paragraph{FN annealing schedule} The temperature follows the Fowler--Nordheim (FN) annealing schedule of NeuroSA and N-HOIM\cite{Chen2025,Ahsan2026},
\begin{equation}
\tau_n=\frac{\tau_\mathrm{FN}}{\ln\!\left(1+\dfrac{n+n_0}{C}\right)},\qquad n=0,\ldots,B-1,
\label{eq:fn-schedule}
\end{equation}
for a budget of $B$ iterations. We fix the constants of the schedule by the start and end temperatures, $\tau_\mathrm{FN}=\tau_\mathrm{end}\ln(1+\eta)$, $n_0=q_\mathrm{s}C$ and $C=(B-1)/(\eta-q_\mathrm{s})$ with $q_\mathrm{s}=\mathrm{e}^{\tau_\mathrm{FN}/\tau_\mathrm{start}}-1$, so that the schedule starts at $\tau_\mathrm{start}$ and ends at $\tau_{B-1}=\tau_\mathrm{end}$. The shape parameter is $\eta=0.1$ for every run of our machine except in the three-color comparison of Fig.~\ref{fig:toy}e, which uses the schedule of N-HOIM. The penalty baselines use $\eta=1$, as in N-HOIM (Supplementary Section~\ref{supp:settings}). Because $(n+n_0)/C\leq\eta$, every run of our machine stays in the early regime of the schedule, where $\tau_n\approx\tau_\mathrm{FN}C/(n+n_0)$ falls approximately as $1/n$, rapidly at the beginning of the run and slowly towards its end. The slow, logarithmic regime, for which convergence to the ground state is guaranteed in the limit of slow annealing\cite{Hajek1988}, is reached only for $n\gg C$. The budget $B$ rescales the same schedule to a larger or smaller number of iterations and therefore sets the annealing speed. The schedule and the thresholds are generated digitally. FN tunneling supplies their physical motivation\cite{Zhou2017,Mehta2020,Mehta2022}.

\paragraph{Automatic temperatures} The two temperatures are set for each instance and move library (for the surface code of Fig.~\ref{fig:event}, for its one distance and $p_\mathrm{err}$) from the energy changes of the machine's own proposed transitions, by the acceptance-ratio rule\cite{Kirkpatrick1983}. From a random feasible state, 1,024 transitions are proposed without being applied, and $U_\mathrm{start}$ is the set of their positive energy changes. A short anneal of 8,192 iterations then reaches a low-energy state, from which another 1,024 proposals give the set $U_\mathrm{end}$. The calibrated temperatures solve
\begin{equation}
\frac{1}{|U_\mathrm{start}|}\sum_{\Delta E\in U_\mathrm{start}}\mathrm{e}^{-\Delta E/\hat\tau_\mathrm{start}}=0.8,\qquad
\frac{1}{|U_\mathrm{end}|}\sum_{\Delta E\in U_\mathrm{end}}\mathrm{e}^{-\Delta E/\hat\tau_\mathrm{end}}=0.01,
\label{eq:calibration}
\end{equation}
so that uphill transitions are accepted with average probability 0.8 at the start of the run and 0.01 near low-energy states at its end. The schedule then uses
\begin{equation}
\tau_\mathrm{start}=m\,\hat\tau_\mathrm{start},\qquad \tau_\mathrm{end}=\min\!\left(\hat\tau_\mathrm{end},\,\tau_\mathrm{start}/\kappa\right),
\label{eq:temperatures}
\end{equation}
with $m=1$ and $\kappa=2$ unless these factors are optimized. Because $\hat\tau_\mathrm{end}$ is measured at a different state from $\hat\tau_\mathrm{start}$, $\hat\tau_\mathrm{end}$ can be the larger of the two. The cap at $\tau_\mathrm{start}/\kappa$ guarantees that the temperature of every run decreases by at least a factor of $\kappa$. The calibration requires about $10^4$ proposals per instance and move library. All replicas and budgets of that instance and move library use the same calibrated temperatures, and the calibration is not counted in the iterations. Supplementary Algorithm~\ref{alg:calibrate} gives the procedure.

\paragraph{Settings of the reported results} Apart from the three-color comparison, the event-driven measurements and the decoding results on the FPGA, the annealing settings of our machine differ between results only in the budget and, for the results with optimized schedules, in $m$ and $\kappa$. The automatically calibrated schedule is used in Figs.~\ref{fig:matching}e, \ref{fig:library}a, d, e, g and Supplementary Figs.~\ref{fig:supp-matching}b, c, \ref{fig:supp-control} and \ref{fig:supp-library}a. Every compared variant is calibrated separately on each instance, with its own move library, so that a difference between variants is attributable to the library or the update process and not to schedule optimization. The batching comparison (Fig.~\ref{fig:library}g) is the exception, with temperatures shared between its variants (Section~\ref{res:library}). The surface-code decoders on the FPGA do not anneal. They sample at fixed temperatures with replica exchange (Supplementary Sections~\ref{supp:realtime} and~\ref{supp:oneshot}). For the results with optimized schedules (Figs.~\ref{fig:matching}c, d, f, Supplementary Fig.~\ref{fig:supp-library}b and Supplementary Table~\ref{tab:supp-its10}), $m$ and $\kappa$ are selected for each method, problem set and size on separate training instances, jointly with the budget, to give the smallest median iterations to solution. Only the selected $m$ and $\kappa$ are carried over to the test instances. The iterations to solution on the test instances are again minimized over the budgets, following the standard definition of the time to solution at the optimal run length\cite{Ronnow2014}. The penalty machines are optimized in the same way through their penalty strength and temperature range. The event-driven measurements use one slow anneal per workload, to 1/1,000 of the calibrated start temperature (Supplementary Section~\ref{supp:event}). Supplementary Section~\ref{supp:settings} gives the reason for each setting, the search ranges and the selected values.

\paragraph{Other update protocols} The penalty-based N-HOIM baseline, in the three-color comparison (Fig.~\ref{fig:toy}e) and in graph matching, uses the global arbitration of N-HOIM\cite{Ahsan2026}. Every spin tested counts as one iteration. In the three-color comparison, both machines use the same FN threshold schedule of N-HOIM over 18,000 iterations, with an anneal coefficient of 8. N-HOIM uses the penalty strength 4 on the row and column constraints (Supplementary Section~\ref{supp:matching-formulation}, with $L=3$). Both values were selected for N-HOIM on separate runs (Supplementary Section~\ref{supp:settings}). In all 200 evaluation runs, both machines reach the feasible optimum within this budget. The NeuroSA baseline tests one uniformly drawn spin per iteration\cite{Chen2025}. Colored updates propose all moves of a color class and apply every accepted spin transition in the same step (Supplementary Section~\ref{supp:fpga}).

\subsection{Benchmark problems and simulation methodology}\label{meth:problems}

\paragraph{Surface-code decoding} The decoding instances are generated: for each distance $d$ of the rotated surface code\cite{Bombin2007} and each physical error rate $p_\mathrm{err}$, errors are sampled independently for every shot. For one syndrome per shot, the noise is the depolarizing code-capacity noise of Section~\ref{res:surface}, with $d=3$--15, $p_\mathrm{err}=0.08$--0.19 and 8,000 shots per point. For real time, a shot has $4d$ rounds of the phenomenological noise of Section~\ref{res:surface}, followed by one round measured without error. In the more common bit-flip convention, data and measurement errors both occur at the rate $p$, and the threshold is 2.9\% for minimum-weight matching\cite{Wang2003} and 3.3\% for optimal decoding\cite{Ohno2004}. In our convention, a qubit acquires an $X$ component (from an $X$ or a $Y$ error), and likewise a $Z$ component, with probability $2p_\mathrm{err}/3$ per round, whereas measurement errors occur with probability $p_\mathrm{err}$. Our crossings therefore lie at a higher $p_\mathrm{err}$ than these thresholds. Real-time shots were decoded at $d=5$--11 and $p_\mathrm{err}=0.025$--0.05 (to 0.04 at $d=11$), with 3,000 shots per point and 9,000 near the crossing of $d=7$ and 9 ($p_\mathrm{err}=0.0425$--0.0475). To measure the crossing on shots that were not used to choose the number of sweeps (Supplementary Section~\ref{supp:qec-results}), 6,000 further shots per point were decoded at $p_\mathrm{err}=0.04$ and 0.05 for $d=7$ and 9. All decoders decode the same shots. Supplementary Sections~\ref{supp:surface-formulation} and~\ref{supp:realtime} specify the check matrices, the objectives, the moves, the windows and the failure criteria. The reference decoders come from open-source Python packages. PyMatching (version 2.4.0)\cite{Higgott2022,Higgott2025} is a minimum-weight perfect-matching decoder\cite{Dennis2002} and is used with uniform weights for one syndrome per shot. BP-OSD\cite{Panteleev2021} and union-find\cite{Delfosse2021} come from ldpc (version 2.4.1)\cite{Roffe2020}. BP-OSD uses min-sum belief propagation with scaling factor 0.625 and 50 iterations, followed by combination-sweep ordered-statistics decoding of order 10, on a joint check matrix of the $X$, $Y$ and $Z$ errors. The tensor-network ML decoder\cite{Bravyi2014} of qecsim (version 1.0b9) uses bond dimension 16. In real time, PyMatching, correlated PyMatching\cite{Fowler2013,Paler2023}, BP+matching\cite{Higgott2023} and BP-OSD decode in the same sliding windows as our decoder (Supplementary Section~\ref{supp:realtime}). The settings of the reference decoders were fixed in advance, with one exception: the real-time BP-OSD was rerun with the scaling factor 0.625 used for one syndrome per shot instead of the library default, 1.0. The rerun gave fewer logical errors at every point, and we report it. Published decoders enter the comparisons with their published times and with the LERs that their algorithms reach on our shots. Decoders based on matching, and the neural decoder, are assigned the LER of PyMatching, those based on union-find and clustering that of union-find, and the FPGA implementation of BP+OSD that of BP-OSD. The published annealers are assigned the LER of the lowest-energy correction found by our decoder. In real time, union-find decodes all rounds of a shot at once rather than in windows.

\paragraph{Graph matching} Real instances come from the public WILLOW-ObjectClass dataset\cite{Cho2013}, which provides five object classes with 10 annotated keypoints per image. We use 40 test pairs and 20 training pairs of images, each pair from one class, with disjoint sets of images for test and training, and only the keypoint coordinates. Synthetic instances are generated by deforming random sets of keypoints, with 20 test and 6 training instances at each size $L=10$, 15, 20, 25 and 30 (and also 40 and 50 for the path library only). Supplementary Section~\ref{supp:matching-formulation} gives the objective, the third-order similarity of corresponding keypoint triangles\cite{Duchenne2011,Nguyen2017}, the expansion of the objective into spin products, the generation of the instances and the penalty formulations of the baselines. The reference value $\Psi_\mathrm{ref}$ is the optimum at $L=10$, certified by enumerating all $10!$ permutations. At larger $L$, it is the best value found by three hypergraph-matching solvers with 500 restarts each (200 at $L=40$ and 50). BCAGM3 and Adapt-BCAGM3\cite{Nguyen2017} were run with the 2017 code release of Nguyen et al., and RRWHM\cite{Lee2011} with release 1.1 of the code of Lee et al. (code sources in Supplementary Section~\ref{supp:matching-formulation}). On WILLOW, the fraction of restarts of each reference solver that reach the certified optimum, averaged over the 40 pairs, gives the success probability of one restart in Fig.~\ref{fig:matching}c. Our machine runs 256 replicas per instance, each from an independent random permutation. The penalty-based baselines, N-HOIM\cite{Ahsan2026} and NeuroSA\cite{Chen2025}, add the row and column penalty with strength $\lambda=\bar\lambda \Psi^\ast/L$ to $E$, where $\Psi^\ast$ is the certified optimum and the factor $\bar\lambda$ is optimized (Supplementary Section~\ref{supp:settings}). Setting $\lambda$ through the optimum can only favor them. They run 256 replicas per instance (64 for NeuroSA on WILLOW) from random spin states. NeuroSA reduces the third-order terms with one auxiliary spin per pair of assignment variables that a term uses\cite{Boros2002}. At $L=10$, this adds about 1,000 auxiliary spins per instance to the 100 assignment spins (Supplementary Section~\ref{supp:matching-formulation}).

\paragraph{Performance measures and statistics} For graph matching, a replica succeeds when its best feasible assignment reaches the target stated with each result, $\Psi\geq0.98\,\Psi_\mathrm{ref}$ or the certified optimum. We denote by $P_\mathrm{s}(B)$ the fraction of the replicas of one instance that succeed in a run of budget $B$. The iterations to solution are
\begin{equation}
\mathrm{ITS}=\min_{B}\,\operatorname*{median}_{\text{instances}}\,B\,\frac{\ln0.01}{\ln[1-P_\mathrm{s}(B)]},
\label{eq:its}
\end{equation}
where the term in the median is $B$ for $P_\mathrm{s}(B)=1$ and $\infty$ for $P_\mathrm{s}(B)=0$\cite{Ronnow2014}. Runs that do not reach the target are included in every statistic. Confidence intervals of means over instances, and of the median iterations to solution, are 95\% percentile bootstrap intervals over instances. Two move libraries or variants run on the same instances are compared through the ratio of their iterations to solution or the difference of their mean success probabilities. The 95\% bootstrap interval of this ratio or difference resamples the same instances for the two libraries or variants. For decoding, LERs carry 95\% Wilson intervals. A threshold crossing is the physical error rate at which the LERs of two distances are equal, interpolated linearly between grid points. Every point was decoded in independent batches, each with its own random seeds (1,000 shots per batch for one syndrome per shot, 250 in real time). The 95\% interval of a crossing is a percentile bootstrap interval that resamples the batches of every point with repetition and recomputes the crossing from the pooled shots. Ratios of LERs and the factors by which the LER falls with the distance have their intervals computed in the same way. Two decoders that are compared are resampled with the same batches (Supplementary Section~\ref{supp:qec-results}).

\subsection{Implementation and measurements}\label{meth:fpga}

For the three-color example, graph matching and the operation counts of Fig.~\ref{fig:event}, the machine is emulated on clocked digital hardware. Each replica, or each decoded shot, runs the process of Section~\ref{meth:benchmarks} (Supplementary Algorithm~\ref{alg:solver}) independently, so many of them run in parallel. The code runs on CPUs in double precision, is compiled by Numba and draws random numbers with the Philox generator\cite{Salmon2011}. The input of the proposed \textit{move} neuron is either maintained by the local updates of Supplementary Section~\ref{supp:local-updates} or recomputed from the clause outputs (Section~\ref{res:event}). The penalty baselines use separate CPU implementations of the N-HOIM and NeuroSA updates. The reported quantities of these experiments are iterations and operation counts, not run times.

Supplementary Section~\ref{supp:fpga} describes the FPGA decoder. Each code distance and decoding mode has its own design, generated from its moves, built with Vivado 2022.1 for the XCZU48DR chip of an RFSoC 4x2 board and run with PYNQ 3.0.1 (Supplementary Table~\ref{tab:supp-designs}). For every design, a software model reproduces each clock cycle: the clause bits, the random numbers, the acceptance tests, the replica exchanges and the decisions. Every design decoded at least 300 shots on the board, and its decisions, class counts, lowest energies and cycle counts matched those of the model on every shot. The LERs of our decoder therefore come from the model.

Decode times come from cycle counters on the decoder clock, read for every shot or window, and from the measured clock frequency. A time per shot is measured from the arrival of the syndrome at the decoder to its decision and excludes the transfers between host and board. A time per round is the decode time of a window, from its start to its commit, divided by the $d$ rounds that the window commits. Power is read from the sensor of the core rail of the chip (Supplementary Section~\ref{supp:board}). The CPU decoders were timed on one core of an otherwise idle computing node with two Intel Xeon Gold 6548Y+ processors (2.5\,GHz base and 4.1\,GHz maximum clock frequency). Every software package was limited to one thread. Each decoder was built once and run before the timed calls, so that set-up costs are excluded. BP+matching, however, builds a new matching graph for every shot inside the timed call, because its weights come from belief propagation on that shot. For one syndrome per shot, the CPU time is the mean over 200 shots at $p_\mathrm{err}=0.12$ (100 for tensor-network ML). The mean board times with early stopping come from 20,000 board shots at $p_\mathrm{err}=0.12$. In real time, the CPU time per round is the mean time of a non-final window over 60 shots at $p_\mathrm{err}=0.03$, divided by $d$. The time per round of the board is defined in the same way.

\section{Data availability}\label{sec:data}
The generated problem instances, the raw outputs of all runs and the FPGA bitstreams are available at Zenodo, \url{https://doi.org/10.5281/zenodo.24182700}.

\section{Code availability}\label{sec:code}
The software package, the scripts that run the experiments and reproduce every figure and table, and the FPGA designs are available at \url{https://github.com/aimlab-wustl/Nimbus} (release v1.0), with documentation at \url{https://aimlab-wustl.github.io/Nimbus}, and are archived with the data at Zenodo, \url{https://doi.org/10.5281/zenodo.24182700}.

\FloatBarrier
\bibliography{references}

\section{Acknowledgements}\label{sec:ack}
This work was supported by the National Science Foundation grants ECCS - 2332166 and FET - 2551604. Claude and Gemini AI tools were used for correcting typos, structuring sentences and proof-checking some of the mathematical models and analysis presented in the Supplementary Information.

\section{Author contributions}\label{sec:contributions}
F.A. and S.C. conceived the architecture. F.A. developed the compiler, the emulation and the FPGA decoder and performed the experiments. F.A. and S.C. analyzed the results and wrote the manuscript.

\section{Competing interests}\label{sec:interests}
The authors declare no competing interests.

\clearpage
\setcounter{page}{1}%
\hypersetup{pageanchor=false}%
\setcounter{section}{0}
\setcounter{equation}{0}
\setcounter{figure}{0}
\setcounter{table}{0}
\setcounter{algorithm}{0}
\renewcommand{\thesection}{S\arabic{section}}
\renewcommand{\theequation}{S\arabic{equation}}
\renewcommand{\thefigure}{S\arabic{figure}}
\renewcommand{\thetable}{S\arabic{table}}
\renewcommand{\thealgorithm}{S\arabic{algorithm}}
\providecommand{\theHsection}{}
\renewcommand{\theHsection}{supp.\arabic{section}}
\providecommand{\theHequation}{}
\renewcommand{\theHequation}{supp.\arabic{equation}}
\providecommand{\theHfigure}{}
\renewcommand{\theHfigure}{supp.\arabic{figure}}
\providecommand{\theHtable}{}
\renewcommand{\theHtable}{supp.\arabic{table}}
\providecommand{\theHalgorithm}{}
\renewcommand{\theHalgorithm}{supp.\arabic{algorithm}}
\makeatletter\def\@extra@b@citeb{@SI}\let\NATwroutmain\NAT@wrout
\def\NAT@wrout#1#2#3#4#5{\NATwroutmain{#1}{#2}{#3}{#4}{#5\@extra@b@citeb}}\makeatother
\pdfbookmark[0]{Supplementary Information}{supp-title}
\begin{center}
{\LARGE\bfseries Supplementary Information}\par\vspace{14pt}
{\Large\bfseries Compiling Constraints into\\[2pt] Neuromorphic Ising Machines}\par\vspace{14pt}
{\large Faiek Ahsan and Shantanu Chakrabartty}
\end{center}
\vspace{18pt}
\noindent{\large\bfseries Contents}\par\vspace{6pt}
{\newcommand{\suppentry}[1]{\noindent\hangindent=2.4em\hangafter=1\makebox[2.4em][l]{\ref{#1}}\nameref{#1}\dotfill\pageref{#1}\par\vspace{1pt}}
\suppentry{supp:notation}
\suppentry{supp:closure}
\suppentry{supp:algorithm}
\suppentry{supp:matching-formulation}
\suppentry{supp:settings}
\suppentry{supp:matching}
\suppentry{supp:control}
\suppentry{supp:resources}
\suppentry{supp:librarysupp}
\suppentry{supp:shortcut-theory}
\suppentry{supp:surface-formulation}
\suppentry{supp:oneshot}
\suppentry{supp:realtime}
\suppentry{supp:fpga}
\suppentry{supp:qec-results}
\suppentry{supp:board}
\suppentry{supp:local-updates}
\suppentry{supp:event}
\suppentry{supp:barriers}
\suppentry{supp:onoff}
}
\clearpage

\section{\texorpdfstring{Notation and conventions}{Notation and conventions}}\label{supp:notation}

The table lists the notation of the main text and Methods by topic. Each symbol has one meaning throughout the paper. Index letters (for example $i$, $j$, $l$ and $a$, $b$, $c$) and symbols used in only one Supplementary section are defined where they are used.
{\renewcommand{\theHtable}{supp.notation}
\begin{longtable}{@{}p{0.27\linewidth}p{0.66\linewidth}@{}}
\toprule
Symbol & Meaning\\
\midrule
\endfirsthead
\toprule
Symbol & Meaning\\
\midrule
\endhead
\multicolumn{2}{@{}l}{\textit{States, clauses and energy}}\\
$\svec=(s_1,\ldots,s_N)$ & Spin variables of $N$ spins, $s_i\in\{-1,+1\}$.\\
$\muvec$, $\nuvec$ & Particular feasible states, for example the two states of a spin transition $\muvec\to\nuvec$. The start state is $\muvec_0$, and $\muvec_n$ is the state after iteration $n$.\\
$\Omega$ & Set of feasible states.\\
$H_{ki}$, $S_k$ & Binary matrix of clause membership, $H_{ki}=1$ when spin $i$ appears in stored clause $k$, and the set $S_k=\{i:H_{ki}=1\}$ of the spins of clause $k$.\\
$T_k$ & Clause output, the stored sign $T_k=\prod_i s_i^{H_{ki}}$.\\
$J_k$ & Clause weight; auxiliary clauses have zero weight.\\
$M,M_\mathrm{add},\widetilde M$ & Number of objective clauses, number of auxiliary clauses added by compilation, and total number of stored clauses after compilation, $\widetilde M=M+M_\mathrm{add}$.\\
$E$, $E_\mathrm{const}$ & Energy $E=E_\mathrm{const}-\sum_kJ_kT_k$ and its constant part.\\
$\Delta E_{\muvec\to\nuvec}$, $\Delta E_{G_r}$ & Energy change $E(\nuvec)-E(\muvec)$ of a spin transition, and energy change of the spin transition of move $G_r$ from the current state.\\
\multicolumn{2}{@{}l}{\textit{Spin transitions, moves and \textup{move} neurons}}\\
$G_r$ & The $r$th move, a map of the feasible states into themselves, $\nuvec=G_r(\muvec)$, which produces one spin transition $\muvec\to\nuvec$ from each feasible state ($G_r(\muvec)=\muvec$: a self-loop). A move is reversible when $G_r(\muvec)=\nuvec$ implies $G_r(\nuvec)=\muvec$, and its spin transitions are then listed by unordered pairs $\muvec\leftrightarrow\nuvec$. The same symbol denotes the \textit{move} neuron that represents the move.\\
$R$ & Number of moves in the move library.\\
$\alpha_{k,\muvec\leftrightarrow\nuvec}$ & Equals one when clause $k$ changes sign between the two states of a spin transition. It takes the same value for both directions.\\
$\alpha_{k,G_r}$ & Switch of the connection from clause $k$ to \textit{move} neuron $G_r$: $\alpha_{k,\muvec\leftrightarrow\nuvec}$ for the spin transition of $G_r$ from the current state $\muvec$, with $\nuvec=G_r(\muvec)$, and zero for a self-loop.\\
$T_j$ & Output of the objective or auxiliary clause that controls a state-dependent switch, $\alpha_{k,G_r}=(1\pm T_j)/2$.\\
$v_{G_r}$ & Input of \textit{move} neuron $G_r$, $v_{G_r}=\Delta E_{G_r}/2$.\\
$q_r$, $\xi_r$ & Output of \textit{move} neuron $G_r$, one when its spin transition from the current state is accepted, and its stochastic threshold.\\
$\chi_k$ & Clause toggle, which reverses the clause output $T_k$ when $\chi_k=1$.\\
\multicolumn{2}{@{}l}{\textit{Stochastic dynamics and annealing}}\\
$n$, $B$ & Iteration index and budget, the number of iterations of a run.\\
$r_n$, $u_n$, $\xi_n$ & Index of the \textit{move} neuron proposed at iteration $n$, uniform random number in $(0,1)$, and stochastic acceptance threshold $\xi_n=-(\tau_n/2)\ln u_n$, exponentially distributed with mean $\tau_n/2$.\\
$\mathcal L_n$, $w_r$ & Set of moves loaded at iteration $n$, and proposal weight of move $r$.\\
$\tau_n$, $\tau$ & Annealing temperature at iteration $n$, and a fixed temperature.\\
$\pi(\svec)$ & Boltzmann distribution on $\Omega$ at a fixed temperature $\tau$, $\pi(\svec)\propto\mathrm{e}^{-E(\svec)/\tau}$ (Supplementary Sections~\ref{supp:shortcut-theory} and~\ref{supp:barriers}).\\
$\tau_\mathrm{start},\tau_\mathrm{end}$ & Start and end temperatures of the schedule.\\
$\eta$, $\tau_\mathrm{FN}$, $q_\mathrm{s}$, $C$, $n_0$ & Shape parameter and constants of the Fowler--Nordheim schedule.\\
$\hat\tau_\mathrm{start},\hat\tau_\mathrm{end}$ & Calibrated temperatures, at which uphill transitions are accepted with mean probability 0.8 and 0.01, respectively.\\
$U_\mathrm{start},U_\mathrm{end}$ & Sets of positive energy changes sampled for the calibration.\\
$m,\kappa$ & Factor applied to the calibrated start temperature, and minimum ratio of start to end temperature.\\
\multicolumn{2}{@{}l}{\textit{Library programming and batching}}\\
$K$ & Number of shortcuts added to the backbone (shortcut exchanges in graph matching). In batched libraries, the number of shortcuts in one batch.\\
$w_\mathrm{B}$ & Proposal weight of each backbone move.\\
$\psi_{ia}$, $\phi_i(a)$ & Sum of the similarity weights $\psi_{ijl;abc}$ of the terms that contain the assignment of keypoint $i$ to target $a$, and the resulting distribution over targets, used to choose informed shortcuts.\\
$D$, $N_\mathrm{b}$, $b(n)$, $R_b$ & Replacement interval (iterations for which a batch stays loaded), number of batches, batch loaded at iteration $n$, and number of moves in batch $b$.\\
$h$, $h^\ast$ & Depth of a local minimum (the smallest energy rise needed to reach a state of lower energy from it) and critical depth (the largest depth over all nonoptimal local minima).\\
$\beta$ & Backbone share: the fraction of all proposals given to backbone moves (Supplementary Section~\ref{supp:shortcut-theory}).\\
\multicolumn{2}{@{}l}{\textit{Event-driven updates}}\\
$a$ & Acceptance: accepted state changes per iteration.\\
$N_\mathrm{w}$, $N_\mathrm{acc}$ & Number of iterations in a window of a run, and number of accepted state changes in it; $a=N_\mathrm{acc}/N_\mathrm{w}$.\\
$f_\mathrm{out}$, $f_\mathrm{in}$ & Local updates per accepted state change, and connection reads per iteration.\\
$\rho$ & Sharing ratio $f_\mathrm{out}/f_\mathrm{in}$.\\
$\delta v_{G_r}^{(T,k)}$, $\delta v_{G_r}^{(\alpha,k)}$ & Local updates of $v_{G_r}$ when the clause output $T_k$ changes and when the switch $\alpha_{k,G_r}$ changes (Eqs.~\eqref{eq:event-clause-local} and~\eqref{eq:event-gate-local}, Supplementary Section~\ref{supp:local-updates}).\\
$\ell$ & Order (number of spins) of an interaction term.\\
\multicolumn{2}{@{}l}{\textit{Benchmarks and statistics}}\\
$d$, $p_\mathrm{err}$ & Code distance of the surface code and physical error rate. The code distance is the smallest number of qubits that a logical operator flips (Supplementary Section~\ref{supp:surface-formulation}), and the physical error rate is the probability with which each qubit suffers an error.\\
$s^X_i$, $s^Z_i$; $s^X_{i,t}$, $s^Z_{i,t}$, $s^\mathrm{m}_{j,t}$ & Spins of the $X$ and $Z$ components of a correction on qubit $i$, $-1$ when the correction applies that component. In real time, the same for an error that occurs before round $t$, and the spin that is $-1$ when the outcome of check $j$ in round $t$ is a measurement error (Section~\ref{res:surface}).\\
$\gamma$, $\Omega_\gamma$, $P(\gamma)$, $F_\gamma$ & Logical class, its set of feasible corrections, its probability given the syndrome and its free energy (Section~\ref{res:surface}).\\
$\mathrm{syn}_Z$, $\mathrm{syn}_X$; $H_Z$, $H_X$ & Syndromes of the surface code, the $Z$ checks that the $X$ errors violate and the $X$ checks that the $Z$ errors violate, and the check matrices of the $Z$ and $X$ checks (Supplementary Section~\ref{supp:surface-formulation}).\\
$\omega_3$, $\omega_\mathrm{m}$ & Weight that each qubit with an error adds to the negative log-probability of the correction under depolarizing noise (Supplementary Section~\ref{supp:surface-formulation}), and weight of a measurement error in real time (Supplementary Section~\ref{supp:realtime}).\\
$L$ & Number of positions in each row of a one-hot encoding. In graph matching, $L$ is the number of keypoints, and a permutation has $L$ rows of $L$ positions (Supplementary Sections~\ref{supp:closure} and~\ref{supp:event}).\\
$N_\uparrow$ & Number of $+1$ spins fixed by an exact-cardinality constraint (Supplementary Sections~\ref{supp:closure} and~\ref{supp:barriers}).\\
$x_{ia}$, $s_{ia}$ & Assignment variable, one when keypoint $i$ is assigned to target $a$, and its spin, $x_{ia}=(1+s_{ia})/2$.\\
$\psi_{ijl;abc}$ & Similarity weight of triangle $(i,j,l)$ of image 1 and triangle $(a,b,c)$ of image 2 in the matching objective (Supplementary Section~\ref{supp:matching-formulation}).\\
$\Psi$, $\Psi_\mathrm{ref}$, $\Psi^\ast$ & Matching objective, reference value and certified optimum.\\
$\sigma$ & Standard deviation of the keypoint deformation in synthetic matching instances.\\
$P_\mathrm{pen}$, $\lambda$, $\bar\lambda$ & Penalty of the assignment constraints, its strength $\lambda=\bar\lambda \Psi^\ast/L$, and the strength in units of $\Psi^\ast/L$.\\
$P_\mathrm{s}(B)$, $\mathrm{ITS}$ & Fraction of the replicas of one instance that reach the target in a run of $B$ iterations, and iterations to solution (Methods Eq.~\eqref{eq:its}).\\
$\ind\{\cdot\}$ & Indicator: one when the condition in braces holds and zero otherwise.\\
\bottomrule
\end{longtable}}
\addtocounter{table}{-1}

\section{Number of auxiliary clauses added by compilation}\label{supp:closure}

\begin{table}[!tb]
\centering
\footnotesize
\caption{\textbf{Auxiliary clauses added by compilation for five classes of constraints and moves.} Top: the classes, with the product of spins that a switch requires. Bottom: the bound on the number $M_\mathrm{add}$ of auxiliary clauses under each condition, and its growth with the number of spins $N$ for the largest library ($N=VL$ in (iii) and (iv), $N=L^2$ in (v)). Largest library: all $R=\tbinom N2$ pair exchanges in (ii), all $R=V\tbinom L2$ exchanges within a row in (iii) and all $R=\tbinom V2$ exchanges of two rows in (iv) and (v). In (iv) and (v), the growth assumes objective clauses of bounded order, so that $\ell_2$ is a constant and $\Lambda(L,\ell_2)=O(L^{\ell_2})$. $R$: number of moves in the move library. $N_\uparrow$: number of $+1$ spins among $N$. $V$: number of rows. $L$: number of spins per row. $s_{ia}$: the spin at position $a$ of row $i$. $A$: the positions at which a clause contains the spin of exactly one of the two exchanged rows. $\Lambda(L,\ell_2)$: the number of such position sets that give distinct spin products. $\ell_2$: the largest number of spins that an objective clause has in any two rows together. Many of these problems have standard Ising or QUBO formulations\protect\citeSI{Lucas2014}.}\label{tab:supp-closure}
\begin{tabularx}{\textwidth}{@{}>{\raggedright\arraybackslash}p{0.2\textwidth}>{\raggedright\arraybackslash}X>{\raggedright\arraybackslash}p{0.15\textwidth}>{\raggedright\arraybackslash}p{0.15\textwidth}@{}}
\toprule
Constraint & Example problems & Moves & Product required by a switch\\
\midrule
(i) Parity checks: fixed products of specified sets of spins (a coset of a binary linear code), or no constraint & With parity checks: surface-code decoding (this work) and decoding of other stabilizer and LDPC codes. With no constraint: Max-Cut, spin glasses and MAX-SAT & Fixed reversals: codeword additions with parity checks, single-spin flips with no constraint & None (constant switch)\\
\addlinespace
(ii) Exact cardinality: $N_\uparrow$ of $N$ spins are $+1$ & Graph bisection, densest subgraph of fixed size, maximum coverage, cardinality-constrained portfolio selection, maximum diversity & Pair exchanges & $s_is_j$\\
\addlinespace
(iii) One-hot groups: $V$ rows of $L$ spins, one $+1$ per row & Graph coloring with $L$ colors, Max-$L$-Cut, frequency assignment, timetabling, clustering into $L$ clusters & Exchanges within a row (recoloring one item) & $s_{ia}s_{ib}$\\
\addlinespace
(iv) Partition with prescribed part sizes: as (iii), with a fixed number of $+1$ spins per column & Balanced $L$-way graph partitioning, equitable coloring, load-balanced task assignment, clustering with fixed cluster sizes & Item exchanges (exchange of two rows) & $\prod_{a\in A}s_{ia}s_{ja}$\\
\addlinespace
(v) Permutation: $V=L$, one $+1$ in every row and column & Graph matching (this work), quadratic assignment, graph isomorphism, traveling salesman, linear ordering, single-machine scheduling with sequence-dependent set-up times & Row exchanges & As (iv)\\
\bottomrule
\end{tabularx}

\vspace{10pt}
\begin{tabularx}{\textwidth}{@{}>{\raggedright\arraybackslash}p{0.16\textwidth}>{\raggedright\arraybackslash}X>{\raggedright\arraybackslash}p{0.2\textwidth}>{\raggedright\arraybackslash}p{0.32\textwidth}@{}}
\toprule
Class & Condition & Bound on $M_\mathrm{add}$ & Growth in $N$\\
\midrule
(i) Parity checks & General & $0$ & $0$\\
\addlinespace
(ii) Cardinality & General & $R$ & $O(N^2)$\\
\addlinespace
(iii) One-hot & General & $R$ & $O(N^2)$; $O(N)$ for fixed $L$\\
 & $L\le3$ & $0$ & $0$\\
\addlinespace
(iv) Partition & $\ell_2\le2$ & $R\,\Lambda(L,\ell_2)$ & $O(N^2)$\\
 & $\ell_2>2$ & $R\,\Lambda(L,\ell_2)$ & $O(N^{\ell_2})$\\
 & Cut objectives & $LR$ & $O(N^2)$\\
\addlinespace
(v) Permutation & General & $R\,\Lambda(L,\ell_2)$ & $O(N^{1+\ell_2/2})$\\
 & $\ell_2\le2$ and $L\ge5$ & $R\bigl(L+\tbinom L2\bigr)$ & $O(N^2)$\\
 & Traveling salesman & $2LR$ & $O(N^{3/2})$\\
\bottomrule
\end{tabularx}
\end{table}

We bound the number $M_\mathrm{add}$ of auxiliary clauses that compilation adds to supply switches (Methods Section~\ref{meth:model}) for five classes of constraints and moves (Supplementary Table~\ref{tab:supp-closure}). For many standard and practically relevant problems, $M_\mathrm{add}$ grows only polynomially with the problem size when the move library is suitably chosen. The classes include the fixed spin reversals of surface-code decoding and the row exchanges of graph matching.

The total number of stored clauses is $\widetilde M=M+M_\mathrm{add}$, and $S_k$ is the set of spins of clause $k$.

We consider moves that permute or reverse spins, as in both applications. Such a move maps the spins $\svec$ to $G_r(\svec)$ with
\begin{equation*}
[G_r(\svec)]_i=\begin{cases}-s_{g(i)},&i\in\mathcal F,\\ \phantom{-}s_{g(i)},&i\notin\mathcal F,\end{cases}
\end{equation*}
where $g$ is a fixed permutation of the spins and $\mathcal F$ is a fixed set of spins. The value of clause $k$ after the move is then $T_k(G_r(\svec))=\epsilon\prod_{i\in S_k}s_{g(i)}$, with the constant sign $\epsilon=(-1)^{|S_k\cap\mathcal F|}$. In $T_k(G_r(\svec))T_k(\svec)$, a spin in both $S_k$ and $g(S_k)=\{g(i):i\in S_k\}$ drops out, because $s_i^2=1$, so
\begin{equation}
T_k(G_r(\svec))T_k(\svec)=\epsilon\prod_{i\in S_k\triangle g(S_k)}s_i.
\label{eq:supp-perm-response}
\end{equation}
Here $S_k\triangle g(S_k)$ is the symmetric difference, the spins that belong to exactly one of the two sets. The switch $\alpha_{k,G_r}=[1-T_k(G_r(\svec))T_k(\svec)]/2$ (Methods Eq.~\eqref{eq:alpha-definition}) is therefore set by the product over $S_k\triangle g(S_k)$. It is constant when this set is empty. Otherwise, the product must be available as a stored clause output. The function \textsc{ClauseChange} of Supplementary Algorithm~\ref{alg:compile} evaluates the two special cases in which $\mathcal F$ is empty or $g$ is the identity.

Compilation generates the required products iteratively (Supplementary Algorithm~\ref{alg:compile}). It starts from the products that the switches of the stored clauses require and stores each new product as an auxiliary clause, whose switches may require further products. If all products that a family of moves requires in this way lie in a collection whose products require nothing outside it, $M_\mathrm{add}$ is at most the size of this collection. Compilation always ends, because there are finitely many spin products. Reuse can only lower $M_\mathrm{add}$. A required product may already be stored, or it may equal a stored clause output, up to sign, in every feasible state. Such equalities follow from identities of the feasible set, relations among spin products that every feasible state satisfies. In the example of Methods Section~\ref{meth:model}, exactly one color occupies $p_1$, so $s_1s_4s_7=1$, and the product $s_1s_4$ is replaced by the stored clause $s_7$.

\subsection{Parity constraints: fixed reversals}

Parity constraints fix the product of the spins in each of several specified sets, the checks of a code. A codeword is a set of spins that shares an even number of spins with every check. Reversing the spins of a codeword, which we call a codeword addition, leaves every check product unchanged and therefore maps feasible states to feasible states. The codewords form a linear code, a set of binary patterns closed under addition modulo 2. Every feasible state is one fixed feasible state with the spins of some codeword reversed, so the feasible states form a coset of this code.

A move that always reverses the same set $\mathcal F$ of spins has $g$ equal to the identity, so $S_k\triangle g(S_k)$ is empty. Eq.~\eqref{eq:supp-perm-response} then gives, for every stored clause and every feasible state,
\begin{equation}
T_k(G_r(\svec))T_k(\svec)=(-1)^{|S_k\cap\mathcal F|}.
\label{eq:supp-fixed-response}
\end{equation}
The right-hand side does not depend on the state, so a clause changes sign either in every transition of the move or in none. The switch is therefore constant, $\alpha_{k,G_r}=1$ when $|S_k\cap\mathcal F|$ is odd and $\alpha_{k,G_r}=0$ otherwise. No product is required, and $M_\mathrm{add}=0$. This case covers the stabilizer and logical-operator moves of the surface code and the codeword additions of other stabilizer and LDPC codes. A single-spin flip is the fixed reversal of one spin, so unconstrained problems need no auxiliary clauses either.

\subsection{Exact cardinality and one-hot groups: pair exchanges}

An exact-cardinality constraint fixes the number $N_\uparrow$ of $+1$ spins among $N$. An exchange of the values of two spins $s_i$ and $s_j$ preserves this number. It is the permutation $g$ that swaps $i$ and $j$, with $\mathcal F$ empty. The set $S_k\triangle g(S_k)$ is empty for a clause that contains both spins or neither, and $\{i,j\}$ for a clause that contains exactly one of them. The switch of clause $k$ is therefore
\begin{equation}
\alpha_{k,G_r}=\begin{cases}0,&H_{ki}=H_{kj},\\(1-s_is_j)/2,&H_{ki}\ne H_{kj}.\end{cases}
\label{eq:supp-cardinality-response}
\end{equation}
Each exchange of $i$ and $j$ thus requires at most the product $s_is_j$. If $s_is_j$ is stored as an auxiliary clause, an exchange of $i$ or $j$ with a third spin $l$ requires $s_is_l$ or $s_js_l$, respectively, the product of that exchange itself. Every other exchange leaves this auxiliary clause unchanged. Every required product is therefore the pair product of one exchange of the move library. For a library of $R$ pair exchanges,
\begin{equation}
M_\mathrm{add}\le R.
\label{eq:supp-cardinality-bound}
\end{equation}
The bound does not depend on the order of the objective clauses. It is $\binom N2$ for the complete library and $N-1$ for exchanges along a spanning tree of the spins, a set of $N-1$ exchanges that connects all spins without forming a cycle.

One-hot groups encode categorical variables. Each of $V$ items takes one of $L$ values and is represented by a row of $L$ spins $s_{i1},\ldots,s_{iL}$, with $s_{ia}=+1$ when item $i$ takes value $a$. The constraint is an exact cardinality of one in every row. An exchange of $s_{ia}$ and $s_{ib}$ moves the $+1$ of row $i$ between positions $a$ and $b$, which recolors one item. The argument above gives $M_\mathrm{add}\le R\le VL(L-1)/2$, the number of all recolorings. Identities of the feasible set remove the need for auxiliary clauses in small groups. For $L=2$, $s_{i1}s_{i2}=-1$ in every feasible state, so the switch is constant. For $L=3$, $s_{ia}s_{ib}=s_{ic}$, where $c$ is the third position: both sides are $+1$ exactly when the item takes value $c$. The switch is then set by the single spin $s_{ic}$, which the machine already stores (Methods Section~\ref{meth:model}). Hence $M_\mathrm{add}=0$ for $L\le3$.

\subsection{Partitions and permutations: row exchanges}

In a partition with prescribed part sizes, each of $V$ items is placed in one of $L$ parts, and the number of items in each part is fixed. In the one-hot encoding, $s_{ia}=+1$ when item $i$ is in part $a$. Every row then contains one $+1$, and every column contains a fixed number of them. A permutation is the case $V=L$ with one $+1$ in every row and every column. In graph matching, row $i$ is a keypoint of image 1 and position $a$ a keypoint of image 2. An exchange of rows $i$ and $j$ swaps the parts, or targets, of items $i$ and $j$ and preserves the counts of every row and every column.

The exchange of rows $i$ and $j$ is the permutation $g$ that swaps $s_{ia}$ and $s_{ja}$ at every position $a$. Let $H_{k,ia}=1$ when clause $k$ contains $s_{ia}$. A position at which the clause contains both spins, or neither, leaves its factor unchanged, so the required product is fixed by the positions
\begin{equation}
A=\{a:H_{k,ia}\oplus H_{k,ja}=1\}.
\label{eq:supp-row-subset}
\end{equation}
Here $\oplus$ denotes addition modulo 2. At each position in $A$, the exchange replaces the spin of one row by that of the other, which multiplies $T_k$ by $s_{ia}s_{ja}$. Eq.~\eqref{eq:supp-perm-response} gives, for every feasible state,
\begin{equation}
T_k(G_r(\svec))T_k(\svec)=\prod_{a\in A}s_{ia}s_{ja},\qquad \alpha_{k,G_r}=\frac{1-\prod_{a\in A}s_{ia}s_{ja}}{2}.
\label{eq:supp-row-response}
\end{equation}
This cancellation does not use the one-hot constraint.

Suppose the product $\prod_{a\in A}s_{ia}s_{ja}$ becomes an auxiliary clause. It contains only the spins of rows $i$ and $j$ at the positions of $A$. Exchanging row $i$ or $j$ with a third row $l$ therefore requires the same position set $A$, with row $l$ in place of row $j$ or $i$, respectively. Every other exchange leaves the auxiliary clause unchanged. The recursion can thus change the pair of rows but never the position set. Every required product is fixed by one position set generated by a stored clause and by one exchange of the move library. Let $\mathcal A_0$ be the set of distinct position sets that the objective clauses generate, together with the $L$ sets $\{a\}$ of one position, without those whose product is the same in every feasible state. For a library of $R$ row exchanges,
\begin{equation}
M_\mathrm{add}\le R\,|\mathcal A_0|.
\label{eq:supp-row-basic-bound}
\end{equation}
A clause contributes to $A$ only through its spins in rows $i$ and $j$. Let $\ell_2\ge1$ be the largest number of spins that an objective clause has in any two rows. Then $|A|\le\ell_2$, so $|\mathcal A_0|\le\sum_{\ell'=1}^{\min(\ell_2,L)}\binom L{\ell'}$. Because $\binom L{\ell'}\le L^{\ell'}$, this sum is at most $\ell_2L^{\ell_2}$. The bound is therefore polynomial in $L$ at fixed $\ell_2$ for a polynomial number of exchanges, even before any identity is used.

In a partition, as in a permutation, every row contains one $+1$ and $L-1$ spins $-1$, so $\prod_{a=1}^{L}s_{ia}s_{ja}=1$ for $i\ne j$. Complementary position sets therefore give the same product. Counting only the smaller set of each complementary pair, and half of the sets of size $L/2$ when $L$ is even, gives
\begin{equation}
\Lambda(L,\ell_2)=\sum_{\ell'=1}^{\min(\ell_2,\lfloor(L-1)/2\rfloor)}\binom L{\ell'}+\ind\{L\text{ even},\ \ell_2\ge L/2\}\frac12\binom L{L/2}.
\label{eq:supp-row-B}
\end{equation}
Thus $M_\mathrm{add}\le R\,\Lambda(L,\ell_2)$.

When every clause has at most one spin in each row, $\ell_2\le2$, whatever the order of the objective. This holds for graph matching, quadratic assignment, graph isomorphism and the traveling salesman problem, in which every term couples different items. Third-order graph matching therefore needs no more auxiliary clauses than a quadratic objective. For $L\ge5$, $\Lambda(L,2)=L+\binom L2$, and the compiled graph-matching machines attain the bound $M_\mathrm{add}=R\bigl(L+L(L-1)/2\bigr)$ exactly (Fig.~\ref{fig:supp-size}c), as did random quadratic-assignment objectives in our numerical tests. Since $\Lambda(L,2)$ grows as $L^2/2$, the bound is $O(L^4)=O(N^2)$ for the complete library of $\binom L2$ exchanges and $O(L^3)=O(N^{3/2})$ for a connected library of $L-1$ exchanges, with $N=L^2$ spins. In a partition with a cut objective, every term couples items in the same part, so all spins of a clause lie in one column. Then $|A|\le1$, and $M_\mathrm{add}\le LR$.

In the standard permutation encoding of the traveling salesman problem, row $i$ is a city and position $a$ its place in the tour. The tour length sums the distances between the cities at positions $a$ and $a+1$, with position $L+1$ being position 1. Expanded into spin products, it has the clauses $s_{ia}$ and $s_{ia}s_{j,a+1}$. By Eq.~\eqref{eq:supp-row-subset}, this objective generates only the at most $2L$ sets $\{a\}$ and $\{a,a+1\}$, which gives the tighter bound
\begin{equation}
M_\mathrm{add}^\mathrm{TSP}\le 2LR.
\label{eq:supp-tsp-gate-bound}
\end{equation}
This is $O(L^3)$ for the complete row-exchange library and $O(L^2)$ for a connected library of $L-1$ exchanges. Single-machine scheduling with sequence-dependent set-up times has the same structure, with jobs in place of cities, and the same bound. The bounds count auxiliary clauses, not connections.

A cyclic exchange of three rows, which permutes the targets of three keypoints, is also a fixed permutation of the spins, which changes the row of every spin but not its position $a$. Let $P_i$ be the set of positions of the spins of a clause in row $i$. For the cycle in which row $i$ receives the content of row $j$, row $j$ that of row $l$, and row $l$ that of row $i$, the required product has the position sets $P_i\triangle P_l$, $P_j\triangle P_i$ and $P_l\triangle P_j$ in rows $i$, $j$ and $l$, and no spin in other rows. Every required product therefore lies in at most three rows, with position sets formed by symmetric differences of those of one stored clause. Their number is polynomial in $L$ at fixed objective order. With a quadratic-assignment objective and all $2\binom L3$ such cycles, the compiled machines had exactly $3L(L-1)\binom L3+\binom L2\Lambda(L,2)$ auxiliary clauses for $L=5$ to 10, a number that grows as $L^5/2$.

\section{Algorithmic implementation}\label{supp:algorithm}

\begin{algorithm}[!tbp]
\caption{Compilation of a move library}\label{alg:compile}
\small
\begin{algorithmic}[1]
\Require spins $s_1,\ldots,s_N$; objective clauses $T_k=\prod_{i\in S_k}s_i$ with spin sets $S_k$ and weights $J_k$, $k=1,\ldots,M$; moves $G_1,\ldots,G_R$, each a map $\svec\mapsto G_r(\svec)$ of the feasible states into themselves; identities of the feasible set $\Omega$, spin products with the same value in every feasible state (possibly none)
\Ensure stored clauses $T_1,\ldots,T_{\widetilde M}$ with $J_k=0$ for $k>M$; switch rules $\alpha_{k,G_r}$ (a connection exists where $\alpha_{k,G_r}\not\equiv0$)
\Function{ClauseChange}{$k,r$} \Comment{$T_k(G_r(\svec))T_k(\svec)=\epsilon\prod_{i\in A}s_i$ for all $\svec\in\Omega$}
  \If{$[G_r(\svec)]_i=-s_i$ for $i$ in a fixed spin set $\mathcal F$ and $[G_r(\svec)]_i=s_i$ otherwise}
    \State \Return $\epsilon=(-1)^{|S_k\cap\mathcal F|}$, $A=\emptyset$
  \ElsIf{$[G_r(\svec)]_i=s_{g(i)}$ for a fixed permutation $g$ of $1,\ldots,N$}
    \State $S'_k\gets\{g(i):i\in S_k\}$ \Comment{$T_k(G_r(\svec))=\prod_{i\in S'_k}s_i$}
    \State \Return $\epsilon=1$, $A=S_k\triangle S'_k$ \Comment{spins in exactly one of the two sets}
  \Else
    \State \Return $\epsilon$ and $A$ derived from the map $G_r$
  \EndIf
\EndFunction
\State add the singleton clause $s_i$ with weight 0 for every spin $i$ for which $s_i$ is not an objective clause
\State $\alpha_{k,G_r}\gets0$ for all $k,r$; $\mathcal Q\gets$ queue of all stored clauses
\While{$\mathcal Q\neq\emptyset$}
  \State remove a clause $k$ from $\mathcal Q$
  \For{each move $r$ that changes a spin in $S_k$}
    \State $(\epsilon,A)\gets$ \Call{ClauseChange}{$k,r$}
    \If{$\prod_{i\in A}s_i$ takes the same value $\theta\in\{\pm1\}$ in every feasible state}
      \State \textbf{if} $\epsilon\theta=-1$ \textbf{then} $\alpha_{k,G_r}\gets1$ \Comment{clause always changes}
    \Else
      \If{no stored clause $T_j$ equals $\pm\prod_{i\in A}s_i$ in every feasible state}
        \State add the auxiliary clause $T_j=\prod_{i\in A}s_i$ with $J_j=0$; append $j$ to $\mathcal Q$
      \EndIf
      \State $\zeta\gets$ the sign with $\prod_{i\in A}s_i=\zeta T_j$ in every feasible state
      \State $\alpha_{k,G_r}\gets(1-\epsilon\zeta T_j)/2$ \Comment{on when clause $k$ changes}
    \EndIf
  \EndFor
\EndWhile
\end{algorithmic}
\end{algorithm}
\begin{algorithm}[!tbp]
\caption{Annealing of one replica}\label{alg:solver}
\small
\begin{algorithmic}[1]
\Require clauses $T_k=\prod_{i\in S_k}s_i$ with weights $J_k$ and switch rules $\alpha_{k,G_r}$, each equal to 0, 1 or $(1\pm T_j)/2$, with a sign fixed for the connection, for one stored clause $T_j$ (Algorithm~\ref{alg:compile}); constant $E_\mathrm{const}$; feasible start state $\muvec_0$; proposal probabilities over the loaded moves (all of them unless the move library is loaded in batches); temperatures $\tau_0,\ldots,\tau_{B-1}$
\Ensure lowest-energy state visited, $\muvec_\mathrm{best}$, and its energy $E_\mathrm{best}$
\State $T_k\gets\prod_{i\in S_k}\mu_{0,i}$ for all $k$; evaluate all switches $\alpha_{k,G_r}$
\State $v_{G_r}\gets\sum_kJ_k\alpha_{k,G_r}T_k$ for every loaded move $r$ \Comment{$2v_{G_r}$ = energy change}
\State $E\gets E_\mathrm{const}-\sum_kJ_kT_k$; $E_\mathrm{best}\gets E$; $\muvec_\mathrm{best}\gets\muvec_0$
\For{$n=0,1,\ldots,B-1$}
  \State draw a move $r$ from the proposal probabilities
  \State draw $u\sim\mathrm{Uniform}(0,1)$; $\xi\gets-(\tau_n/2)\ln u$
  \If{$v_{G_r}\leq\xi$} \Comment{accept}
    \State $\mathcal K\gets\{k:\alpha_{k,G_r}=1\}$ \Comment{clauses that change}
    \State $E\gets E+2v_{G_r}$
    \For{each $k\in\mathcal K$}
      \State $T_k\gets-T_k$
      \State $v_{G_{r'}}\gets v_{G_{r'}}+2J_kT_k$ for every loaded move $r'$ with $\alpha_{k,G_{r'}}=1$
    \EndFor
    \For{each connection $(k,r')$ whose switch rule depends on a clause $T_j$ with $j\in\mathcal K$}
      \State $\alpha'\gets$ switch rule of $(k,r')$ evaluated with the new $T_j$
      \State $v_{G_{r'}}\gets v_{G_{r'}}+J_kT_k(\alpha'-\alpha_{k,G_{r'}})$; $\alpha_{k,G_{r'}}\gets\alpha'$
    \EndFor
    \If{$E<E_\mathrm{best}$}
      \State $E_\mathrm{best}\gets E$; $\muvec_\mathrm{best}\gets$ values of the singleton clauses $s_1,\ldots,s_N$
    \EndIf
  \EndIf
\EndFor
\State \Return $\muvec_\mathrm{best}$, $E_\mathrm{best}$
\end{algorithmic}
\end{algorithm}
\begin{algorithm}[!tbp]
\caption{Automatic start and end temperatures}\label{alg:calibrate}
\small
\begin{algorithmic}[1]
\Require compiled library; random feasible state $\muvec$; factors $m>0$ and $\kappa\geq1$
\Ensure temperatures $\tau_\mathrm{start}$ and $\tau_\mathrm{end}$
\Function{AcceptanceTemperature}{$\muvec,a_\mathrm{target}$} \Comment{$a_\mathrm{target}$: target mean acceptance}
  \State draw 1,024 moves $r$ from the proposal probabilities; evaluate $v_{G_r}$ at $\muvec$ as in Algorithm~\ref{alg:solver}, line~2
  \State $\Delta E_{G_r}\gets2v_{G_r}$ for each drawn move $r$ \Comment{energy change; not applied}
  \State $U\gets\{\Delta E_{G_r}:\Delta E_{G_r}>0\}$; $N_+\gets|U|$ \Comment{uphill changes and their number}
  \State \textbf{if} $N_+<8$ \textbf{then} add the uphill changes of 1,024 further draws at $\muvec$ to $U$; update $N_+$
  \State \textbf{if} $N_+<8$ \textbf{then} \Return the fallback temperature described below the algorithm
  \State \Return $\tau$ such that $\frac{1}{N_+}\sum_{\Delta E\in U}\mathrm{e}^{-\Delta E/\tau}=a_\mathrm{target}$ \Comment{mean acceptance; bisection in $\ln\tau$}
\EndFunction
\State $\hat\tau_\mathrm{start}\gets$ \Call{AcceptanceTemperature}{$\muvec,0.8$}
\State $\muvec'\gets$ final state of Algorithm~\ref{alg:solver} from $\muvec$, run for $B=8{,}192$ iterations with the Fowler--Nordheim schedule from $\hat\tau_\mathrm{start}$ to $0.02\,\hat\tau_\mathrm{start}$
\State $\hat\tau_\mathrm{end}\gets$ \Call{AcceptanceTemperature}{$\muvec',0.01$}
\State $\tau_\mathrm{start}\gets m\,\hat\tau_\mathrm{start}$
\State $\tau_\mathrm{end}\gets\min(\hat\tau_\mathrm{end},\ \tau_\mathrm{start}/\kappa,\ \tau_\mathrm{start})$
\State \Return $\tau_\mathrm{start}$, $\tau_\mathrm{end}$
\end{algorithmic}
\end{algorithm}

This section gives the three algorithms of the machine: the compilation of a move library (Supplementary Algorithm~\ref{alg:compile}; Methods Section~\ref{meth:model}), the annealing of one replica (Supplementary Algorithm~\ref{alg:solver}; Methods Section~\ref{meth:benchmarks}) and the automatic choice of the start and end temperatures (Supplementary Algorithm~\ref{alg:calibrate}). A replica is one run of the machine from its own start state. Replicas are independent except in the decoders, where they exchange temperatures.

Supplementary Algorithm~\ref{alg:compile} connects each move to the stored clauses that its spin transitions change and gives each connection its switch. Its inputs include the moves and the identities of the feasible set. A required product that equals a stored clause output, up to sign, in every feasible state needs no auxiliary clause (Supplementary Section~\ref{supp:closure}). Because $s_i^2=1$, multiplying a product by identities can cancel spins, so a new auxiliary clause is stored as an equivalent product with as few spins as the identities allow. In this work the identities follow from the structure of each problem. The surface-code moves reverse fixed sets of spins, and no identity is used. For graph matching, the identities state that the product of the spins of every row and of every column equals $(-1)^{L-1}$, because each holds one spin $+1$ and $L-1$ spins $-1$.

Supplementary Algorithm~\ref{alg:solver} anneals one replica from a feasible start state. The zero-weight singleton clauses $s_i$ added by compilation store the spin state, from which the lowest-energy state is read out. The inputs can also be recomputed by Methods Eq.~\eqref{eq:encoder} when a move is drawn instead of being maintained, with the same acceptance decisions (Methods Section~\ref{meth:fpga}). In the batched libraries of Fig.~\ref{fig:library}g, the loaded moves change every $D$ iterations. The auxiliary clauses, switches and inputs of the newly loaded moves are evaluated from the current state as in lines~1 and~2 of the algorithm. In colored updates (Supplementary Section~\ref{supp:fpga}), one step proposes every move of one color class at one temperature, each with its own $u$. It then applies the accepted spin transitions and advances $n$ by the size of the color class.

Supplementary Algorithm~\ref{alg:calibrate} sets the two temperatures of an instance and move library by the acceptance-ratio rule (Methods Eq.~\eqref{eq:calibration}). The mean acceptance in \textsc{AcceptanceTemperature} rises monotonically with $\tau$, from 0 towards 1, because every term $\mathrm{e}^{-\Delta E/\tau}$ with $\Delta E>0$ does, so bisection finds the unique temperature that gives $a_\mathrm{target}$. Supplementary Section~\ref{supp:settings} lists the results whose temperatures are not set by this algorithm. For the surface code, the term $\tau_\mathrm{start}/\kappa$ is dropped from the minimum that sets $\tau_\mathrm{end}$.

If the end sample still holds fewer than eight positive changes, it is pooled with the start sample. If a sample remains too small, the temperature is set from the clause weights to $\tau=\Delta E_J/\ln(1/a_\mathrm{target})$, the temperature at which a single uphill change $\Delta E_J$ is accepted with probability $a_\mathrm{target}$. For the start temperature, $\Delta E_J$ is the median over moves $G_r$ of $2\sum_{k:\alpha_{k,G_r}\not\equiv0}|J_k|$, the sum over the clauses connected to $G_r$. Because $|v_{G_r}|$ cannot exceed half this sum, the sum is an upper bound on the energy change of any spin transition of $G_r$. For the end temperature, $\Delta E_J$ is twice the smallest nonzero $|J_k|$, the smallest energy change that one clause can cause. This fallback set the start temperature in 9 of the 1,546 calibrations underlying the reported results, all in graph matching, and the end temperature in one of these nine. In these nine calibrations, none of the $2\times1{,}024$ sampled transitions at the random state raised the energy. Because the fallback uses an upper bound rather than sampled changes, these runs started at much higher temperatures than the other graph-matching runs.

\FloatBarrier
\section{Graph matching: objective, instances and penalty formulations}\label{supp:matching-formulation}

\paragraph{Objective} Keypoint $i$ of image 1 is assigned to keypoint $a$ of image 2 when $x_{ia}=(1+s_{ia})/2=1$. The $L^2$ spins thus form an assignment matrix, and in a feasible state each of its rows and columns contains one $+1$. The objective is the third-order similarity of corresponding keypoint triangles\citeSI{Duchenne2011,Nguyen2017},
\begin{equation}
\Psi(\mathbf{x})=\sum_{(i,j,l;\,a,b,c)}\psi_{ijl;abc}\,x_{ia}\,x_{jb}\,x_{lc},
\label{eq:matching-objective}
\end{equation}
summed over pairs of triangles whose interior angles are similar, with weights $\psi_{ijl;abc}\in(0,1]$. The product $x_{ia}x_{jb}x_{lc}$ equals one only when the assignment maps triangle $(i,j,l)$ of image 1 onto triangle $(a,b,c)$ of image 2. The objective $\Psi$ therefore sums the similarities of all triangle pairs that the assignment puts into correspondence.

\paragraph{Triangle similarity} As in refs~\citeSI{Duchenne2011,Lee2011}, we sample triangles of keypoints of image 1 without replacement and link each to the triangles of image 2 whose features are nearest to its own. The feature of a triangle is the vector of the sines of its interior angles. This vector is invariant under rotation, translation and uniform scaling, so corresponding triangles keep similar features under the deformations of the instances. A linked pair of triangles $(i,j,l)$ and $(a,b,c)$ receives the weight $\psi_{ijl;abc}=\exp(-\Delta_\mathrm{f}^2/\bar\Delta_\mathrm{f}^2)$, with $\Delta_\mathrm{f}$ the distance between the two features and $\bar\Delta_\mathrm{f}$ the mean distance over all linked pairs\citeSI{Duchenne2011,Nguyen2017}. We sample $L^2$ triangles and link $3L$ triangles to each sampled triangle. Every keypoint then lies in about $3L$ sampled triangles, and Eq.~\eqref{eq:matching-objective} has $3L^3$ entries, a sparse sample of all triangle pairs that every solver receives unchanged. Every triangle is formed from three distinct keypoints, so each entry multiplies three distinct assignment variables.

\paragraph{Objective as spin products} With $x_{ia}=(1+s_{ia})/2$, each entry of Eq.~\eqref{eq:matching-objective} becomes
\begin{equation}
\psi\,x_{ia}x_{jb}x_{lc}=\frac{\psi}{8}\left(1+s_{ia}+s_{jb}+s_{lc}+s_{ia}s_{jb}+s_{ia}s_{lc}+s_{jb}s_{lc}+s_{ia}s_{jb}s_{lc}\right).
\label{eq:supp-matching-expansion}
\end{equation}
Collecting the terms of $E=-\Psi$ that contain the same spin product gives
\begin{equation}
E(\svec)=E_\mathrm{const}-\sum_{k=1}^{M}J_kT_k(\svec),\qquad J_k=\frac{1}{8}\sum_{\text{entries containing }T_k}\psi,
\end{equation}
in the form of Methods Eq.~\eqref{eq:energy-clauses}, with clauses $T_k$ of first, second and third order. All weights are positive, so every clause favors the output $T_k=+1$. The assignment constraints are not part of $E$.

\paragraph{Instances and reference values} Synthetic instances place $L$ keypoints uniformly in the unit square and map them to image 2 by a random rotation (0--45$^\circ$), a random scale (0.8--1.2) and Gaussian noise of standard deviation $\sigma=0.05$ ($\sigma=0.10$ in Supplementary Section~\ref{supp:matching}). In each of the five object classes of WILLOW-ObjectClass\citeSI{Cho2013}, every third image in file order is reserved for training, and pairs are drawn uniformly at random, with a fixed seed, within one subset (training or test): 4 training and 8 test pairs per class. In all instances, the order of the target keypoints is shuffled. At $L>10$, $\Psi_\mathrm{ref}$ is the best value of Eq.~\eqref{eq:matching-objective} found over all restarts of three hypergraph-matching solvers (Methods Section~\ref{meth:problems}). Hypergraph matching is graph matching with third-order terms, so called because each term links three correspondences rather than two. The three solvers use fixed settings. BCAGM3 and Adapt-BCAGM3 increase a multilinear form of the objective by block-coordinate ascent, with at most 50 iterations per restart\citeSI{Nguyen2017}. RRWHM ranks candidate correspondences by reweighted random walks on the hypergraph of triangle pairs, with at most 300 iterations per restart\citeSI{Lee2011}. The code of BCAGM3 and Adapt-BCAGM3\citeSI{Nguyen2017} was taken from a public mirror (\url{https://github.com/smallsmartboy/An-Efficient-Multilinear-Optimization-Framework-for-Hypergraph-Matching}, commit dd746b1). RRWHM was run with the same mirror's version of release 1.1 of the code of Lee et al.\citeSI{Lee2011} (\url{https://cv.snu.ac.kr/research/~RRWHM/}). This version differs from the release of Lee et al. only in accepting a starting assignment for random restarts.

\paragraph{Move libraries} The row exchange of keypoints $i$ and $j$ exchanges $s_{ia}$ with $s_{ja}$ for every $a$, a fixed permutation of spins. Its switches are state dependent (main text Section~\ref{res:matching}), and the compiler supplies them through $L+L(L-1)/2$ auxiliary clauses per exchange (Supplementary Section~\ref{supp:closure}). Informed shortcuts (main text Section~\ref{res:library}) exchange the pairs of keypoints $i$ and $j$ with the largest overlap $\sum_a\min[\phi_i(a),\phi_j(a)]$, where $\phi_i(a)=\psi_{ia}/\sum_b\psi_{ib}$. The summed similarity $\psi_{ia}$ is the sum of the weights $\psi_{ijl;abc}$ over all terms of Eq.~\eqref{eq:matching-objective} that contain $x_{ia}$. A large $\psi_{ia}$ means that many triangles through keypoint $i$ resemble triangles through target $a$. The overlap is the probability mass that the two distributions have in common, 1 when they are equal and 0 when they share no target. Summing over $a$ gives one value per pair of keypoints, as an exchange swaps whatever targets its two keypoints hold. Computing all overlaps takes $O(L^3)$ operations, which is negligible next to compilation. Informed and random shortcuts are drawn from the same pairs outside the backbone, so libraries with equal $K$ differ only in which shortcuts they add. Supplementary Section~\ref{supp:librarysupp} gives the resulting backbone shares.

\paragraph{Penalty formulation of the assignment constraints} A single spin flip turns a permutation matrix into an infeasible assignment, so single-spin-flip machines enforce the constraints through the penalty
\begin{equation}
P_\mathrm{pen}(\mathbf{x})=\sum_{i=1}^{L}\Bigl(\sum_{a=1}^{L}x_{ia}-1\Bigr)^2+\sum_{a=1}^{L}\Bigl(\sum_{i=1}^{L}x_{ia}-1\Bigr)^2,
\label{eq:supp-assignment-penalty}
\end{equation}
which is zero exactly on permutation matrices and at least one otherwise. For one row (or column) with variables $x_1,\ldots,x_L$, $x_a^2=x_a$ gives $(\sum_ax_a-1)^2=2\sum_{a<b}x_ax_b-\sum_ax_a+1$, and with $x_a=(1+s_a)/2$,
\begin{equation}
\Bigl(\sum_{a=1}^{L}x_a-1\Bigr)^2=\frac12\sum_{a<b}s_as_b+\frac{L-2}{2}\sum_{a=1}^{L}s_a+\frac{L^2-3L+4}{4}.
\end{equation}
Every spin belongs to one row and one column, so
\begin{equation}
P_\mathrm{pen}(\svec)=\frac12\sum_{\{ia,jb\}\in\mathcal{N}}s_{ia}s_{jb}+(L-2)\sum_{i,a}s_{ia}+\mathrm{const},
\label{eq:supp-penalty-spins}
\end{equation}
where $\mathcal{N}$ contains the $L^2(L-1)$ pairs of spins that share a row or a column. In the clause form of Methods Eq.~\eqref{eq:energy-clauses}, the penalty term $\lambda P_\mathrm{pen}$ of the penalized energy $E+\lambda P_\mathrm{pen}$ therefore adds one second-order clause $s_{ia}s_{jb}$ of weight $J=-\lambda/2$ for every pair in $\mathcal{N}$ and lowers the weight of every first-order clause $s_{ia}$ by $\lambda(L-2)$. With $\lambda=\bar\lambda \Psi^\ast/L$ (Methods Section~\ref{meth:problems}), $\bar\lambda$ is expressed in units of the objective per keypoint. If $\lambda$ is too small, an infeasible assignment can have a lower penalized energy than the optimum. If it is too large, the infeasible states through which every route between two permutations must pass become high barriers. The penalty strength $\lambda$ is therefore optimized through $\bar\lambda$.

\paragraph{N-HOIM with penalty} The N-HOIM baseline\citeSI{Ahsan2026} minimizes $E+\lambda P_\mathrm{pen}$ with the third-order clauses of $E$ kept, by single spin flips with the global arbitration of N-HOIM (Methods Section~\ref{meth:benchmarks}). Its spins start from independent random values, so the run starts from an infeasible state. Its schedule and optimized settings are given in Supplementary Section~\ref{supp:settings}.

\paragraph{NeuroSA with quadratized penalty} NeuroSA\citeSI{Chen2025} accepts only pairwise interactions, so every third-order term is reduced to second order with auxiliary variables\citeSI{Boros2002}. In the Boolean form of $E+\lambda P_\mathrm{pen}$, the third-order terms are the similarity terms $-\psi\,x_{ia}x_{jb}x_{lc}$. In each such term, the pair of variables that occurs in the most third-order terms, say $(x_{ia},x_{jb})$, is replaced by an auxiliary variable $z_{ia,jb}\in\{0,1\}$, so that one auxiliary variable serves as many terms as possible. The replacement turns the term into $-\psi\,z_{ia,jb}x_{lc}$, and the penalty
\begin{equation}
\Gamma_{ia,jb}\left(x_{ia}x_{jb}-2x_{ia}z_{ia,jb}-2x_{jb}z_{ia,jb}+3z_{ia,jb}\right)
\label{eq:supp-rosenberg}
\end{equation}
is added. This penalty is zero when $z_{ia,jb}=x_{ia}x_{jb}$ and at least $\Gamma_{ia,jb}$ otherwise. With $\Gamma_{ia,jb}=1+\sum|\psi|$, summed over the terms that use $z_{ia,jb}$, a wrong auxiliary value changes those terms by at most $\sum|\psi|$ but adds at least $\Gamma_{ia,jb}$. It therefore cannot lower the energy, and the minima are unchanged. One auxiliary variable is introduced per pair used, 780 to 1,191 per instance at $L=10$. Substituting $x=(1+s)/2$ and $z=(1+s)/2$, with $s$ the spin of each variable, turns the quadratic polynomial into a pairwise Ising energy over the $L^2$ assignment spins and the auxiliary spins. NeuroSA tests one uniformly drawn spin per iteration (Methods Section~\ref{meth:benchmarks}), with the same schedule form and optimized quantities as N-HOIM. The multiplier of $\Gamma_{ia,jb}$ was also searched, and 1 was selected.

\paragraph{Three-color example} The N-HOIM comparison of Fig.~\ref{fig:toy}e adds the assignment penalty $P_\mathrm{pen}$ with $L=3$ and $\lambda=4$ to the objective of Methods Eq.~\eqref{eq:ex-spins}. That objective has first- and second-order clauses only, so $E+4P_\mathrm{pen}$ is quadratic. At every step, the global arbitration of N-HOIM tests all nine spin neurons and applies one spin flip that passes its test, so one step is nine iterations.

\section{Annealing settings}\label{supp:settings}

\begin{table}[!tbp]
\centering
\small\setlength{\tabcolsep}{4pt}
\caption{\textbf{Selected schedule settings.} Settings selected on the training instances for the iterations-to-solution results and used in Fig.~\ref{fig:matching}c. This work: factor $m$ on the calibrated start temperature and minimum ratio $\kappa$ of start to end temperature (Methods Eq.~\eqref{eq:temperatures}). Penalty machines: penalty strength $\bar\lambda$ and end temperature $\tau_\mathrm{end}$, both in units of $\Psi^\ast/L$, and ratio of start to end temperature. Complete: the backbone with all shortcuts. Path: the backbone alone.}\label{tab:supp-settings}
\begin{tabular}{@{}lll@{}}\toprule
Problem set & Method & Selected setting\\\midrule
Synthetic, $L=10$ & This work, complete & $m=0.125$, $\kappa=2$\\
 & This work, path & $m=8$, $\kappa=2$\\
 & N-HOIM, penalty & $\bar\lambda=1$, $\tau_\mathrm{end}=0.3$, ratio 30\\
 & NeuroSA, quadratized penalty & $\bar\lambda=2$, $\tau_\mathrm{end}=0.3$, ratio 100\\\midrule
WILLOW, $L=10$ & This work, complete & $m=2$, $\kappa=32$\\
 & This work, path & $m=0.5$, $\kappa=1.5$\\
 & N-HOIM, penalty & $\bar\lambda=1$, $\tau_\mathrm{end}=0.3$, ratio 30\\
 & NeuroSA, quadratized penalty & $\bar\lambda=2$, $\tau_\mathrm{end}=0.3$, ratio 10\\\midrule
Synthetic, $L=15$, 20, 25 & This work, complete & $m=0.5$, $\kappa=2$\\
Synthetic, $L=30$ & This work, complete & $m=0.25$, $\kappa=2$\\\midrule
Synthetic, $L=20$ & Informed, $K=40$ & $m=2$, $\kappa=2$\\
(library experiments) & Random, $K=40$ & $m=4$, $\kappa=2$\\
 & Informed or random, $K=80$ & $m=1$, $\kappa=2$\\
 & Complete, uniform proposals & $m=0.5$, $\kappa=2$\\
 & Complete, backbone share 0.5 & $m=0.5$, $\kappa=2$\\\midrule
Synthetic, $L=10$ & Path & $m=4$, $\kappa=8$\\
(library experiments, & Informed or random, $K=4$ & $m=2$, $\kappa=8$\\
target $\Psi^\ast$) & Informed, $K=9$ & $m=2$, $\kappa=2$\\
 & Random, $K=9$ & $m=8$, $\kappa=32$\\
 & Informed, $K=18$ & $m=0.5$, $\kappa=4$\\
 & Random, $K=18$ & $m=1$, $\kappa=4$\\
 & Informed, $K=27$ & $m=0.25$, $\kappa=2$\\
 & Random, $K=27$ & $m=0.5$, $\kappa=2$\\
 & Complete, backbone share 0.5 & $m=0.5$, $\kappa=4$\\\bottomrule
\end{tabular}
\end{table}

All results of our machine use one annealing procedure with automatically set start and end temperatures (Methods Section~\ref{meth:benchmarks}), except the three-color comparison of Fig.~\ref{fig:toy}e and the decoding results on the FPGA (Supplementary Sections~\ref{supp:realtime} and~\ref{supp:oneshot}). No step uses the solution, and no setting is adjusted manually for an instance. Apart from the budget, settings change only where $m$ and $\kappa$ are selected on training instances (below) and in the slow anneals of the event-driven measurements.

\paragraph{Three-color comparison} For Fig.~\ref{fig:toy}e, the anneal coefficient (8, 12 or 16) and the penalty strength (2.5, 3 or 4) of N-HOIM were selected on 200 separate runs from each of the five nonoptimal arrangements. The criterion was the fraction of runs that reach the optimum, with ties broken by the mean number of iterations until the first visit of the optimum. The run shown in Fig.~\ref{fig:toy}e is one of the 200 evaluation runs from state 6 of the main text, not one of these selection runs.

\paragraph{Cooling schedule}
The schedule is named after Fowler--Nordheim (FN) tunneling of electrons through a thin insulating barrier under a strong electric field, which motivates it physically. It is generated digitally (Methods Section~\ref{meth:benchmarks}). In terms of the fraction $n/(B-1)$ of a run of $B$ iterations completed at iteration $n$, the schedule of Methods Eq.~\eqref{eq:fn-schedule}\citeSI{Chen2025,Ahsan2026} reads
\begin{equation}
\tau_n=\frac{\tau_\mathrm{FN}}{\ln\!\left(1+q_\mathrm{s}+(\eta-q_\mathrm{s})\,n/(B-1)\right)},\qquad
\tau_\mathrm{FN}=\tau_\mathrm{end}\ln(1+\eta),\qquad
q_\mathrm{s}=\mathrm{e}^{\tau_\mathrm{FN}/\tau_\mathrm{start}}-1 .
\label{eq:supp-fn}
\end{equation}
The definitions of $\tau_\mathrm{FN}$ and $q_\mathrm{s}$ make the schedule pass exactly through $\tau_0=\tau_\mathrm{start}$ and
$\tau_{B-1}=\tau_\mathrm{end}$. With $\eta=0.1$ (Methods Section~\ref{meth:benchmarks}), the quantity added to 1 inside the logarithm never exceeds 0.1, so the logarithm is close to this quantity. When $\tau_\mathrm{start}/\tau_\mathrm{end}$ is large, $q_\mathrm{s}$ is small, and $\tau_n$ then falls approximately as $1/n$. Such a temperature halves each time $n$ doubles, so most of the cooling happens early in the run and most of the iterations are spent at low temperature.

\paragraph{Automatic choice of the two temperatures}
The two temperatures are set for each instance and move library by the acceptance-ratio rule of Methods Eq.~\eqref{eq:calibration} (Supplementary Algorithm~\ref{alg:calibrate}). The end of the run is spent near low-energy states, where uphill steps differ in size from those at random states. The end temperature is therefore measured after a short anneal, not at a random feasible state.

\paragraph{Why the end temperature is capped ($\kappa=2$)}
Because the two temperatures are measured at different states, the measured end temperature can exceed the start
temperature. At a random feasible assignment, almost no third-order term is satisfied (all three of its keypoint correspondences present), so exchanging two rows changes
the energy very little. At a low temperature, 80\% of these small uphill steps are already accepted. Near a low-energy assignment (high $\Psi$)
many terms are satisfied and an exchange removes several satisfied terms at once. A temperature at which only 1\% of these large steps are accepted can still
be relatively high. For uphill steps of one typical size $\Delta E$, an average acceptance $a_\mathrm{target}$ corresponds to $\tau\approx\Delta E/\ln(1/a_\mathrm{target})$. Let $\Delta E_\mathrm{start}$ and $\Delta E_\mathrm{end}$ be the typical sizes of the uphill steps at the random and at the low-energy assignment. The start temperature is then about $\Delta E_\mathrm{start}/\ln(1/0.8)$, and the measured end temperature about $\Delta E_\mathrm{end}/\ln100$. Since $\ln(1/0.8)\approx0.22$ and $\ln 100\approx4.6$, the measured end temperature exceeds the start temperature when the uphill steps near low-energy
assignments are more than about 20 times larger than those at random assignments. For graph matching, this often happens. In
Eq.~\eqref{eq:supp-fn}, $\tau_\mathrm{end}=\tau_\mathrm{start}$ gives $q_\mathrm{s}=\eta$, so the term $(\eta-q_\mathrm{s})\,n/(B-1)$
vanishes and $\tau_n=\tau_\mathrm{start}$ for the whole run. A measured end temperature above $\tau_\mathrm{start}$
would even make the temperature rise, which is excluded. Such a constant-temperature run performs little better than a random walk
(Supplementary Section~\ref{supp:control}). The cap keeps the calibrated $\tau_\mathrm{start}$ and guarantees cooling by at least a
factor $\kappa$. With $\tau_\mathrm{end}=\tau_\mathrm{start}/\kappa$, Eq.~\eqref{eq:supp-fn} gives
$q_\mathrm{s}=(1+\eta)^{1/\kappa}-1=0.049<\eta$ for $\kappa=2$. Because $\mathrm{e}^{-\Delta E/(\tau/2)}=(\mathrm{e}^{-\Delta E/\tau})^2$, halving the temperature squares the acceptance probability of every
uphill step. The cap acts only when the measured end temperature exceeds $\tau_\mathrm{start}/2$. Uphill transitions near the low-energy assignment are then accepted at the end of the run with average probability below 1\%, the value at the measured end temperature. A factor of 2 can therefore take a run from the hot regime to a nearly greedy one. In the control runs at $L=20$ (Supplementary Section~\ref{supp:control}), the cap acted on all three instances, and on two of them every replica reaches $0.98\,\Psi_\mathrm{ref}$ although the anneal cooled by only this factor. The control runs that fail started too hot. The value $\kappa=2$ was chosen before any test run, on six training instances at $L=10$ and four at $L=20$, from $\kappa=1.25$, 1.5, 2 and 4. It was the largest of these values that did not lower the
success probability of any move library. A larger $\kappa$ (a wider temperature range) raised the success probability on some instances and lowered it on others. A later check on the training
instances confirmed the choice. We measured the fraction of replicas that reach $0.98\,\Psi_\mathrm{ref}$, averaged over the budgets of each instance and then over 50 combinations of training instance and move library. This fraction was not higher, beyond the sampling uncertainty, than with $\kappa=2$ for any of the larger values tested ($\kappa=4$, 8 and 16).

\paragraph{Iterations to solution: each method at its own optimum}
The iterations to solution (Methods Eq.~\eqref{eq:its}) are the iterations needed to reach a target with 99\% probability by repeating independent runs. With success probability $P_\mathrm{s}(B)$ per run of $B$ iterations, $N_\mathrm{run}=\ln0.01/\ln[1-P_\mathrm{s}(B)]$ runs all fail with probability 0.01. The iterations to solution depend on the run length and on the schedule: a run that is too short
rarely succeeds, and one that is too long wastes iterations. A comparison at one fixed schedule and budget would therefore measure
how well that choice happens to suit each method, not the methods themselves. As in Methods Section~\ref{meth:benchmarks}, the iterations to solution are defined at their minimum over run
length on the test instances and at the schedule selected on training instances, separately for each method, problem set, size and move library. The training instances, six synthetic instances per size or the 20 WILLOW-ObjectClass training pairs, are not used for testing. Every training instance is run with 32 replicas for every method, at every combination of settings and budget. The selected combination of settings is the one with the smallest iterations
to solution on the training instances. This minimum is taken jointly over settings and budgets, but only the settings are kept. The range of $m$ and of the penalty settings was widened until the selection lay inside it. For this work the same automatic
calibration is used, and only the factors $m$ and $\kappa$ of Methods Eq.~\eqref{eq:temperatures} are optimized.
The factor $m$ sets how hot the run starts relative to the calibrated value ($m<1$ colder, $m>1$ hotter), and $\kappa$ is the smallest factor by which the run must cool. The penalty
machines are optimized through their penalty strength and temperature range. Every method is thus evaluated at its own best setting, found by the same criterion on the same training instances. Figs.~\ref{fig:matching}d, f report iterations to solution obtained in this way, as do Supplementary Sections~\ref{supp:librarysupp} and~\ref{supp:shortcut-theory} for move libraries at $L=20$ and at $L=10$. The WILLOW comparison of Fig.~\ref{fig:matching}c and its re-analysis in Supplementary Section~\ref{supp:matching} show these settings at every budget of the test grid, so that no method is shown at an
arbitrary schedule. In the per-size selection for Fig.~\ref{fig:matching}d only, the best setting with $\kappa=2$ was selected where its training iterations to solution lay inside the 95\% bootstrap interval of the best setting overall ($L=10$, 15 and 30). The automatic rule is thus changed only where the data require it.

\paragraph{Selected values}
Supplementary Table~\ref{tab:supp-settings} lists the selected settings. The target was the certified optimum for WILLOW, for synthetic $L=10$ in Fig.~\ref{fig:matching}f and for the $L=10$ library experiments. It was $0.98\,\Psi_\mathrm{ref}$ for Fig.~\ref{fig:matching}d and the $L=20$ library experiments. At synthetic $L=10$, both targets select the same setting for the complete library. The searched ranges differed between experiments and together span $m=1/32$ to 32 and $\kappa=1$ to 128 (for the move libraries at $L=10$, see Supplementary Section~\ref{supp:shortcut-theory}). In the per-size and $L=20$ library selections, $\kappa$ was searched from 2 to 16, so the selected $\kappa=2$ lies at the lower end of that range. The penalty machines use the same
schedule, Methods Eq.~\eqref{eq:fn-schedule}, with the shape parameter $\eta=1$ of N-HOIM. Their settings $\bar\lambda$, $\tau_\mathrm{end}$ (in units of $\Psi^\ast/L$)
and $\tau_\mathrm{start}/\tau_\mathrm{end}$ were searched over $\bar\lambda=0.5$ to 32, $\tau_\mathrm{end}=0.1$ to 3 and ratios of 3 to 1,000. The budget grids span $2^{6}$ to $2^{24}$ iterations.

\paragraph{Results at a stated budget: one rule for every variant}
The other comparisons between variants of our own method use the automatically calibrated schedule unchanged ($m=1$, $\kappa=2$), calibrated on each instance and move library (Methods Section~\ref{meth:benchmarks}). Most are made at budgets run with both variants, shown on the horizontal axis or stated. Figure~\ref{fig:library}d instead takes each library at its own best budget, and Fig.~\ref{fig:library}e uses one common budget. Optimizing each variant separately would mix the effect of the move library or the update process with that of the schedule, and the best schedule would change with the budget.

In the event-driven measurements (Fig.~\ref{fig:event} and Supplementary Section~\ref{supp:event}), the ratio of the operations of event-driven updating to those of recomputation depends on the acceptance. The schedule only sets when in the run a given acceptance is reached. One slow anneal per workload, which passes through the full range of acceptance, is therefore used (Supplementary Section~\ref{supp:event}).

The selected settings, the full search ranges and all runs are provided with the code.

\section{Graph matching: agreement with human annotations and a harder deformation level}\label{supp:matching}

\begin{figure}[!tbp]
\centering
\includegraphics[width=\linewidth]{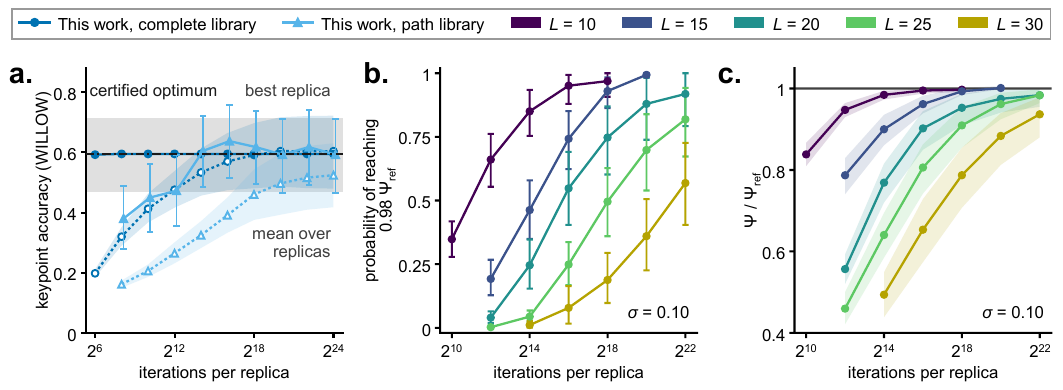}
\caption{\textbf{Graph matching: agreement with human annotations and a harder deformation level.}
\textbf{a}, Keypoint accuracy (fraction of keypoints assigned as in the human annotation) versus iterations per replica on the 40 WILLOW-ObjectClass test pairs. Curves show this work with all shortcuts (complete library) and with the backbone alone (path library). These are the runs of Fig.~\ref{fig:matching}c, with 256 replicas per pair. The factors $m$ and $\kappa$ of each library were selected on separate training pairs (Supplementary Section~\ref{supp:settings}). Solid lines: the replica with the highest objective value $\Psi$. Dotted lines: mean over replicas, with shaded confidence bands. Solid and dotted lines are averaged over pairs. Dash-dot line: accuracy of the certified optimum (0.595). Grey band: its 95\% confidence interval (CI) over pairs. This band is also the CI of the complete library's best replica, since that replica's assignment is the certified optimum at every budget from $2^{8}$ on.
\textbf{b}, Probability that a replica reaches $0.98\,\Psi_\mathrm{ref}$ on synthetic instances with stronger deformation ($\sigma=0.10$, against $\sigma=0.05$ in the main text). Runs use the complete library with $L=10$--30 keypoints. Means are over 20 test instances $\times$ 256 replicas.
\textbf{c}, $\Psi/\Psi_\mathrm{ref}$ for the same runs, where $\Psi$ is the best value found by each replica. It is averaged over replicas, then over instances (as in Fig.~\ref{fig:matching}e). $\Psi_\mathrm{ref}$ is the certified optimum for $L=10$. For $L\geq15$, it is the best value found by three domain solvers, so values slightly above 1 are possible. The runs of \textbf{b} and \textbf{c} used the automatically calibrated schedule (Methods Section~\ref{meth:benchmarks}), as in Fig.~\ref{fig:matching}e, not schedules optimized for each size.
Error bars (in \textbf{a}, only for the best replica of the path library) and bands: 95\% bootstrap CIs over pairs (\textbf{a}) or instances (\textbf{b}, \textbf{c}).}\label{fig:supp-matching}
\end{figure}

The objective of Eq.~\eqref{eq:matching-objective} measures the geometric similarity of triangles, which only approximates the human annotation. Once the best replica of the complete library has found the certified optimum (main text Section~\ref{res:matching}), its accuracy equals that of the optimum, 0.595 (Fig.~\ref{fig:supp-matching}a). An assignment with a slightly lower objective value can match the annotation on more keypoints. The best replica of the path library reaches an accuracy of 0.64 at $2^{16}$ iterations, within the confidence interval of the optimum.

With twice the deformation of the main text ($\sigma=0.10$), the complete library reaches about the same success probability as at $\sigma=0.05$ at $L=20$ and 25 and a lower one at $L=30$ (Fig.~\ref{fig:supp-matching}b, c). At $L=30$, both the probability of reaching $0.98\,\Psi_\mathrm{ref}$ and the mean $\Psi/\Psi_\mathrm{ref}$ still rise between the two largest budgets ($2^{20}$ and $2^{22}$ iterations), so these runs have not converged. The stronger deformation makes the similarity terms less distinctive. The target $a$ that maximizes $\phi_i(a)$ (Supplementary Section~\ref{supp:matching-formulation}) is the generating target of keypoint $i$ for a median of 13 to 30\% of keypoints at $\sigma=0.10$, against 23 to 60\% at $\sigma=0.05$ ($L=10$ to 30, test instances). At $L=10$, the certified optimum matches the generating correspondence on a median of 50\% of keypoints at $\sigma=0.10$, against 100\% at $\sigma=0.05$.

\section{\texorpdfstring{The matching solutions result from annealing, not from random sampling}{The matching solutions result from annealing, not from random sampling}}\label{supp:control}

\begin{figure}[!tbp]
\centering
\includegraphics[width=\linewidth]{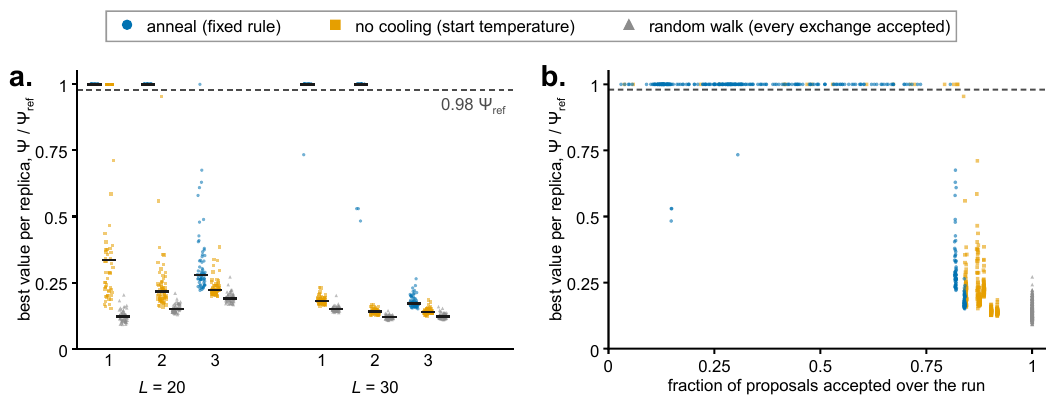}
\caption{\textbf{The matching solutions result from annealing, not from random sampling.}
Six synthetic test instances of graph matching: instances 1--3 at $L=20$ ($2^{18}$ iterations) and instances 1--3 at $L=30$ ($2^{20}$
iterations). Each uses the complete library and 64 replicas. The three variants use the same moves, budget, starting assignments and
random numbers. Anneal: the automatically calibrated schedule ($m=1$, $\kappa=2$), as in Fig.~\ref{fig:matching}e. No cooling: held at the calibrated start temperature for
the whole run. Random walk: every proposed spin transition accepted. \textbf{a}, Best value found by each replica, $\Psi/\Psi_\mathrm{ref}$
(dots), and the median over replicas (bar). Dashed line: $0.98\,\Psi_\mathrm{ref}$. \textbf{b}, The same values against the fraction of proposals each replica accepted
over the run. $\Psi_\mathrm{ref}$: best value found by three domain solvers.}\label{fig:supp-control}
\end{figure}

For graph matching the calibrated temperature range can be narrow: the end temperature is often only half the start temperature
(Supplementary Section~\ref{supp:settings}). One may therefore ask whether the solver reaches assignments with $\Psi\geq0.98\,\Psi_\mathrm{ref}$ by
annealing, or merely by visiting many assignments at random. Three variants with the same moves, budget, random feasible starting assignments and random numbers test this on six synthetic test instances (Fig.~\ref{fig:supp-control}). They are the anneal with the automatically calibrated schedule, a random walk that accepts every proposed spin transition, and a run held at the calibrated start temperature without cooling. The best assignment that the random walk visits shows
what random exploration achieves with the same number of iterations.

Random exploration remains far below the reference value (Fig.~\ref{fig:supp-control}a). The best assignment visited by any replica of
the random walk reaches at most 0.27 of $\Psi_\mathrm{ref}$ on every instance, although each replica visits up to $10^6$ assignments.
Without cooling, the result is only slightly better: the mean best value stays below half of $\Psi_\mathrm{ref}$, and replicas reach $0.98\,\Psi_\mathrm{ref}$ on only one instance. With the anneal, 95 to 100\% of replicas reach 0.98\,$\Psi_\mathrm{ref}$ on each of four of the
six instances.

The fraction of proposed transitions that are accepted over the run shows why (Fig.~\ref{fig:supp-control}b). Nearly all replicas (anneal or no cooling) that accepted fewer than 75\% of their proposals reached $0.98\,\Psi_\mathrm{ref}$, and nearly all that accepted more stayed below $0.75\,\Psi_\mathrm{ref}$. The acceptance reflects both the temperature and the current assignment. Near a low-energy assignment (high $\Psi$), uphill steps are large, so a replica that has reached such an assignment accepts fewer proposals. At a low enough temperature, the replica stays near that assignment. On the four instances on which the anneal succeeds, the anneal accepts 13 to 53\% of its proposals over the run: it cools into this low-acceptance regime. On the other two instances, at most 2\% of the anneal's replicas reach 0.98\,$\Psi_\mathrm{ref}$. On these instances, the anneal accepts nearly as many proposals as the runs without cooling, and its best values are close to theirs. The calibrated temperatures were too high: the runs cooled by at least the guaranteed factor $\kappa=2$ (Methods Eq.~\eqref{eq:temperatures}) but stayed too hot. The calibration samples proposals at only one random feasible state and one annealed state (Methods Eq.~\eqref{eq:calibration}), so its temperatures depend on these states. The start temperature varies about four-fold between such trial runs. Runs whose calibrated temperatures were too high are included in all reported averages. A calibration from several trial runs is a possible refinement.

Reaching the reference value therefore depends on the cooling, not on the number of assignments visited. The schedule must end in a nearly greedy regime, in which almost all accepted transitions lower the energy or leave it unchanged, and the calibration guarantees a minimum cooling range for this reason (Supplementary Section~\ref{supp:settings}).

\section{Size of the compiled machine}\label{supp:resources}
\begin{figure}[!tbp]
\centering
\includegraphics[width=\linewidth]{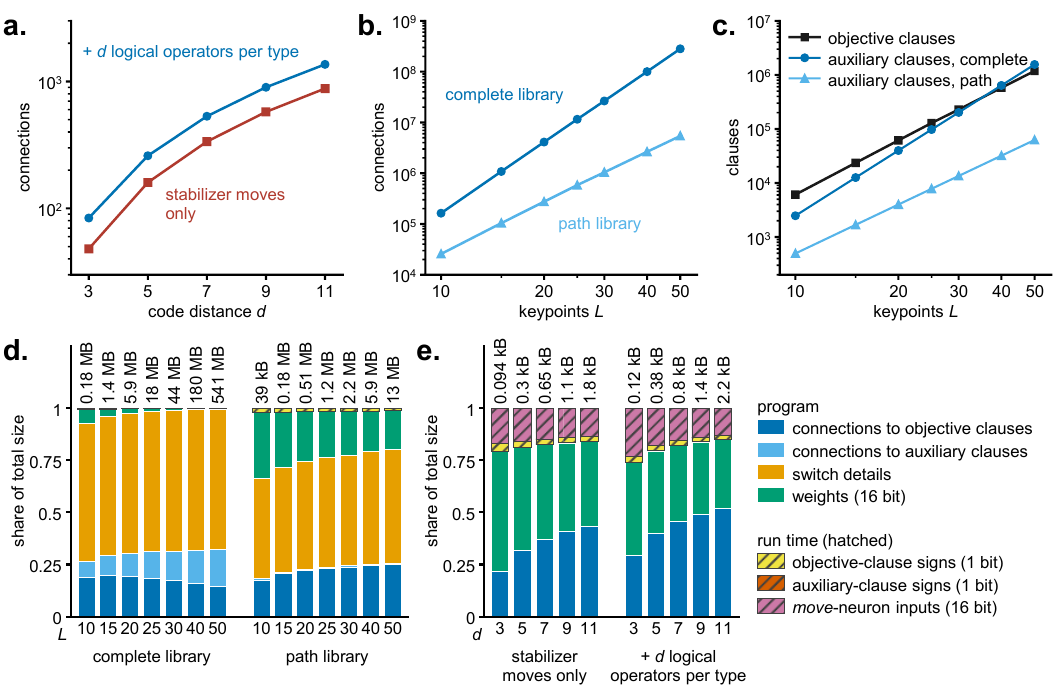}
\caption{\textbf{Size of the compiled machine.}
\textbf{a}, Number of connections (non-zero entries of the clause-to-neuron connection pattern) of the surface-code machine for one syndrome per shot versus code distance $d$ (depolarizing noise). It is shown for stabilizer moves only and for the backbone with all shortcuts, which adds all $d$ straight-line logical operators of each type.
\textbf{b}, Connections of the matching machine versus the number of keypoints $L$, for the complete library (backbone with all shortcuts) and the path library (backbone alone). Each $L$ uses one synthetic instance.
\textbf{c}, Numbers of objective clauses (the same for both libraries) and auxiliary clauses of the matching machine. Each move requires its own $L+L(L-1)/2$ auxiliary clauses, so a library of $R$ moves has exactly $R(L+L(L-1)/2)$ of them.
\textbf{d}, \textbf{e}, Information needed to program and run the machine, as shares of the total, for matching (\textbf{d}) and the surface code with one syndrome per shot (\textbf{e}). Number above each bar: total. Program: connections to objective and to auxiliary clauses, switch details (which clause controls each switched connection) and 16-bit objective weights. Run time (hatched): one bit per clause output and one 16-bit maintained input per \textit{move} neuron. Connections and switch details are counted at their information minimum (Supplementary Section~\ref{supp:resources}), so the totals are a lower bound on the information stored, not a circuit or area estimate.}\label{fig:supp-size}
\end{figure}

The compiler output determines what a hardware implementation must store (Fig.~\ref{fig:supp-size}). The program consists of the connections between stored clauses and \textit{move} neurons, the details of every state-dependent switch (which clause controls it) and the objective weights. A connection exists for every pair of a clause and a move whose switch is not identically zero. The run-time state consists of one clause output (one bit) per stored clause and one maintained input per \textit{move} neuron. We count connections at their information minimum. When $N_\mathrm{conn}$ connections are placed among $N_\mathrm{pair}$ possible clause--neuron pairs of the same clause type (objective or auxiliary clause), the pattern is one of $\binom{N_\mathrm{pair}}{N_\mathrm{conn}}$ possible patterns. No encoding can therefore specify it with fewer than $\log_2\binom{N_\mathrm{pair}}{N_\mathrm{conn}}$ bits. Switch details are counted as the information needed to name the controlling clause of every switched connection. Weights and maintained inputs use 16 bits each. The spins contained in each clause are used once by the host to initialize the clause outputs and are not stored on the machine. The totals are therefore a lower bound on the information stored, not a circuit or area estimate.

The surface-code machine has no auxiliary clauses. For one syndrome per shot, its size grows with the number of moves and their connections (Fig.~\ref{fig:supp-size}a). A window of the real-time decoder holds $2d$ rounds with the measurement errors of every round and compiles to a larger machine, which is not counted here. The matching machine is dominated by its auxiliary clauses (Fig.~\ref{fig:supp-size}b, c). Their number grows as $L^4$ for the complete library (backbone with all shortcuts) and as $L^3$ for the path library (backbone alone). With the complete library, the auxiliary clauses outnumber the objective clauses from $L=40$ on. The largest share of the information stored by the matching machine specifies which auxiliary clause controls each switch: 66 to 67\% at every $L$ with the complete library and 48 to 55\% with the path library (Fig.~\ref{fig:supp-size}d). The run-time state, by contrast, takes less than 3\%. The surface-code machine has no state-dependent switches, so its maintained inputs take a larger share of its smaller total, 13 to 23\% (Fig.~\ref{fig:supp-size}e). Selecting the shortcuts when the number of \textit{move} neurons is limited (Fig.~\ref{fig:library}) therefore reduces the dominant part of the program.

The number of connections whose switches are on depends on the state, and an auxiliary clause can control several switches. The number of auxiliary clauses alone therefore does not determine the switching activity after an accepted spin transition, which the operation counts of Fig.~\ref{fig:event} measure.

\FloatBarrier
\section{\texorpdfstring{Exchange library at $L=30$ and batching statistics}{Exchange library at L = 30 and batching statistics}}\label{supp:librarysupp}

\begin{table}[!tbp]
\centering
\small
\caption{\textbf{Batched versus static libraries (Fig.~\ref{fig:library}g).} Graph matching with $L = 20$, 20 synthetic test instances and 256 replicas per instance. Static: the backbone (19 path exchanges) and the first batch of $K$ informed shortcuts (those with the highest overlap) stay loaded for the whole run. Batched: the same number of loaded shortcuts, with the batch replaced every $D$ iterations. The batches of all 171 shortcuts are thus loaded in turn, as far as the budget allows. Both use the automatically calibrated schedule ($m=1$, $\kappa=2$), calibrated on the backbone with the first batch, so they have the same temperature at every iteration. Columns give the probability that a replica reaches 0.98\,$\Psi_{\mathrm{ref}}$ within the budget (mean over instances) and the paired mean difference (batched minus static) with its 95\% bootstrap confidence interval over instances (Methods Section~\ref{meth:problems}).}
\label{tab:supp-dwell}
\begin{tabular}{rcccccc}
\toprule
$K$ & budget & $D$ & static & batched & difference & 95\% CI \\
\midrule
20 & $2^{18}$ & $2^{12}$ & 0.08 & 0.86 & ${+}0.78$ & $[{+}0.68,\,{+}0.86]$ \\
 &  & $2^{14}$ & 0.08 & 0.69 & ${+}0.61$ & $[{+}0.50,\,{+}0.71]$ \\
 &  & $2^{16}$ & 0.08 & 0.21 & ${+}0.13$ & $[{+}0.08,\,{+}0.17]$ \\
\addlinespace
 & $2^{20}$ & $2^{12}$ & 0.24 & 0.93 & ${+}0.70$ & $[{+}0.59,\,{+}0.80]$ \\
 &  & $2^{14}$ & 0.24 & 0.92 & ${+}0.68$ & $[{+}0.56,\,{+}0.78]$ \\
 &  & $2^{16}$ & 0.24 & 0.82 & ${+}0.59$ & $[{+}0.46,\,{+}0.70]$ \\
\midrule
40 & $2^{18}$ & $2^{12}$ & 0.34 & 0.92 & ${+}0.58$ & $[{+}0.45,\,{+}0.70]$ \\
 &  & $2^{14}$ & 0.34 & 0.89 & ${+}0.55$ & $[{+}0.41,\,{+}0.67]$ \\
 &  & $2^{16}$ & 0.34 & 0.57 & ${+}0.23$ & $[{+}0.11,\,{+}0.34]$ \\
\addlinespace
 & $2^{20}$ & $2^{12}$ & 0.57 & 0.97 & ${+}0.40$ & $[{+}0.27,\,{+}0.53]$ \\
 &  & $2^{14}$ & 0.57 & 0.96 & ${+}0.40$ & $[{+}0.27,\,{+}0.53]$ \\
 &  & $2^{16}$ & 0.57 & 0.94 & ${+}0.38$ & $[{+}0.25,\,{+}0.51]$ \\
\midrule
80 & $2^{18}$ & $2^{12}$ & 0.68 & 0.91 & ${+}0.22$ & $[{+}0.12,\,{+}0.33]$ \\
 &  & $2^{14}$ & 0.68 & 0.90 & ${+}0.22$ & $[{+}0.12,\,{+}0.33]$ \\
 &  & $2^{16}$ & 0.68 & 0.86 & ${+}0.18$ & $[{+}0.09,\,{+}0.28]$ \\
\addlinespace
 & $2^{20}$ & $2^{12}$ & 0.87 & 0.94 & ${+}0.07$ & $[{+}0.02,\,{+}0.15]$ \\
 &  & $2^{14}$ & 0.87 & 0.94 & ${+}0.07$ & $[{+}0.01,\,{+}0.15]$ \\
 &  & $2^{16}$ & 0.87 & 0.94 & ${+}0.07$ & $[{+}0.02,\,{+}0.15]$ \\
\bottomrule
\end{tabular}
\end{table}
\begin{figure}[!tbp]
\centering
\includegraphics[width=\linewidth]{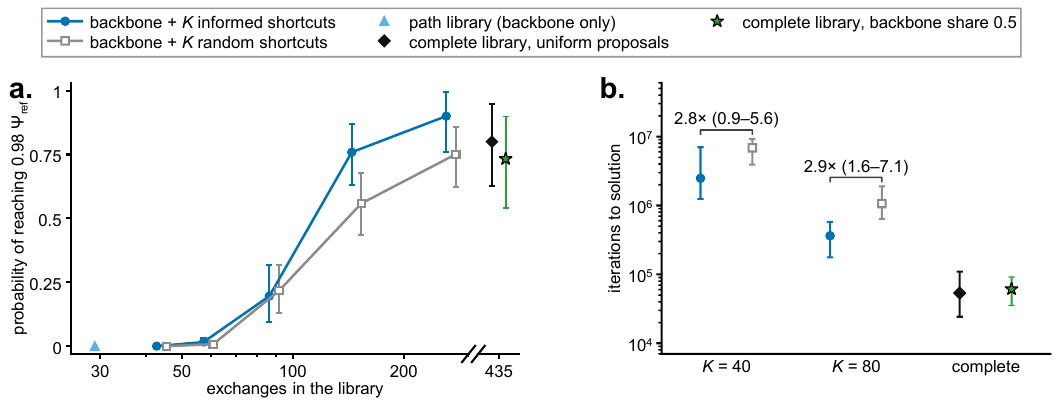}
\caption{\textbf{Informed and random shortcuts at $L=30$ and with optimized schedules.}
\textbf{a}, Probability that a replica reaches $0.98\,\Psi_\mathrm{ref}$ within $2^{22}$ iterations versus the number of exchanges in the
move library, for graph matching with $L=30$ keypoints (same design as Fig.~\ref{fig:library}a at $L=20$). Libraries hold the backbone of 29 path
exchanges plus $K=15$--240 informed or random shortcuts. The backbone receives about half of all proposals (0.66 at $K=15$). References: the backbone alone
(path library), and the backbone with all 406 shortcuts (the complete library of 435 exchanges, shown after the axis break) with uniform proposals or with backbone share 0.5. All move libraries use the automatically calibrated schedule.
Points are offset slightly horizontally for legibility. \textbf{b}, Iterations to solution (target $0.98\,\Psi_\mathrm{ref}$; Methods Section~\ref{meth:problems}) at $L=20$ for informed and random shortcuts at $K=40$ and 80 and for the two complete references, each with
the factors $m$ and $\kappa$ selected on six separate training instances. Budgets on the test instances: $2^{8}$ to $2^{22}$. Brackets: ratio (random/informed) of the iterations to solution, with its 95\% bootstrap interval, which resamples the
same instances for both libraries. Both panels: 20 synthetic test instances and 256
replicas per instance. Error bars: 95\% bootstrap confidence intervals over instances. $\Psi_\mathrm{ref}$: best value found by three domain
solvers.}\label{fig:supp-library}
\end{figure}

The backbone share (main text Section~\ref{res:library}) is 0.49 for every $K$ used from 20 on at $L=20$ (0.66 at $K=10$) and from 30 on at $L=30$ (0.66 at $K=15$). It is 0.5 for every $K$ used from 9 on at $L=10$ (0.69 at $K=4$). The complete library with backbone share 0.5 uses the same construction. At $L=20$, the ranked shortcuts form 9, 5 and 3 batches for $K=20$, 40 and 80 ($171/K$ rounded up), and the last batch is shorter. Because the shared temperatures of the batching comparison are calibrated on the backbone with the first batch, batches loaded later may run at less suited temperatures. Such a mismatch is expected to lower, not raise, the success probability of the batched libraries. When the budget $B$ is shorter than $D$ times the number of batches, not every batch is loaded.

At $L=30$ (Fig.~\ref{fig:supp-library}a), informed and random shortcuts give similar success probabilities up to $K=60$, and from $K=120$ on, the success probability of informed shortcuts lies above that of random ones. The backbone with 240 informed shortcuts reaches $0.98\,\Psi_\mathrm{ref}$ about as often as the complete library with either proposal rule. With schedules optimized on separate training instances at $L=20$ (Fig.~\ref{fig:supp-library}b), random shortcuts need 2.8 times as many iterations to solution as informed shortcuts at $K=40$ (95\% CI 0.9--5.6, which includes equal iterations) and 2.9 times as many at $K=80$ (1.6--7.1).

Table~\ref{tab:supp-dwell} lists every comparison of Fig.~\ref{fig:library}g. In every combination of $K$, budget and $D$, the batched library reaches $0.98\,\Psi_\mathrm{ref}$ more often than the static library and less often than the complete library with backbone share 0.5 and the same temperatures. The gain over the static library is largest with the fewest loaded shortcuts and the shortest interval. There, the static library holds the smallest fraction of the shortcuts, and every batch is loaded many times during the run. At $K=20$ and 40, the gain is smallest with the budget $2^{18}$ and $D=2^{16}$, at which only four batches are loaded (four of nine at $K=20$).

\FloatBarrier
\section{\texorpdfstring{How shortcuts reduce the annealing time}{How shortcuts reduce the annealing time}}\label{supp:shortcut-theory}

\begin{table}[!tbp]
\centering
\footnotesize\setlength{\tabcolsep}{3.2pt}
\caption{\textbf{Iterations to solution at $L=10$ with the automatically calibrated schedule and with optimized schedules (Fig.~\ref{fig:library}d).} Graph matching with 20 synthetic test instances and 256 replicas per instance. Target: the certified optimum $\Psi^\ast$. ITS: iterations to solution (Methods Section~\ref{meth:problems}). For random shortcuts, the median over the ten draws of each instance is taken first. 95\% CI: percentile bootstrap over instances (5,000 resamples). Optimized schedule: $m$ and $\kappa$ selected on six separate training instances (Supplementary Section~\ref{supp:settings}). Path: the backbone of 9 path exchanges alone. Informed and random: the backbone plus $K$ informed or random shortcuts. Complete: the backbone with all 36 shortcuts. Ratios: ratio of the iterations to solution of the two libraries, with a 95\% bootstrap interval that resamples the same instances for both.}
\label{tab:supp-its10}
\begin{tabular}{@{}lccc@{}}
\toprule
 & \begin{tabular}[b]{@{}c@{}}Automatically calibrated\\schedule\end{tabular} & \multicolumn{2}{c}{Optimized schedule}\\\cmidrule(l){3-4}
Library & ITS [95\% CI] & $(m,\kappa)$ & ITS [95\% CI]\\
\midrule
Path & $2.4\times10^{6}$ $[8.0\times10^{5},\,6.4\times10^{6}]$ & $(4,\,8)$ & $1.4\times10^{6}$ $[6.8\times10^{5},\,2.4\times10^{6}]$\\
Informed, $K=4$ & $9.2\times10^{4}$ $[5.4\times10^{4},\,1.5\times10^{5}]$ & $(2,\,8)$ & $9.9\times10^{4}$ $[5.8\times10^{4},\,1.5\times10^{5}]$\\
Random, $K=4$ & $2.1\times10^{5}$ $[1.2\times10^{5},\,2.6\times10^{5}]$ & $(2,\,8)$ & $2.3\times10^{5}$ $[1.5\times10^{5},\,2.8\times10^{5}]$\\
Informed, $K=9$ & $2.5\times10^{4}$ $[1.3\times10^{4},\,5.9\times10^{4}]$ & $(2,\,2)$ & $2.7\times10^{4}$ $[1.7\times10^{4},\,5.4\times10^{4}]$\\
Random, $K=9$ & $6.1\times10^{4}$ $[4.1\times10^{4},\,8.0\times10^{4}]$ & $(8,\,32)$ & $7.1\times10^{4}$ $[4.4\times10^{4},\,7.5\times10^{4}]$\\
Informed, $K=18$ & $9{,}062$ $[4{,}388,\,1.4\times10^{4}]$ & $(0.5,\,4)$ & $6{,}902$ $[4{,}917,\,1.4\times10^{4}]$\\
Random, $K=18$ & $1.5\times10^{4}$ $[1.2\times10^{4},\,2.3\times10^{4}]$ & $(1,\,4)$ & $1.3\times10^{4}$ $[1.1\times10^{4},\,1.8\times10^{4}]$\\
Informed, $K=27$ & $5{,}381$ $[3{,}262,\,9{,}017]$ & $(0.25,\,2)$ & $4{,}501$ $[3{,}271,\,5{,}778]$\\
Random, $K=27$ & $7{,}471$ $[4{,}485,\,1.3\times10^{4}]$ & $(0.5,\,2)$ & $5{,}135$ $[4{,}015,\,8{,}113]$\\
Complete, backbone share 0.5 & $3{,}965$ $[2{,}562,\,6{,}590]$ & $(0.5,\,4)$ & $2{,}779$ $[2{,}093,\,4{,}642]$\\
\midrule
Ratio & \begin{tabular}[b]{@{}c@{}}Automatically calibrated\\schedule\end{tabular} & & Optimized schedule\\
\midrule
Path / complete & $608$ $[178,\,1819]$ & & $503$ $[184,\,921]$\\
Informed / random, $K=4$ & $0.44$ $[0.25,\,0.82]$ & & $0.44$ $[0.25,\,0.96]$\\
Informed / random, $K=9$ & $0.40$ $[0.22,\,1.10]$ & & $0.37$ $[0.26,\,0.91]$\\
Informed / random, $K=18$ & $0.59$ $[0.32,\,0.89]$ & & $0.52$ $[0.36,\,1.00]$\\
Informed / random, $K=27$ & $0.72$ $[0.48,\,1.09]$ & & $0.88$ $[0.64,\,1.14]$\\
\bottomrule
\end{tabular}
\end{table}

This section gives the theory, the computations and the statistics behind main text Section~\ref{res:library} and Fig.~\ref{fig:library}b--e.

\paragraph{A bound that holds for any shortcuts} A shorter graph distance does not by itself shorten the time that a run needs to mix over the states. A run cannot mix in fewer iterations than the diameter of the state graph, the largest number of spin transitions needed to join two feasible states. Because it proposes its moves at random and revisits states, it usually needs many more. At low temperature, a shortcut whose spin transition leads to a state much higher in energy is rejected and only reduces the proposal rate of the backbone. The following bound shows that this cost is at most a constant factor. Let the backbone receive a fixed fraction $\beta$ of all proposals, as set by the weight $w_\mathrm{B}$ in the library experiments ($\beta$ between 0.49 and 0.69). One iteration is then the mixture $\mathcal P=\beta\mathcal P_\mathrm{B}+(1-\beta)\mathcal P_\mathrm{S}$ of a step that proposes only backbone moves and a step that proposes only shortcuts. Both steps are reversible with respect to the same Boltzmann distribution $\pi(\svec)\propto\mathrm{e}^{-E(\svec)/\tau}$ (Supplementary Section~\ref{supp:barriers}). That is, in equilibrium, the probability flow of each step from $\svec$ to $\svec'$ equals the reverse flow: $\pi(\svec)\mathcal P(\svec,\svec')=\pi(\svec')\mathcal P(\svec',\svec)$. For a function $f$ of the state, the Dirichlet form
\begin{equation}
\mathcal{E}_{\mathcal P}(f)=\frac12\sum_{\svec,\svec'}\pi(\svec)\mathcal P(\svec,\svec')\left[f(\svec')-f(\svec)\right]^2
\end{equation}
is half the mean squared change of $f$ in one step started from equilibrium. A function that one step changes only slightly relative to its variance $\mathrm{Var}_\pi(f)$ under $\pi$ identifies a slowly relaxing mode of the chain. An example is the indicator of a set of states that the run rarely leaves. The spectral gap is the smallest ratio of the Dirichlet form to this variance,
\begin{equation}
\mathrm{gap}(\mathcal P)=\min_{f}\frac{\mathcal{E}_{\mathcal P}(f)}{\mathrm{Var}_\pi(f)},
\end{equation}
over all functions $f$ that are not constant. The spectral gap equals one minus the second-largest eigenvalue of $\mathcal P$. The eigenvector of that eigenvalue is the slowest-decaying deviation from equilibrium. Each iteration multiplies this deviation by $1-\mathrm{gap}(\mathcal P)$, so the relaxation time $1/\mathrm{gap}(\mathcal P)$ is the number of iterations over which it shrinks by a factor of about $\mathrm{e}$. The Dirichlet form is linear in the step, so the Dirichlet forms of the two steps add: $\mathcal{E}_{\mathcal P}=\beta\,\mathcal{E}_{\mathcal P_\mathrm{B}}+(1-\beta)\,\mathcal{E}_{\mathcal P_\mathrm{S}}\geq\beta\,\mathcal{E}_{\mathcal P_\mathrm{B}}$ for every $f$, because $\mathcal{E}_{\mathcal P_\mathrm{S}}\geq0$. Since both steps share $\pi$, and hence $\mathrm{Var}_\pi(f)$, the spectral gap of $\mathcal P$ is at least $\beta$ times the spectral gap of $\mathcal P_\mathrm{B}$. At every fixed temperature and for any choice of shortcuts, the relaxation time of $\mathcal P$ is therefore at most $1/\beta$ times that of the backbone alone. Shortcuts only add transitions between states, while the backbone keeps a fraction $\beta$ of all proposals.

\paragraph{High temperature} When every proposal is accepted, a library of row exchanges performs the interchange process on its exchange graph. The vertices of this graph are the $L$ rows, and its edges are the exchanges, weighted by their proposal probabilities. Each row carries its current target, and each exchange swaps the targets carried by its two rows. A single target then performs a random walk on the exchange graph, stepping from row $i$ to row $j$ whenever the exchange of $i$ and $j$ is applied. This walk is described by the weighted Laplacian of the exchange graph, the $L\times L$ matrix with off-diagonal entries $-W_{ij}$ and diagonal entries $\sum_{j}W_{ij}$, where $W_{ij}$ is the proposal probability of the exchange of rows $i$ and $j$ (zero if it is not in the move library). The smallest eigenvalue of the Laplacian is zero. Its second-smallest eigenvalue $\nu_2$ is the decay rate of the slowest deviation of the walk from equal occupation of all rows. This eigenvalue is small when the graph has a bottleneck, that is, a set of rows joined to the other rows only by few exchanges or by rarely proposed ones. The whole permutation, with its $L!$ states, relaxes as fast as one target: the spectral gap of the interchange process equals $\nu_2$\citeSI{Caputo2010}. The relaxation time $1/\nu_2$ of a library therefore follows from an $L\times L$ eigenvalue problem. In the path library (the backbone alone), each of the $L-1$ exchanges is proposed with probability $1/(L-1)$. This gives $\nu_2=2[1-\cos(\pi/L)]/(L-1)\approx\pi^2/L^3$, so the relaxation time of the path library grows as $L^3$ (772 iterations at $L=20$). The value $1/\nu_2$ is the relaxation time of the continuous-time process. In that process, each exchange is applied at random times, at a rate equal to its proposal probability per iteration, so one unit of time corresponds to one iteration on average. The discrete chain at infinite temperature has the same spectral gap, but its distribution does not converge. Every exchange changes the parity of the permutation (an even or odd number of exchanges from the starting assignment), so the run alternates between even and odd permutations. This periodicity is absent in the continuous-time process, in which the number of exchanges applied in a given time is random, and at any finite temperature, at which rejected proposals leave the state unchanged. The relaxation time of the continuous-time process therefore measures the relaxation rate of the run at high temperature. Figure~\ref{fig:library}b evaluates this relaxation time on the 20 test instances for the libraries of Fig.~\ref{fig:library}a ($L=20$) and for the same construction at $L=30$ (Fig.~\ref{fig:supp-library}a). For the path and complete libraries, the known mixing times, after which the assignment is close to uniformly distributed from any start, exceed the relaxation time by a factor of order $\ln L$ only\citeSI{Diaconis1981,Wilson2004}. Informed and random shortcuts are compared here and below through the ratio of their medians over the instances, with its 95\% bootstrap interval (Methods Section~\ref{meth:problems}). With informed shortcuts, the relaxation is up to 1.7 times slower than with the same number of random shortcuts, with the largest slowdown at intermediate numbers of shortcuts (Fig.~\ref{fig:library}b). The 95\% interval of the ratio of the median relaxation times, informed over random, lies above 1 from $K=40$ on at $L=20$ and from $K=30$ on at $L=30$.

\paragraph{Low temperature} A local minimum is a feasible state from which no spin transition of the move library leads to a state of lower energy, so the local minima depend on the library. The depth $h$ of a local minimum is the smallest rise above its energy that a route of transitions must climb to reach a state of lower energy. The energy of the minimum plus $h$ is its escape level, the lowest energy that the run must reach to leave the minimum. Because the Metropolis rule accepts an uphill transition with probability $\mathrm{e}^{-\Delta E/\tau}$, a run at the minimum reaches a state $h$ above it only with a probability of order $\mathrm{e}^{-h/\tau}$ per attempt. At low temperature, the time to leave the minimum therefore grows as $\mathrm{e}^{h/\tau}$, as in thermally activated escape over an energy barrier. The critical depth $h^\ast$ is the largest depth over all nonoptimal local minima. It is the smallest constant $\tau_\mathrm{FN}$ for which the logarithmic schedule $\tau_n=\tau_\mathrm{FN}/\ln n$ reaches the optimum with probability tending to one\citeSI{Hajek1988}. The probability of leaving a minimum of depth $h$ at iteration $n$ is then of order $\mathrm{e}^{-h/\tau_n}=n^{-h/\tau_\mathrm{FN}}$. The sum of these probabilities over all iterations diverges only when $\tau_\mathrm{FN}\geq h$, and the run then leaves the minimum with probability one. The deepest nonoptimal minimum therefore sets how slowly the temperature must fall. The Catoni difficulty $\mathcal{D}$ is the largest ratio, over nonoptimal local minima, of the depth to the energy above the optimum. With the best schedule for a budget of $B$ iterations, the probability of not ending at the optimum falls approximately as $B^{-1/\mathcal{D}}$\citeSI{Catoni1992}. The reason is a trade-off at the end of the run. To leave a minimum of depth $h$ within $B$ iterations, the temperature must stay high enough that $\mathrm{e}^{h/\tau}\lesssim B$, that is, $\tau\gtrsim h/\ln B$. At such a temperature, a minimum that lies $\Delta E$ above the optimum is still occupied with a probability of order $\mathrm{e}^{-\Delta E/\tau}\approx B^{-\Delta E/h}$. The minimum with the largest ratio $h/\Delta E$ therefore limits the run: a deep minimum just above the optimum raises the failure probability more than an equally deep one far above it. A spin transition of a shortcut lowers the depth of a minimum if and only if it joins two states that meet two conditions. First, both lie below the present escape level of the minimum. Second, one is reachable from the minimum and a state of lower energy than the minimum is reachable from the other, each through transitions of the old library and without rising to the escape level. Such a transition skips the highest state of the escape route. The conditions follow because every route of the old library rises to the escape level. A route from the minimum to a lower state that stays below this level must therefore use the added transition, with its parts before and after that transition below the level. The transition itself may go uphill or downhill (Fig.~\ref{fig:concept}e). Adding moves while keeping the old ones cannot create a new local minimum or raise the depth of any minimum, because every previous route remains available. The critical depth falls when shortcuts lower the depth of every minimum of the largest depth. No property of the graph alone guarantees this fall, because the energy function determines where the escape routes rise. For any enlarged library in which some two feasible states are still not joined by one transition, some choice of the energy leaves the deepest minimum as deep as it was without the shortcuts. For example, let one of the two states be the optimum and the other a nonoptimal minimum, and let all remaining states share one higher energy. Every escape route from the minimum, with or without the shortcuts, must therefore climb to that energy. Lowering a depth by $\Delta h$ shortens the escape time by the factor $\mathrm{e}^{\Delta h/\tau}$, whereas a smaller backbone share costs at most the factor $1/\beta$. The gain exceeds the cost once $\tau<\Delta h/\ln(1/\beta)$, so shortcuts that lower the depth of the deepest minima shorten annealing at low temperature.

\paragraph{Exact landscapes at $L=10$} For each of the 20 test instances with $L=10$, all $10!=3{,}628{,}800$ assignments were enumerated. For each move library, the states were added in order of increasing energy, and their basins were merged along the spin transitions of the library, as when a landscape is flooded by a rising water level. Each added state joins the basins of those already added states to which one of these transitions connects it. When a basin merges into one with a lower minimum, its minimum can for the first time reach a lower state without rising above the energy of the joining state. That energy is therefore the escape level of the minimum, and the depth is this level minus the energy of the minimum. This procedure gives every nonoptimal local minimum, its depth, the critical depth and the difficulty. It was checked against an exhaustive search of escape routes on smaller instances. Random shortcuts were drawn ten times per instance and number of shortcuts, and their properties are summarized by the median over the draws. With the backbone alone (path library), the medians over the 20 instances are 74,347 nonoptimal local minima, a critical depth of $0.55\Psi^\ast$ and a difficulty of 5.1 (Fig.~\ref{fig:library}c). All 36 shortcuts together (complete library) reduce these to 382, $0.13\Psi^\ast$ and 0.26. For the same number of shortcuts, informed shortcuts leave as many local minima as random ones, within the sampling uncertainty. They lower the median critical depth to 0.70 to 0.89 times the random value and the median difficulty to 0.39 to 0.62 times. Both reductions exceed the sampling uncertainty except at $K=27$, where any two libraries share at least 18 of their 27 shortcuts.

\paragraph{Annealing at $L=10$} Each move library was annealed with 256 replicas at budgets of $2^6$, $2^7$, \ldots, $2^{16}$ iterations ($2^{20}$ for the path library), with the certified optimum as target. In the iterations to solution (Methods Eq.~\eqref{eq:its}) of random shortcuts, the median over the ten draws of each instance is taken first. For every library, the smallest median lies inside the range of budgets. Figure~\ref{fig:library}d uses the automatically calibrated schedule (Methods Section~\ref{meth:benchmarks}). Its temperatures differ between libraries but follow the same rule ($m=1$, $\kappa=2$), and no setting is chosen by performance. Each library is also taken at its own best budget, so neither the temperatures nor the budget favor any library. Table~\ref{tab:supp-its10} repeats this comparison with $m$ and $\kappa$ selected for each library on the six training instances by the protocol of Supplementary Section~\ref{supp:settings}. The selection is made separately for the path and complete libraries and for informed and random shortcuts at each $K$. Random shortcuts were drawn three times for each training instance, and these draws were used only for training. The selection lies inside the searched ranges ($m=1/16$ to 16, $\kappa=1$ to 32, and to 128 for random shortcuts at $K=9$) for every library.

Informed shortcuts give the smaller iterations to solution at every number of shortcuts with both schedules (Fig.~\ref{fig:library}d and Table~\ref{tab:supp-its10}). The 95\% interval of the ratio, informed over random (Table~\ref{tab:supp-its10}), lies below 1 at $K=4$ and 18 with the automatically calibrated schedule and at $K=4$ and 9 with optimized schedules. At $K=27$, the large number of shared shortcuts limits the possible difference.

Across library sizes, the iterations to solution and all properties of the landscape change together with the number of shortcuts (Fig.~\ref{fig:library}c, d). Comparing libraries of equal size separates the effect of each landscape property from that of the number of shortcuts. The 11 move libraries with $K$ shortcuts of one instance (one informed, ten random) are compared at one budget, the budget at which the $K$ informed shortcuts attain their iterations to solution in Fig.~\ref{fig:library}d. For each move library, the number of iterations needed to reach the optimum with 99\% probability by repeating runs of this budget is compared with each property of its landscape. The Spearman (rank) correlation is 1 when one quantity always increases with the other, whatever the form of the relation. Among libraries of equal size, the number of iterations correlates most strongly with the difficulty and the critical depth and least with the relaxation time at infinite temperature (Fig.~\ref{fig:library}e). With the optimized schedules, at the budget at which the $K$ informed shortcuts attain their iterations to solution in Table~\ref{tab:supp-its10}, the median correlations fall in the same order. The ordering of the four properties is the same when the common budget is the best budget of the random shortcuts, the best budget of all 11 libraries together, or any fixed budget from $2^8$ to $2^{16}$ iterations. These correlations across libraries do not by themselves establish a cause. The argument above explains why the depth, and not the relaxation time at infinite temperature, sets the annealing time at low temperature. The exact landscapes are limited to $L=10$. At $L=20$ and 30, only the relaxation time at infinite temperature can be computed.

\FloatBarrier
\section{\texorpdfstring{Surface-code decoding: objective, moves and reference decoders}{Surface-code decoding: objective, moves and reference decoders}}\label{supp:surface-formulation}

\paragraph{Code and feasible set} The rotated surface code of distance $d$ has $d^2$ data qubits on a $d\times d$ grid, indexed $i=1,\ldots,d^2$, and $d^2-1$ checks of weight four in the bulk and two on the boundary\citeSI{Bombin2007}. Each check measures the parity of the errors on the four (or two) neighboring qubits that it acts on. Half of the checks, the $Z$ checks, detect $X$ (bit-flip) errors, and the other half, the $X$ checks, detect $Z$ (phase-flip) errors. The checks of each type are collected in a binary check matrix, $H_Z$ or $H_X$, with one row per check, one column per qubit and an entry 1 when the check acts on the qubit. The distance $d$ is the smallest number of single-qubit errors that together violate no check yet change the encoded information. An $X$-error vector $e^X\in\{0,1\}^{d^2}$, with $e^X_i=1$ when qubit $i$ has an $X$ error, violates the $Z$ checks marked by the syndrome $\mathrm{syn}_Z=H_Ze^X$ (mod 2). Likewise, a $Z$-error vector $e^Z$ violates the $X$ checks $\mathrm{syn}_X=H_Xe^Z$. A correction $(c^X,c^Z)$ is feasible when it reproduces the measured syndrome, $H_Zc^X=\mathrm{syn}_Z$ and $H_Xc^Z=\mathrm{syn}_X$. The feasible set $\Omega$ is therefore an affine subspace over GF(2), the arithmetic of bits modulo 2. Every feasible correction is one particular solution of these linear equations plus a pattern of flips with zero syndrome, and every move adds such a pattern. With the spins $s^X_i=1-2c^X_i$ and $s^Z_i=1-2c^Z_i$, these equations are the parity constraints of main text Section~\ref{res:surface}.

\paragraph{Objective as spin products} The energy is the negative log-probability of the correction under the noise model. The goal is a correction in the same logical class as the actual error $e$. A correction $c$ that reproduces the measured syndrome explains it with certainty, $\Pr(\mathrm{syn}\mid e=c)=1$, so Bayes' rule gives
\[\Pr(e=c\mid\mathrm{syn})=\frac{\Pr(e=c)}{\Pr(\mathrm{syn})}.
\]
The denominator is the same for every feasible correction. The most probable error given the syndrome is therefore the feasible correction with the largest prior probability $\Pr(c)$, that is, with the smallest $-\ln\Pr(c)$. Under depolarizing noise, the $X$, $Y$ and $Z$ errors each have probability $p_\mathrm{err}/3$, and a $Y$ error has both an $X$ and a $Z$ component. The probability of a correction therefore depends only on the number of qubits that carry any error. Each such qubit contributes a factor $p_\mathrm{err}/3$ instead of $1-p_\mathrm{err}$, and so adds $\omega_3$ to the negative log-probability. With $\vee$ the logical OR,
\begin{equation}
-\ln\Pr(c)=\omega_3\sum_{i=1}^{d^2}\left(c^X_i\vee c^Z_i\right)+\mathrm{const},\qquad \omega_3=\ln\frac{3(1-p_\mathrm{err})}{p_\mathrm{err}}.
\end{equation}
Writing $c^X_i\vee c^Z_i=c^X_i+c^Z_i-c^X_ic^Z_i$ in spins gives $(3-s^X_i-s^Z_i-s^X_is^Z_i)/4$, and therefore
\begin{equation}
E=\omega_3\sum_{i=1}^{d^2}\frac{3-s^X_i-s^Z_i-s^X_is^Z_i}{4}=E_\mathrm{const}-\frac{\omega_3}{4}\sum_{i=1}^{d^2}\left(s^X_i+s^Z_i+s^X_is^Z_i\right).
\label{eq:surface-objective}
\end{equation}
The objective has $M=3d^2$ clauses of weight $\omega_3/4$, the first-order clauses $s^X_i$ and $s^Z_i$ and the second-order clause $s^X_is^Z_i$ of each qubit $i$. The second-order clause makes a $Y$ error cost $\omega_3$, the cost of a single error, instead of the $2\omega_3$ that its $X$ and $Z$ components would cost if counted separately. Minimum-weight matching that decodes the two error types separately, as PyMatching does here, counts a $Y$ error as two errors, one of each type. Because $\omega_3$ multiplies every clause, the lowest-energy correction does not depend on $p_\mathrm{err}$. It is the smallest set of errors that explains the syndrome, with a $Y$ error counted once. Sampling the classes, in contrast, needs the energy on the scale of the negative log-probability, so that the Boltzmann weight at $\tau=1$ is proportional to the probability of the correction.

\paragraph{Moves} An $X$ stabilizer move reverses the spins $s^X_i$ on the four (on the boundary, two) qubits of one $X$ check (main text Section~\ref{res:surface}). An $X$ check and a $Z$ check share either no qubit or two, so the move changes the error count of every $Z$ check by an even number and leaves the syndrome unchanged. A logical operator reverses the spins of a straight line of $d$ qubits from one boundary to the opposite one. The line shares an even number of qubits with every check that detects these flips, so it also keeps the syndrome, but it changes the encoded information. The feasible corrections of one error type therefore fall into two logical classes, and corrections in different classes differ by a logical operator. A correction outside the class of the actual error results in a logical error. The move library of stabilizer moves only cannot change the logical class (main text Fig.~\ref{fig:surface}c). The backbone adds one logical operator per error type and connects the feasible corrections, because any two of them differ by stabilizers and at most one logical operator per type. The backbone with all shortcuts adds the other $d-1$ logical operators per type and has $R=d^2-1+2d$ moves, from 14 at $d=3$ to 254 at $d=15$.

\paragraph{Initial state, output and failure} Each shot starts from a particular solution of the syndrome equations. In the emulation (main text Fig.~\ref{fig:event}), this solution is obtained by Gaussian elimination over GF(2). The FPGA decoder forms it from fixed chains of qubits, one from each violated check to the boundary. Each chain violates only the check that it starts from (Supplementary Section~\ref{supp:fpga}). Both start corrections depend on the syndrome alone, not on the actual error, so every move library and every budget starts a shot from the same correction. For error $e$ and correction $c$, the residual $e+c$ (mod 2) is the pattern of flips that remain after the correction is applied. The residual has zero syndrome, so it is a combination of stabilizers, possibly plus a logical operator. The shot fails when the logical operator is present, because $c$ then lies in a different logical class from $e$. This case is detected when the residual anticommutes with the logical operator of the other type, that is, when it shares an odd number of qubits with that line. Every stabilizer, by contrast, shares an even number of qubits with that line. Under depolarizing noise, a failure of either error type counts as a failure of the shot.

\paragraph{Reference decoders} Versions and settings of the reference decoders are given in Methods Section~\ref{meth:problems}. PyMatching\citeSI{Higgott2022,Higgott2025} is the standard minimum-weight perfect-matching decoder of the surface code\citeSI{Dennis2002}. It decodes the $X$ and $Z$ errors separately. Because a qubit error of one type violates at most two checks, the checks can be drawn as the nodes of a graph whose edges are the qubits. The decoder pairs the violated checks, or joins them to the boundary, by chains of edges with the smallest total weight. With uniform weights, this gives the correction with the fewest flipped qubits of that type. BP-OSD\citeSI{Panteleev2021,Roffe2020} runs min-sum belief propagation followed by ordered-statistics decoding. Belief propagation passes messages between checks and qubits to estimate the probability that each qubit carries an error. A scaling factor damps the min-sum messages. Ordered-statistics decoding is used when these estimates do not give a correction that reproduces the syndrome. It ranks the qubits from most to least likely in error and solves the syndrome equations by Gaussian elimination over GF(2) on the most likely ones. It then tests combinations of flips on the remaining qubits, as many as its order sets, for a more likely correction. Its joint check matrix has one column for each of the $X$, $Y$ and $Z$ errors of every qubit, each with probability $p_\mathrm{err}/3$, so it also uses the correlation between the two error types. Union-find (peeling decoder)\citeSI{Delfosse2021} decodes the $X$ and $Z$ errors separately. It grows clusters of qubits around the violated checks until each cluster contains an even number of violated checks or touches the boundary, so that each cluster can be corrected on its own. Each cluster is then corrected by peeling a spanning tree from its leaves inward. The run time grows almost linearly with the number of qubits. Tensor-network maximum-likelihood (ML) decoding\citeSI{Bravyi2014} writes the sum of the probabilities of all corrections in a class as a network of small arrays (tensors). The tensors are multiplied and summed over their shared indices (contracted). An exact contraction costs time exponential in $d$. The decoder therefore contracts the network column by column and keeps only the largest terms of each intermediate result, a matrix product state whose size is set by the bond dimension. The windowed real-time baselines are described in Supplementary Section~\ref{supp:realtime}.

\FloatBarrier
\section{One syndrome per shot: sampling the logical classes}\label{supp:oneshot}

\paragraph{Why sampling at $\tau=1$ gives the most likely class} Under depolarizing noise, $E(c)=-\ln\Pr(c)+\mathrm{const}$ (Supplementary Eq.~\eqref{eq:surface-objective}), so the fraction of iterations at $\tau=1$ in class $\gamma$ estimates $P(\gamma)$ (main text Eq.~\eqref{eq:class-probability}). This requires that the moves reach every feasible correction: stabilizer moves connect the corrections within a logical class, and logical operators connect the classes. There are four classes, the combinations of the two classes of each error type. Markov-chain Monte Carlo decoders use the same principle\citeSI{Wootton2012}. A decoder that returns the lowest-energy correction, as an annealer does, instead estimates the most likely single correction. When the number of errors is small, the sum over a class is dominated by its largest term, so both rules usually choose the same class.

\paragraph{Replica exchange} Each design runs between 16 and 64 replicas at temperatures spaced geometrically from 1 to 10 (Supplementary Table~\ref{tab:supp-designs}). At $\tau=10$, even the largest energy rise of a logical operator at $d=15$ and $p_\mathrm{err}=0.12$, $d\,\omega_3$, is accepted with probability $\mathrm{e}^{-4.6}\approx0.01$. The exchanges of main text Eq.~\eqref{eq:replica-exchange} are attempted after every second sweep, alternately between the even and the odd pairs of neighboring replicas (Supplementary Section~\ref{supp:fpga}). Neither the temperature range nor the exchange interval was optimized. On 16,000 shots at $p_\mathrm{err}=0.12$ outside the reported shots, decoded with the same random-number seeds, we raised the top temperature to 20 or lowered it to 5. We also exchanged after every sweep, with fewer sweeps so that the number of clock cycles stayed the same. Two LERs measured on the same shots are compared by Newcombe's 95\% interval for paired proportions\citeSI{Newcombe1998}, here and in the rest of this section. None of these changes lowered the logical error rate (LER) beyond the sampling uncertainty at $d=11$ or 15. At $d=11$, the top temperature of 20 and the exchange after every sweep raised the LER slightly.

\paragraph{Budgets and replicas} The fixed budget of each design is 46, 128, 216, 304, 430, 512 and 1,722 sweeps per shot at $d=3$ to 15. For $d=7$ to 15, we measured the LER at increasing budgets on 16,000 shots that are not among the reported shots, and fitted these LERs. The fixed budget is the smallest budget at which the fit lies within 0.005 of the converged LER, or within 5\% of it if that is smaller. These budgets were set at $p_\mathrm{err}=0.15$ for $d=7$ and 9 and at 0.12 for $d\geq11$, so the budgets at $d=7$ and 9 are conservative at 0.12. At $d=3$, the budget was set against ML on 4,000 shots. At $d=5$, it was raised to 128 sweeps, because a smaller budget left the LER above that of ML. Near the threshold, at $p_\mathrm{err}\geq0.14$, these budgets leave the LER at $d\geq5$ up to 0.017 above that of ML. The results there therefore use $2^{12}$ sweeps at $d=5$ to 15, set at run time on the same designs. For each distance, we compared two numbers of replicas: 16, and the largest number that fits within 70\% of the logic resources of the chip, up to 64. We tested both numbers on the same 16,000 shots and kept the number with the lower decode time at the same LER.

\paragraph{Counting, early stopping and confidence} The counts of the first half of the run (burn-in) are discarded, so that the start correction has little influence on the decision. The class counters record the class of the replica at $\tau=1$ in every clock cycle of the second half, and the decision is the class with the largest count. With early stopping, the decoder checks the counts at checkpoints after 2, 4, 8, \ldots\ sweeps, from a smallest checkpoint $g_\mathrm{min}$ on. It stops when, among the counts since the previous checkpoint, the largest class count is at least a fraction $f_\mathrm{stop}$ of their total, and it decides from those counts. For each $d$, $g_\mathrm{min}$ and $f_\mathrm{stop}$ were chosen as the fastest setting whose LER differs by at most 0.001 from that of the fixed budget of each design on 20,000 board shots at $p_\mathrm{err}=0.12$. The chosen values are $g_\mathrm{min}=16$, 32, 64, 64, 64, 128 and 256 and $f_\mathrm{stop}=30/32$, 29/32, 29/32, 29/32, 31/32, 29/32 and 29/32 at $d=3$ to 15. The 8,000 shots of each point of main text Fig.~\ref{fig:surface}d at $p_\mathrm{err}=0.12$ were not used to choose these settings. On these shots, early stopping leaves the LER unchanged within the sampling uncertainty, except at $d=11$, where the LER rises slightly, from 0.0464 to 0.0478. The fraction of the counts that falls on the chosen class is the confidence of the decision (Supplementary Section~\ref{supp:board}).

\paragraph{Accuracy and reference} The variants of main text Fig.~\ref{fig:surface}c without shortcuts or without replica exchange exist only in the software model of the designs (Methods Section~\ref{meth:fpga}). The board itself decoded 20,000 further shots per distance for early stopping and the confidence. It also decoded the 8,000 shots at $p_\mathrm{err}=0.16$ and 0.18 again with $2^{12}$ sweeps at $d=5$ to 15, and its LERs there are within 0.007 of those of ML. For the tensor-network decoder\citeSI{Bravyi2014}, bond dimension 32 gives the same decision as bond dimension 16 on all 4,000 shots tested at each of $d=15$, $p_\mathrm{err}=0.12$ and 0.18. The reference at bond dimension 16 is therefore converged.

\FloatBarrier
\section{Real-time decoding: noise model, moves and sliding windows}\label{supp:realtime}

\paragraph{Noise model and memory experiment} A memory experiment stores one logical qubit for a number of rounds of check measurements and then tests whether the decoded logical state still equals the stored one. At code distance $d$, our memory experiment has $4d$ noisy rounds followed by one round measured without error, which represents the final readout of the data qubits. Before every round, each qubit suffers an $X$, $Y$ or $Z$ error, each with probability $p_\mathrm{err}/3$, and these errors accumulate. In each of the $4d$ noisy rounds, every measured check outcome is also flipped with probability $p_\mathrm{err}$. This is a depolarizing variant of the phenomenological noise model\citeSI{Dennis2002}. Methods Section~\ref{meth:problems} relates it to the more common bit-flip convention. It adds measurement errors to the code-capacity noise of Supplementary Section~\ref{supp:surface-formulation} but does not model the gates of the measurement circuit. A data error that occurs before round $t$ changes the outcome of its checks in round $t$ and in every later round, whereas a measurement error changes one outcome only. The decoder therefore operates on detection events. A detection event marks a change of the outcome of check $j$ between rounds $t-1$ and $t$, $\mathrm{syn}(j,t)\oplus\mathrm{syn}(j,t-1)=1$, where $\mathrm{syn}(j,t)$ is the measured outcome and $\oplus$ is the exclusive OR.

\paragraph{Variables, objective and moves} A configuration assigns to every qubit $i$ and round $t$ the bits $x_{i,t}=(1-s^X_{i,t})/2$ and $z_{i,t}=(1-s^Z_{i,t})/2$, and to every check $j$ and noisy round $t$ the measurement-error bit $m_{j,t}=(1-s^\mathrm{m}_{j,t})/2$, with the spins of main text Eq.~\eqref{eq:realtime-energy}. A configuration is feasible when it explains every detection event. Its negative log-probability is, up to a constant,
\begin{equation}
E=\omega_3\sum_{i,t}\bigl(x_{i,t}\vee z_{i,t}\bigr)+\omega_\mathrm{m}\sum_{j,t}m_{j,t},\qquad
\omega_3=\ln\frac{3(1-p_\mathrm{err})}{p_\mathrm{err}},\qquad \omega_\mathrm{m}=\ln\frac{1-p_\mathrm{err}}{p_\mathrm{err}},
\label{eq:supp-rt-objective}
\end{equation}
by the argument of Supplementary Section~\ref{supp:surface-formulation}, applied to each round and to each measurement. The stabilizer moves, logical operators and time-like moves of main text Section~\ref{res:surface} keep a configuration feasible. In every round, all $d$ logical operators of each type are used. The bits and the moves form clauses and \textit{move} neurons exactly as for one syndrome per shot, with two weights instead of one (Supplementary Section~\ref{supp:fpga}).

\paragraph{Sliding windows} The decoder operates on windows of $2d$ rounds\citeSI{Dennis2002}. The first window covers rounds 0 to $2d-1$, the second rounds $d$ to $3d-1$, and the final window rounds $2d$ to $4d-1$ together with the round without error. A window is decoded once its last round has arrived and the previous window has been committed. In a non-final window, the detection events of the top round can be formed only against the round below, because the round above has not arrived. The time-like moves of the top round are truncated accordingly, to the error bit of the top round and the measurement-error bits of its checks. A chain of time-like moves on the $d$ qubits of a logical line can shift the line up one round at a time, and the truncated moves of the top round can then remove it. A non-final window therefore cannot distinguish the logical classes, so it commits a configuration rather than a class: the lowest-energy configuration that any replica has found. Its bottom $d$ rounds lie at least $d$ rounds below the open top, so this configuration determines them reliably. At the end of every sweep, the decoder compares the energy of every replica with the lowest energy found so far and stores a replica's configuration whenever its energy is strictly lower. After the last sweep, the window commits the stored configuration. Removing its measurement errors from the outcomes of the last committed round gives the corrected syndrome $\mathrm{syn}^\ast(j)$, the syndrome of the committed correction. The next window compares its outcomes with $\mathrm{syn}^\ast$, which removes the effect of the errors that the committed correction already explains. The start configuration of every replica explains each reported outcome that differs from $\mathrm{syn}^\ast(j)$ by a measurement error, $m_{j,t}=\mathrm{syn}(j,t_0+t)\oplus\mathrm{syn}^\ast(j)$, with no data errors. Here $t_0$ is the first round of the window, and $t$ counts the rounds within the window. The moves then change this explanation, for example by replacing repeated measurement errors with a data error in any round of the window, including its first. The final window also needs data errors in its start configuration, because the round without error contains no measurement errors to which a changed outcome can be attributed. For each check whose outcome in that round differs from $\mathrm{syn}^\ast(j)$, it adds a fixed shortest chain of data errors from that check to the boundary. The start correction for one syndrome per shot is formed in the same way. The final window commits the class that the replica at $\tau=1$ visits most often in the second half of its run, as for one syndrome per shot (Supplementary Section~\ref{supp:oneshot}). A shot fails when the committed correction, accumulated over all windows, differs from the actual errors by a logical operator of either type. The LER is the fraction of failed shots, each shot being one memory experiment of $4d$ noisy rounds, not one round.

\paragraph{Settings} The decoder runs 16 replicas at $d\leq7$ and 24, the largest number that fits on the chip, at $d=9$ and 11, at temperatures spaced geometrically from 1 to 10. This range was adopted from the decoding of one syndrome per shot and not optimized. On 3,000 other shots at $d=9$ and $p_\mathrm{err}=0.03$, a top temperature of 5 or 20 instead of 10 leaves the LER unchanged within the sampling uncertainty (Newcombe's interval, Supplementary Section~\ref{supp:oneshot}). Neighboring replicas attempt to exchange their temperatures after every sweep, alternately the even and the odd pairs. The 16 replicas and the exchange after every sweep are those of the software decoder on which the windows were designed. In software at $d=9$, 24 replicas give the same LER as 32, whereas 16 replicas give 12\% more logical errors than 24 at $p_\mathrm{err}=0.04$. Every window runs $2^{12}$ sweeps (Supplementary Section~\ref{supp:qec-results}). Committing a configuration instead of a class reduces the accuracy, as measured in software against decoding the whole memory experiment at once and returning the most visited class. At $d=7$ and 9, the windowed LER is 6 to 13\% higher than the LER of this whole-experiment decoding. More sweeps (up to $2^{14}$) do not remove this gap. The window size was chosen in software. A window that extends only $d/2$ rounds above its committed rounds was less accurate, and committing only $d/2$ rounds per window doubled the computation per round without lowering the LER.

\paragraph{Windowed CPU baselines} Four CPU decoders decode the same shots in the same windows, with the same commit rule and the same prior probabilities ($p_\mathrm{err}/3$ per error type and $p_\mathrm{err}$ per measurement). PyMatching\citeSI{Higgott2022,Higgott2025} decodes the $X$ and $Z$ errors separately, with the weight $\ln[(1-2p_\mathrm{err}/3)/(2p_\mathrm{err}/3)]$ per data error and $\omega_\mathrm{m}$ per measurement error. The data weight uses $2p_\mathrm{err}/3$, the probability that a qubit acquires an $X$ component (from an $X$ or a $Y$ error). Correlated PyMatching\citeSI{Fowler2013,Paler2023} runs the correlated matching of PyMatching on a detector error model of the window built with Stim\citeSI{Gidney2021}. This model lists every possible error with its probability and the detection events that it causes. In it, a $Y$ error causes the detection events of an $X$ and a $Z$ error together. Correlated matching first matches the two error types separately and then lowers the weights of the errors that are likely given the other type's matching, before it matches again. BP+matching\citeSI{Higgott2023} runs belief propagation (ldpc version 2.4.1, product-sum, 30 iterations) on the joint check matrix of $X$, $Y$ and $Z$ errors. It combines the resulting error probabilities into one probability per error type and passes them to PyMatching as weights, so it builds a new matching graph for every shot. BP-OSD\citeSI{Panteleev2021,Roffe2020} uses the settings of one syndrome per shot on the joint check matrix (Methods Section~\ref{meth:problems}).

\FloatBarrier
\section{The decoder on the FPGA}\label{supp:fpga}

\begin{figure}[!tbp]
\centering
\includegraphics[width=\linewidth]{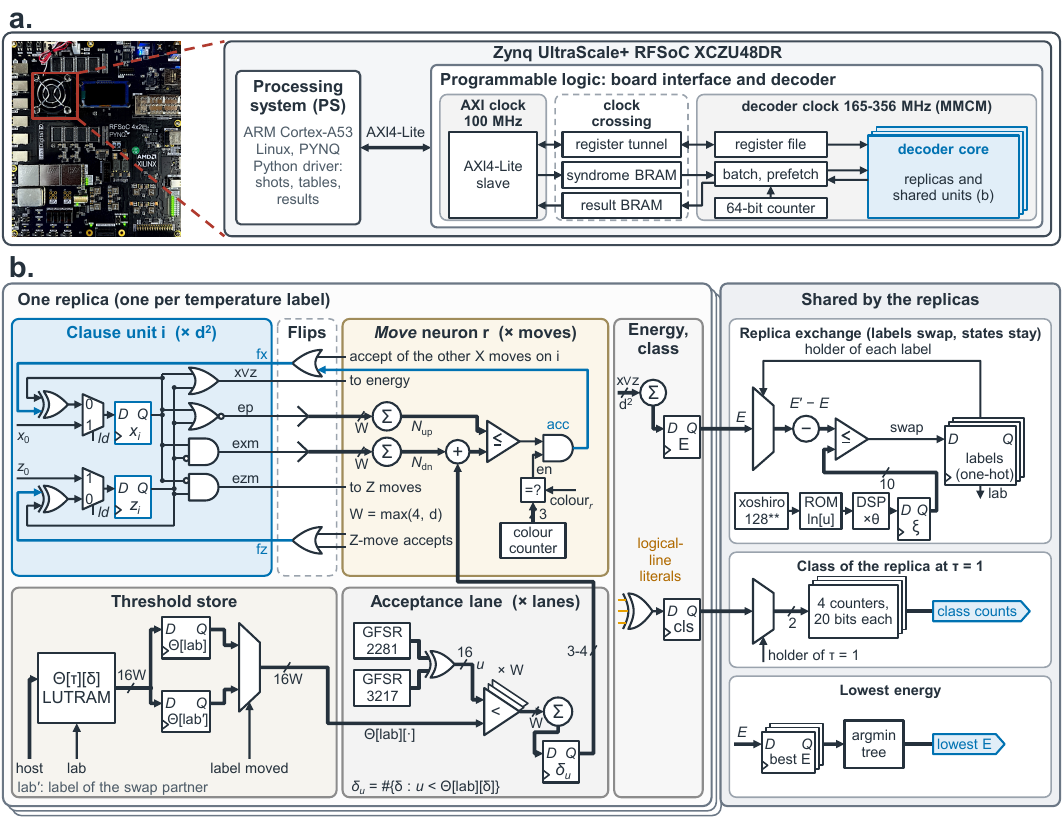}
\caption{\textbf{The decoder on the RFSoC 4x2 board.}
\textbf{a}, The board and the system. The Zynq UltraScale+ RFSoC XCZU48DR (under the heat sink) holds the processing system (PS) and the programmable logic. The PS runs the Python driver. The programmable logic holds the board interface and the decoder core. The decoder core runs on its own clock (165--356\,MHz). AXI4-Lite: the bus between PS and logic. BRAM: block memory. Register tunnel: transfers settings between the two clock domains. MMCM: clock generator. 64-bit counter: time stamps on the decoder clock.
\textbf{b}, The core, shared by both decoding modes. Each replica holds its state in the clause units (blue). The clause unit of qubit $i$ stores the bits $x_i$ and $z_i$ (flip-flops $D$, $Q$), loaded with the start state $x_0$, $z_0$ (ld). Its outputs ep, exm and ezm are $+1$ (the move adds an error), $-1$ (an $X$ move or a $Z$ move clears one) or 0. Flips: fixed wiring that collects the accepts of the moves on qubit $i$ into the flips fx and fz. $x\vee z$: the qubit carries an error. A \textit{move} neuron counts its clause units, $\Delta E/\omega_3=N_\mathrm{up}-N_\mathrm{dn}$ over 4 or $d$ units. It compares this count with the count $\delta_u$ of its acceptance lane. When its color class is active, the neuron flips the bits of its clause units directly (one-cycle loop). The acceptance lane draws a 16-bit random number $u$. It counts the entries above $u$ in the threshold row $\Theta[\mathrm{lab}]$ of the replica's temperature label. $W=\max(4,d)$: thresholds per row and the largest number of units per neuron. LUTRAM: table memory built from lookup tables (LUTs). Shared by the replicas: replica exchange, the energy ($E$) and class (cls) registers of each replica, the class counters of the replica at $\tau=1$ and the lowest-energy search. Replica exchange swaps the temperature labels and leaves the states in place. It uses a random-number generator (xoshiro128**), a table of logarithms (ROM) and a multiplier (DSP). The multiplier scales the logarithm from the table by the constant $\theta$ of each pair of neighboring temperature labels. Blue tags: the two outputs of the core, used in Fig.~\ref{fig:supp-arch2}.}\label{fig:supp-arch1}
\end{figure}
\begin{figure}[!tbp]
\centering
\includegraphics[width=\linewidth]{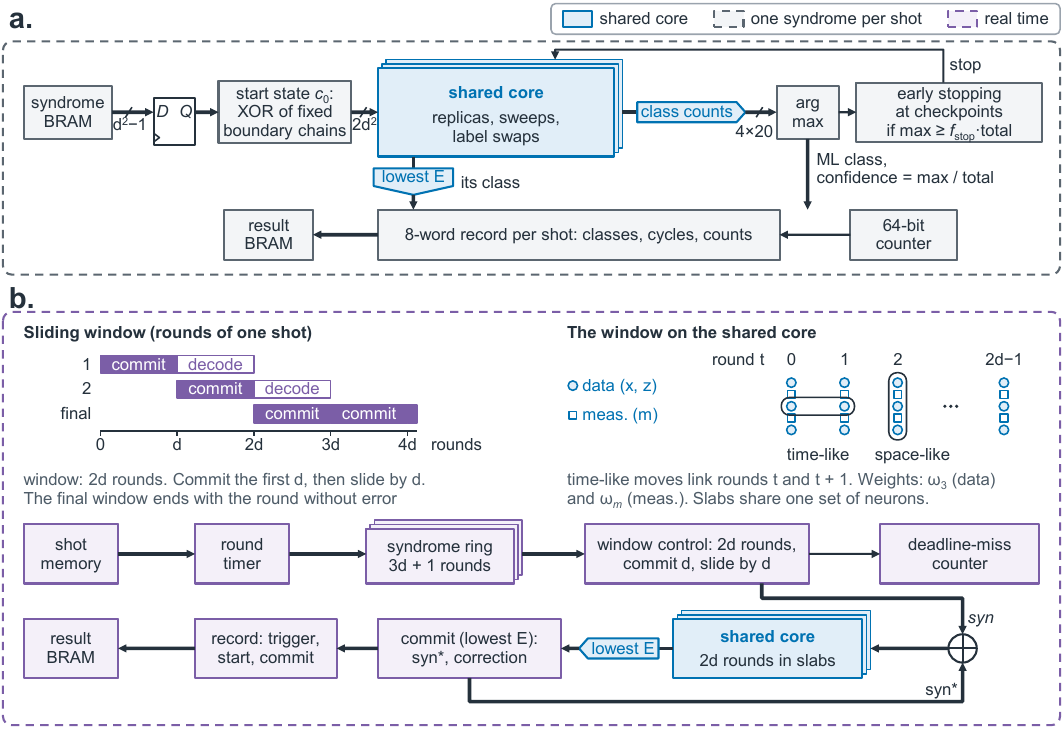}
\caption{\textbf{What each decoding mode adds around the core.} Blue: the shared core of Fig.~\ref{fig:supp-arch1}, with its outputs (tags). Dashed frames: blocks used only in that mode (grey: one syndrome per shot, purple: real time).
\textbf{a}, One syndrome per shot. The start state $c_0$ is the exclusive OR of fixed chains to the boundary for the violated checks ($d^2-1$ syndrome bits in, $2d^2$ bits out). The decision is the most frequent logical class, from the four class counts of 20 bits each, with its confidence. Early stopping: the run ends at a checkpoint when the largest count reaches the fraction $f_\mathrm{stop}$ (Supplementary Section~\ref{supp:oneshot}) of the counts since the previous checkpoint.
\textbf{b}, Real time. A round timer releases the rounds of a stored shot into a ring memory of syndromes. Each window of $2d$ rounds runs on the same core, with the clause bits of the $2d$ rounds stored in slabs from $d=5$ on. Time-like moves link the rounds. Stabilizer moves and logical operators act within one round. A non-final window commits the lowest-energy state and returns the correction of its first $d$ rounds and the corrected syndrome syn$^\ast$ that starts the next window. The final window commits the most frequent class. A counter records missed deadlines.}\label{fig:supp-arch2}
\end{figure}

\paragraph{Binary variables} The decoder runs on the Zynq UltraScale+ RFSoC XCZU48DR chip of an RFSoC 4x2 board (Supplementary Fig.~\ref{fig:supp-arch1}a). As in ref.~\citeSI{Aadit2022}, binary variables are more convenient than bipolar ones for computing on an FPGA. We therefore binarize the architecture and store each first-order clause output $T$ as the bit $\bar T=(1-T)/2$ instead of its bipolar value $\pm1$. The following paragraphs show that the hardware evaluates the same energy changes and applies the same clause toggles as the equations of Methods Section~\ref{meth:model}. Three further choices, colored updates, replica exchange and the start state, differ from the Methods equations, improve speed and accuracy and leave the structure of the architecture unchanged.

\paragraph{Clause units} Under depolarizing noise, each qubit $i$ has three objective clauses, $s^X_i$, $s^Z_i$ and $s^X_is^Z_i$, each of weight $\omega_3/4$ (Supplementary Eq.~\eqref{eq:surface-objective}). In bits, $x_i=(1-s^X_i)/2$ and $z_i=(1-s^Z_i)/2$ are the outputs of the two first-order clauses themselves. The output of the second-order clause, $s^X_is^Z_i=(1-2x_i)(1-2z_i)$, follows from these two bits, so the clause unit of qubit $i$ stores only $x_i$ and $z_i$. The energy of the three clauses of qubit $i$ is $\omega_3(x_i\vee z_i)$ (Supplementary Section~\ref{supp:surface-formulation}), which the unit computes with one gate. Every bit of the state belongs to exactly one clause unit, so the clause units hold the whole state, as the architecture requires.

\paragraph{Colored updates} Methods Eq.~\eqref{eq:single-commit} allows at most one accepted spin transition before the inputs settle. Moves that share no clause, and none of which changes a clause that controls a switch of another, are an exception. Applying one of them changes no input of the others, so applying them together gives the same result as applying them one after another. Such moves form a color class and are proposed, tested and applied in the same clock cycle. We form the color classes with the greedy DSATUR algorithm\citeSI{Brelaz1979}, which does not guarantee the minimum number of color classes. The surface code is sparse, so a few color classes suffice at every distance: six for one syndrome per shot and 12 for a real-time window (15 at $d=7$, where the rounds are arranged differently in hardware). For one syndrome per shot, the six color classes are two of $X$ stabilizer moves, one of $X$ logical operators, and likewise for $Z$. Each error type needs three color classes because a logical operator and two neighboring checks that it crosses conflict pairwise. $X$ and $Z$ moves need separate color classes because they share clause units.

\paragraph{Encoder} Consider a move $G_r$ that reverses the $X$ components of a set $\mathcal F_r$ of qubits. Its switches are constant and on, so by Methods Eq.~\eqref{eq:encoder} its energy change is twice the weighted sum of the outputs of the clauses that it toggles. On qubit $i\in\mathcal F_r$, it toggles the two clauses $s^X_i$ and $s^X_is^Z_i$, each of weight $\omega_3/4$. With the bits $x_i$ and $z_i$ of the clause unit,
\begin{equation}
\begin{aligned}
\Delta E_{G_r}&=2\cdot\frac{\omega_3}{4}\sum_{i\in\mathcal F_r}\left(s^X_i+s^X_is^Z_i\right)
=\omega_3\sum_{i\in\mathcal F_r}\bigl[(1-x_i)(1-z_i)-x_i(1-z_i)\bigr]\\
&=\omega_3\Bigl[\sum_{i\in\mathcal F_r}\neg\left(x_i\vee z_i\right)-\sum_{i\in\mathcal F_r}\left(x_i\wedge\neg z_i\right)\Bigr]
=\omega_3\left(N_\mathrm{up}-N_\mathrm{dn}\right),
\end{aligned}
\label{eq:supp-fpga-de}
\end{equation}
where $\vee$, $\wedge$ and $\neg$ are the logical OR, AND and NOT. $N_\mathrm{up}$ counts the qubits of $\mathcal F_r$ without any error, to which the move adds an error. $N_\mathrm{dn}$ counts the qubits with an $X$ component only, which the move clears. A qubit with a $Z$ component ($z_i=1$) adds to neither, because flipping its $X$ component turns $Z$ into $Y$ or $Y$ into $Z$. A clause unit thus sends $+1$, $-1$ or 0 to each connected neuron, and the neuron adds them. Moves that flip the $Z$ component are handled in the same way, with $x_i$ and $z_i$ exchanged. Because every clause has the same weight, $\omega_3$ is common to all energy changes, and the neuron computes only the integer $\delta=N_\mathrm{up}-N_\mathrm{dn}$. The weight enters the acceptance tables instead. A stabilizer move adds at most four terms, and a logical operator adds $d$ terms. On the FPGA, each neuron recomputes its count from its clause units in every clock cycle.

\paragraph{Acceptance} The Metropolis test of Methods Eqs.~\eqref{eq:firing} and~\eqref{eq:threshold} accepts an energy change $\Delta E=\omega_3\delta$ with probability $\min(1,\mathrm{e}^{-\omega_3\delta/\tau})$ at temperature $\tau$. Each test draws a 16-bit uniform random integer $u$ and accepts when $\delta\leq0$ or $u<\Theta_\tau(\delta)$. The host computes the table $\Theta_\tau(\delta)=\lfloor2^{16}\mathrm{e}^{-\omega_3\delta/\tau}\rfloor$ for every possible count, from 1 to the largest number of qubits that one move flips, $\max(4,d)$. The tables store $p_\mathrm{err}$ and the temperatures, so one design serves every physical error rate. The test resolves probabilities down to $2^{-16}$. An uphill move with a smaller acceptance probability, such as $\delta\geq4$ at $\tau=1$ and $p_\mathrm{err}\leq0.15$, is never accepted, and the agreement with ML decoding (main text Fig.~\ref{fig:surface}d) includes this effect. Because $\Theta_\tau(\delta)$ falls with $\delta$, the hardware counts, one cycle ahead, the table entries above $u$, $\delta_u=\#\{\delta:u<\Theta_\tau(\delta)\}$, and accepts when $\delta\leq\delta_u$. The two tests give the same decision, and the move loop then compares only two small integers. Every move proposed in a clock cycle needs its own random number $u$. Each replica therefore has one random-number source, an acceptance lane, for each move of its largest color class. In each cycle, every lane serves one move of the active color class, so the same lanes serve every color class in turn. Each lane XORs the outputs of two lagged exclusive-OR generators, in which each new word is the exclusive OR of two earlier words.

\paragraph{Decoder} An accepted move toggles its clauses, $T\leftarrow(1-2\chi)T$ in Methods Eq.~\eqref{eq:decoder-clause}, which in bits reads $\bar T\leftarrow\bar T\oplus\chi$. The clause unit flips the bit of the accepted move's component. With one clock cycle per color class and four cycles of replica exchange after every second sweep, a sweep takes eight cycles on average. A shot adds a few cycles to start and to decide (Supplementary Table~\ref{tab:supp-designs}). In each cycle, the clause bits pass through the clause units, the neuron counts and the acceptance tests, and the accepted flips are written back to the clause bits (Supplementary Fig.~\ref{fig:supp-arch1}b).

\paragraph{Replicas and decisions} Each replica is a complete copy of the clause units and \textit{move} neurons. Replica exchange swaps temperature labels between neighboring replicas, and the states stay in place. A swap of two replicas at temperatures $\tau<\tau'$ with energies $E$ and $E'$ is accepted with the probability of main text Eq.~\eqref{eq:replica-exchange}. Both energies are integer multiples of $\omega_3$, and the swap is accepted when $(E'-E)/\omega_3\leq\theta(-\ln u)$, where $u$ is a uniform random number and $\theta=1/[\omega_3(1/\tau-1/\tau')]$. The host writes $\theta$ for each pair of neighboring temperatures. The FPGA looks up $-\ln u$ for a 12-bit random number in a table of logarithms and multiplies it by $\theta$ with one multiplier. Four counters, one for each combination of the $X$ and $Z$ logical classes, record in every clock cycle the class of the replica that holds $\tau=1$. A pipelined search keeps, for each replica, the lowest energy found and the logical class of that configuration. Supplementary Sections~\ref{supp:oneshot} and~\ref{supp:realtime} describe how each decoding mode decides from these outputs.

\paragraph{Start state} The start correction of Supplementary Section~\ref{supp:surface-formulation}, the exclusive OR of fixed shortest chains from the violated checks to the boundary, is formed by exclusive-OR gates in one clock cycle. In a real-time window, the start configuration is copied from the stored outcomes with one exclusive OR per check (Supplementary Section~\ref{supp:realtime}).

\paragraph{Real-time additions} The real-time decoder adds one clause unit per check and round that holds the measurement-error bit, with its own weight $\omega_\mathrm{m}$ (Supplementary Eq.~\eqref{eq:supp-rt-objective}). A time-like move changes the energy by $\Delta E=\Delta_\mathrm{d}\,\omega_3+\Delta_\mathrm{m}\,\omega_\mathrm{m}$, with small integers $\Delta_\mathrm{d}$ and $\Delta_\mathrm{m}$. The ratio $\omega_3/\omega_\mathrm{m}$ depends on $p_\mathrm{err}$, 1.30 to 1.37 over the tested range, so the integer count $\delta$ of the other moves no longer orders the energy changes. A time-like move, however, has only seven possible positive energy changes when two checks detect the error on its qubit, and five when one check on the boundary does. The host sorts them and writes the acceptance thresholds in that order. The \textit{move} neuron of a time-like move outputs the rank of its energy change (3 bits), and the test compares this rank with the lane count. Energy changes that are not positive have rank 0 and are always accepted. The \textit{move} neurons of stabilizer moves and logical operators keep the integer count. Each replica stores, for each temperature, the thresholds of the integer counts and those of the ranks, and all moves of a color class use the same kind of table. The order of the seven ranks changes only where $3\omega_3=4\omega_\mathrm{m}$, at $p_\mathrm{err}=1/28$, and one mode bit selects the order. The order of the five ranks never changes. The test is therefore exact up to the 16-bit resolution of the tables.

A window of $2d$ rounds holds up to 8,448 moves at $d=11$, too many to give every move of every round its own neuron in each replica. From $d=5$ on, the rounds of a window are therefore grouped into slabs that share one set of neurons and are processed one after another in each sweep (Supplementary Table~\ref{tab:supp-designs}). Each move belongs to the slab of the round in which it starts, so a time-like move at the top of a slab also flips a bit in the first round of the next slab. The clause bits of all rounds stay stored, and only the neurons are shared. At $d=9$ and 11, the loop from clause bits to accepted flips is also split into three pipeline stages, which allows a faster clock. A replica's next color class needs the flips of the previous color class, so one replica alone would leave the neurons idle for two of every three cycles. Several replicas are therefore time-multiplexed on each set of neurons, one per clock cycle: 3 at $d=9$ and 6 at $d=11$. Each replica keeps its own clause bits. Replica exchange adds 7 to 9 cycles after every sweep.

A ring memory stores $3d+1$ rounds of outcomes: the $2d$ rounds of the window being decoded, the $d$ rounds that arrive meanwhile and the round without error. A window that would have to wait for the commit of the previous window misses its deadline, and a counter records every such window. Supplementary Fig.~\ref{fig:supp-arch2} shows what each decoding mode adds around the core.

\paragraph{System} The processing system of the chip runs Linux with PYNQ and a Python driver (Supplementary Fig.~\ref{fig:supp-arch1}a). The host writes the acceptance and swap tables, the seeds of the random-number generators and the budget as run-time settings. For every shot, the decoder writes a record of the decisions, the number of sweeps, the cycle counts from the arrival of the syndrome to the decision, the class counts and a time stamp.

\FloatBarrier
\section{Further decoding results}\label{supp:qec-results}

\begin{figure}[!tbp]
\centering
\includegraphics[width=\linewidth]{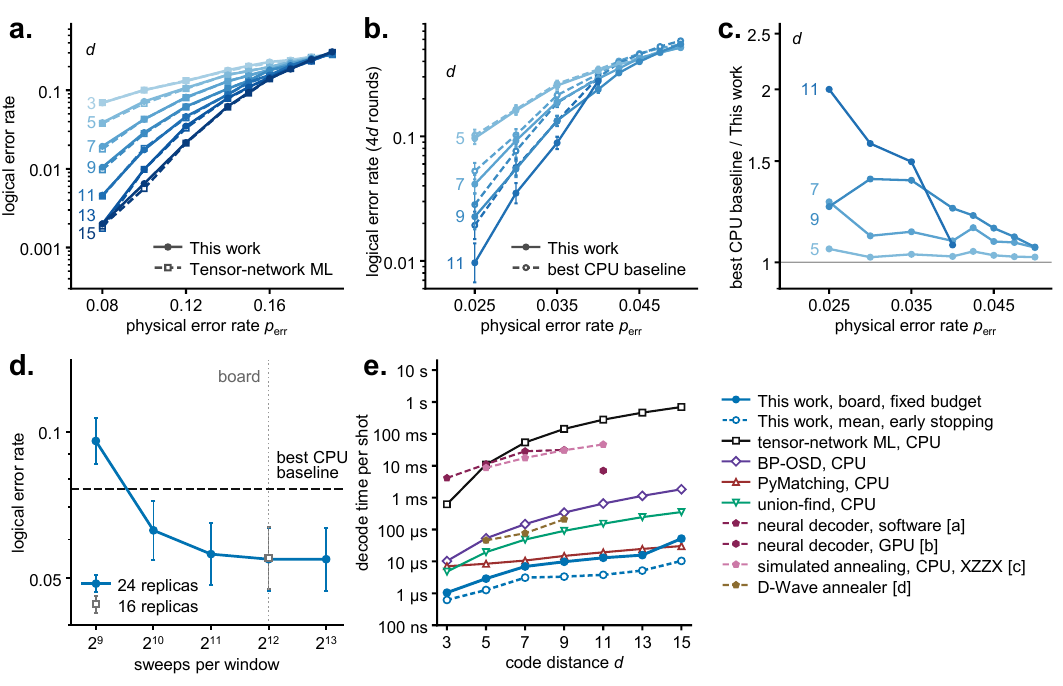}
\caption{\textbf{Further decoding results.} Same designs, shots and decoders as Fig.~\ref{fig:surface}. Labels: code distance $d$. CPU decoders: one core of an Intel Xeon Gold 6548Y+. Error bars: 95\% Wilson confidence intervals.
\textbf{a}, One syndrome per shot (depolarizing noise, measurement without error): logical error rate versus the physical error rate $p_\mathrm{err}$, $d=3$--15, 8,000 shots per point. Filled: this work. Open: tensor-network maximum-likelihood (ML) decoding, bond dimension 16. This work runs the fixed budget of each design, and $2^{12}$ sweeps at $p_\mathrm{err}\geq0.14$ for $d\geq5$.
\textbf{b}, Real time (phenomenological noise): this work (solid) and the best CPU baseline (dashed) versus $p_\mathrm{err}$, $d=5$--11, 3,000 shots per point (9,000 at $p_\mathrm{err}=0.0425$--0.0475 for $d=7$ and 9). Logical error rate: fraction of shots, each a memory experiment of $4d$ noisy rounds and one round without error, that end with a logical error. Best CPU baseline: lowest logical error rate of correlated PyMatching, BP+matching and BP-OSD at each point, in the same sliding windows as this work.
\textbf{c}, Logical error rate of the best CPU baseline divided by that of this work.
\textbf{d}, Real time, $d=9$, $p_\mathrm{err}=0.03$: logical error rate versus sweeps per window for 24 replicas (the board design) and 16 replicas. Dashed line: best CPU baseline. Dotted line: the board's budget.
\textbf{e}, Decode time per shot versus $d$ for decoders of one syndrome per shot, $p_\mathrm{err}=0.12$. Solid lines: decoders run by us on the same shots. This work: board, fixed budget. Dashed line with open circles: this work, mean with early stopping. Dashed lines with pentagons: published times of surface-code decoders. [a]: neural decoder in software. [b]: neural decoder on a GPU, one shot at a time ($d=11$). [c]: simulated annealing on one CPU core (XZZX surface code). [d]: D-Wave quantum annealer running an Ising-model decoder (link and annealing time, cloud latency excluded). References: Supplementary Section~\ref{supp:qec-results}. Noise models and physical error rates differ between works.}\label{fig:supp-dec}
\end{figure}

For one syndrome per shot, over all 63 points of Fig.~\ref{fig:supp-dec}a, our decoder makes 0.7\% more logical errors than ML (95\% confidence interval (CI) 0.4--1.0\%). Neighboring distances differ too little near the threshold to locate their crossing at 8,000 shots per point, so the crossings in main text Section~\ref{res:surface} use $d=11$ and 15. Their CIs, like that of the 0.7\%, come from 4,000 bootstrap resamples of the eight batches of 1,000 shots of each point, the same batches for both decoders (Methods Section~\ref{meth:problems}). Resampling single shots instead gives the same interval.

In real time (Fig.~\ref{fig:supp-dec}b), at $d=5$, 7, 9 and 11, the best windowed CPU baseline makes 1.03 (95\% CI 1.01--1.05), 1.15 (1.11--1.20), 1.32 (1.23--1.42) and 1.51 (1.39--1.66) times as many logical errors as our decoder. Each ratio is the geometric mean over the four physical error rates $p_\mathrm{err}=0.025$--0.04, the range tested at every distance. These CIs come from the paired bootstrap over batches of 250 shots (Methods Section~\ref{meth:problems}). The ratio is largest at the largest distance and the lowest physical error rate (Fig.~\ref{fig:supp-dec}c). BP-OSD is the best baseline at most physical error rates at $d=5$ and at the lower ones at $d=7$ to 11, and BP+matching at the others. The Wilson intervals in Fig.~\ref{fig:supp-dec}b overlap at many points because each of them also contains the variation of the shots, which the two decoders share. The paired bootstrap removes this shared variation.

At $p_\mathrm{err}=0.03$, we convert the LER of a shot into a logical error rate per round, $\varepsilon_d=[1-(1-2\,\mathrm{LER})^{1/4d}]/2$, the rate that, compounded over the $4d$ noisy rounds, gives the LER\citeSI{GoogleQAI2025}. We fit the logarithm of $\varepsilon_d$ versus $d$ over $d=5$--11. For our decoder, $\varepsilon_d$ falls by a factor of 2.27 per step of 2 in $d$ (95\% CI 2.13--2.47). For the best baseline at each distance, $\varepsilon_d$ falls by a factor of 1.92 per step (1.83--2.04). The best baseline is BP-OSD at $d=5$--9 and BP+matching at $d=11$. The ratio of the two factors is 1.18 (1.12--1.28). Against BP+matching, whose rate falls the fastest among the baselines (factor 2.03, 1.93--2.15), the ratio is 1.12 (1.06--1.20). The advantage therefore grows with $d$. We also tested a memory twice as long, with $8d$ noisy rounds, on 3,000 other shots per distance at $d=7$ and 9 and $p_\mathrm{err}=0.03$. BP+matching makes 1.33 times as many logical errors per round as our decoder at $d=7$ (95\% CI 1.23--1.46) and 1.65 times at $d=9$ (1.46--1.91), close to the ratios with $4d$ rounds. From $d=7$ to 9, our logical error rate per round falls by the same factor as with $4d$ rounds. The rate itself is about 20\% higher with $8d$ rounds. The rounds near the start and near the round without error are likely easier to decode, and they make up a smaller share of a longer memory. BP-OSD was not run on these shots.

At $d=9$, the board budget of $2^{12}$ sweeps per window lies where the LER has saturated (Fig.~\ref{fig:supp-dec}d). At $d=11$, the LER still falls beyond it: in the software decoder with 32 replicas, $2^{13}$ sweeps give about 10\% fewer logical errors than $2^{12}$ at $p_\mathrm{err}=0.03$. At $d=9$ and $p_\mathrm{err}=0.03$, the LER lies below that of the best CPU baseline from $2^{10}$ sweeps on, and 16 replicas give the same LER as 24 at $2^{12}$ sweeps. Halving the sweeps to $2^{11}$ would halve the time per round and raise the LER by only 2\%. We kept $2^{12}$ sweeps because, with $2^{11}$, the crossing of $d=7$ and 9 did not exceed that of BP+matching on the first 3,000 of the 9,000 shots per point at $p_\mathrm{err}=0.0425$--0.0475.

The crossings near the threshold are therefore taken on the other shots: the other 6,000 shots per point at $p_\mathrm{err}=0.0425$--0.0475, together with 6,000 new shots at 0.04 and 0.05. On these shots, the curves of $d=7$ and 9 cross at $p_\mathrm{err}=4.74\%$ (95\% CI 4.64--4.85\%) for our decoder and at 4.59\% (4.17--4.77\%) for BP+matching, interpolated linearly between grid points (Methods Section~\ref{meth:problems}). Their difference, 0.16 percentage points ($-0.005$ to 0.60), lies within the sampling uncertainty. Our crossing is higher than that of correlated PyMatching by 0.30 percentage points (0.10--0.63), and the curves of BP-OSD cross below 4\%. Including the shots used to choose the number of sweeps gives nearly the same crossings. The curves of $d=5$ and 7 also cross higher for our decoder than for BP+matching. The curves of $d=9$ and 11 cross lower for our decoder than for BP+matching, so that above about 3.8\% our decoder at $d=11$ makes more logical errors than at $d=9$. At $d=11$, the chip holds only 24 replicas (Supplementary Section~\ref{supp:realtime}).

The same windowed baselines decoded the 1,000 board shots per distance at $p_\mathrm{err}=0.03$. On these shots, the best baseline makes 1.77 times as many logical errors as the board at $d=9$ (95\% CI 1.41--2.30) and 1.56 times as many at $d=11$ (1.14--2.19). These CIs come from a paired bootstrap over single shots. At $d=5$ and 7, the two are equal within the sampling uncertainty. The LERs of the board differ from those of the model, which decoded other shots, by at most about two standard errors.

Published decoders of one syndrome per shot are slower than ours at every distance at which they report a time (Fig.~\ref{fig:supp-dec}e). They include neural decoders in software\citeSI{Varsamopoulos2020} and on a GPU\citeSI{Davaasuren2020}, simulated annealing on one CPU core\citeSI{Sakashita2026} and a D-Wave quantum annealer\citeSI{Guo2025}. Only the neural decoder in software uses our code and noise model. The other three decoders use the unrotated layout, with $d^2+(d-1)^2$ qubits. The two annealers were timed at lower physical error rates (0.02 and 0.01), one of them with bit-flip noise, but their fixed annealing schedules make their times depend little on the error rate. On the unrotated layout, which has 1.8 times as many qubits as the rotated layout at $d=9$, their times would be somewhat longer than on ours, by far less than the gaps in main text Fig.~\ref{fig:surface}f. Another uses the XZZX variant, in which every check measures $X$ on two of its qubits and $Z$ on the other two. In main text Fig.~\ref{fig:surface}f, the annealers are therefore placed at the LER of the lowest-energy correction that any replica of our decoder finds on our shots. At $d=9$ and $p_\mathrm{err}=0.12$, the LER of this lowest-energy correction is 0.070, against 0.062 for the most visited class. The neural decoder is placed at the LER of PyMatching, because its accuracy reaches that of matching\citeSI{Varsamopoulos2020}.

The published real-time decoders of main text Fig.~\ref{fig:surface}i are placed at the LERs that their algorithms reach on our shots (Methods Section~\ref{meth:problems}). The FPGA decoders based on union-find\citeSI{Liyanage2024}, clustering\citeSI{Barber2025,Ziad2025} and matching\citeSI{Wu2025} report the decode time of $d$ rounds divided by $d$, except the matching decoder, which reports its average latency after the last round. We divide this latency by $d$ as well. At $d=9$ and $p_\mathrm{err}=0.03$, union-find\citeSI{Delfosse2021} fails in 49\% of our shots.

\FloatBarrier
\section{Measurements on the board}\label{supp:board}
\begin{table}[!tbp]
\centering
\footnotesize\setlength{\tabcolsep}{3pt}
\caption{\textbf{Board designs.} All designs run on the RFSoC 4x2 board (XCZU48DR chip with 425,280 lookup tables (LUTs), the small programmable logic elements of the FPGA). LUTs: number used, with the percentage of the chip in parentheses. Layout: one neuron per move means that every move has its own neuron in each replica. Slabs means that the $2d$ rounds of a window are split into that many groups, which use one set of \textit{move} neurons in turn during each sweep. Time: decode time per shot at the fixed budget (one syndrome per shot), or per round, the window decode time divided by $d$, at $2^{12}$ sweeps per window (real time). Power: supply rail of the chip's logic (core rail), given as the power while decoding / the power while idle. At $d=3$ and 5 for one syndrome per shot, the decoder finishes each batch of shots faster than the host writes the next batch. The decoder is therefore idle during part of every power reading, and the decoding power is a lower bound.}\label{tab:supp-designs}
\begin{tabular}{@{}lrrlrrrr@{}}\toprule
Mode & $d$ & Replicas & Layout & LUTs (\%) & Clock (MHz) & Time & Power (W)\\\midrule
One syndrome & 3 & 16 & one neuron per move & 17,492 (4.1) & 356.2 & 1.05\,$\mu$s & $\geq$3.0 / 2.2\\
per shot & 5 & 16 & one neuron per move & 27,020 (6.4) & 356.2 & 2.89\,$\mu$s & $\geq$3.2 / 2.2\\
 & 7 & 64 & one neuron per move & 200,580 (47.2) & 250.0 & 6.94\,$\mu$s & 9.2 / 3.2\\
 & 9 & 56 & one neuron per move & 273,677 (64.4) & 250.0 & 9.76\,$\mu$s & 11.8 / 3.6\\
 & 11 & 40 & one neuron per move & 268,331 (63.1) & 265.0 & 13.01\,$\mu$s & 12.8 / 3.6\\
 & 13 & 32 & one neuron per move & 278,022 (65.4) & 260.0 & 15.78\,$\mu$s & 13.2 / 3.6\\
 & 15 & 24 & one neuron per move & 283,859 (66.7) & 265.0 & 52.02\,$\mu$s & 13.2 / 3.6\\\midrule
Real time & 3 & 16 & one neuron per move & 107,518 (25.3) & 290.0 & 89.5\,$\mu$s & 5.8 / 3.2\\
 & 5 & 16 & 2 slabs & 248,614 (58.5) & 240.0 & 105.8\,$\mu$s & 10.4 / 3.6\\
 & 7 & 16 & 5 slabs & 240,365 (56.5) & 165.0 & 290.8\,$\mu$s & 7.8 / 3.2\\
 & 9 & 24 & 9 slabs & 279,577 (65.7) & 233.8 & 646.3\,$\mu$s & 15.0 / 4.6\\
 & 11 & 24 & 11 slabs & 299,210 (70.4) & 225.0 & 1,325.8\,$\mu$s & 14.8 / 5.0\\\bottomrule
\end{tabular}
\end{table}

\begin{figure}[!tbp]
\centering
\includegraphics[width=\linewidth]{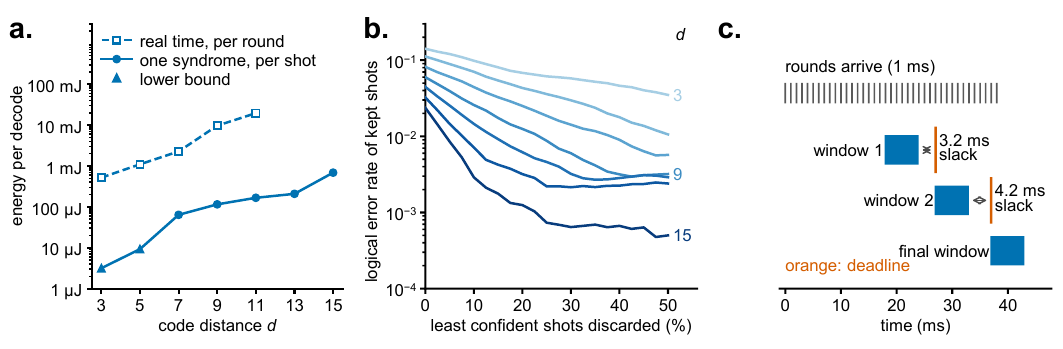}
\caption{\textbf{Measurements on the board.} RFSoC 4x2 board, one design per $d$ and decoding mode (Supplementary Table~\ref{tab:supp-designs}).
\textbf{a}, Energy per decode: core-rail power while decoding times the decode time, per shot at the fixed budget (one syndrome per shot) and per round (real time). Triangles: lower bounds, at $d=3$ and 5 for one syndrome per shot, where the decoder is idle during part of every power reading.
\textbf{b}, One syndrome per shot, $p_\mathrm{err}=0.12$: logical error rate of the shots kept when the least confident are discarded. Confidence: largest class count divided by the total count, read out with every decision. 20,000 board shots per $d$.
\textbf{c}, Real time, timeline of one shot at $d=9$ with rounds every 1\,ms. Ticks: arriving rounds. Bars: each window from its start to its commit. Orange: the start of the next window, which needs the corrected syndrome of the previous window (the deadline).}\label{fig:supp-board}
\end{figure}

Supplementary Table~\ref{tab:supp-designs} lists the logic use and clock of every design (Methods Section~\ref{meth:fpga}). Every design fits within about 70\% of the logic, which leaves routing resources for the interconnect. Within this limit, the numbers of replicas were chosen in software (Supplementary Sections~\ref{supp:realtime} and~\ref{supp:oneshot}).

For one syndrome per shot, one decode costs 0.69\,mJ at $d=15$ and at least 3\,$\mu$J at $d=3$, and one round costs 0.52\,mJ at $d=3$ and 20\,mJ at $d=11$ in real time (Fig.~\ref{fig:supp-board}a). The energy includes the idle power of the chip, which is most of the power of the small designs. The power was read from the sensor of the core rail every 0.1\,s while the board decoded batches of shots for about 10\,s. The idle power is the median reading with the design loaded and no shots. The decoding power is the median of the readings that exceed the idle power by more than 0.3\,W, a margin set by the 0.2 to 0.4\,W steps of the sensor.

The confidence of each decision lets a user discard the least reliable shots (Fig.~\ref{fig:supp-board}b). Discarding the 10\% of shots with the lowest confidence lowers the LER of the remaining shots 1.4-fold at $d=3$. At $d=15$, it lowers the LER 8-fold, and discarding 25\% lowers it 32-fold. The confidence requires no additional hardware, because the class counts are read out with every decision.

The board met every real-time deadline (Fig.~\ref{fig:supp-board}c). A timer on the FPGA released the rounds of 1,000 shots per distance at $p_\mathrm{err}=0.03$ every 1\,ms at $d=3$--9. At $d=11$, the decoder needs 1.33\,ms per round (Supplementary Table~\ref{tab:supp-designs}), so the rounds came every 4.45\,ms, the round time of a neutral-atom experiment\citeSI{Bluvstein2026}. The shots were those of the run without the timer, and the decisions were identical to those of that run. In the shot of Fig.~\ref{fig:supp-board}c ($d=9$), each window takes 5.8\,ms from its start to its commit, and both non-final windows commit more than 3\,ms before their deadlines. The smallest slack over all windows, shots and distances was 2.7\,ms, at $d=3$.

\FloatBarrier
\section{Exact local maintenance of \textit{move}-neuron inputs}\label{supp:local-updates}

Methods Eqs.~\eqref{eq:decoder-sigma-async} and~\eqref{eq:decoder-clause} identify which clause outputs change after an accepted spin transition. Because the auxiliary clauses have zero weight, the input of $G_r$ in Methods Eq.~\eqref{eq:encoder} can be written as a sum over all stored clauses,
\begin{equation}
v_{G_r}=\sum_{k=1}^{\widetilde M}J_k\alpha_{k,G_r}T_k,
\label{eq:maintained-input}
\end{equation}
with $\alpha_{k,G_r}$ and $T_k$ at their current values. Because the clause outputs are stored, each input can be maintained by local updates only through the connections whose clause output or switch has changed, instead of being recomputed (main text Section~\ref{res:event}). By Methods Eq.~\eqref{eq:gate-switch}, a switch changes exactly when the clause output that controls it toggles. A toggled clause output $T_j$ therefore updates the inputs through every connection that it switches and, unless it is an auxiliary clause ($J_j=0$), through its own connections.

If $T_k$ changes from $T_k^{\old}$ to $T_k^{\new}$, the local update of $v_{G_r}$ is
\begin{equation}
\delta v_{G_r}^{(T,k)}=J_k\alpha_{k,G_r}^{\mathrm{current}}\left(T_k^{\new}-T_k^{\old}\right).
\label{eq:event-clause-local}
\end{equation}
If the switch changes from $\alpha_{k,G_r}^{\old}$ to $\alpha_{k,G_r}^{\new}$, the local update is
\begin{equation}
\delta v_{G_r}^{(\alpha,k)}=J_kT_k^{\mathrm{current}}\left(\alpha_{k,G_r}^{\new}-\alpha_{k,G_r}^{\old}\right).
\label{eq:event-gate-local}
\end{equation}
The superscript $\mathrm{current}$ denotes the value when the update is applied. When both $T_k$ and $\alpha_{k,G_r}$ change, the total local update of the connection does not depend on the order of the two updates. Applying the clause update first gives
\begin{equation}
J_k\left[\alpha_{k,G_r}^{\old}\left(T_k^{\new}-T_k^{\old}\right)+T_k^{\new}\left(\alpha_{k,G_r}^{\new}-\alpha_{k,G_r}^{\old}\right)\right].
\label{eq:event-clause-first}
\end{equation}
Applying the switch update first gives
\begin{equation}
J_k\left[T_k^{\old}\left(\alpha_{k,G_r}^{\new}-\alpha_{k,G_r}^{\old}\right)+\alpha_{k,G_r}^{\new}\left(T_k^{\new}-T_k^{\old}\right)\right].
\label{eq:event-gate-first}
\end{equation}
Both reduce to
\begin{equation}
J_k\left(\alpha_{k,G_r}^{\new}T_k^{\new}-\alpha_{k,G_r}^{\old}T_k^{\old}\right).
\label{eq:event-route-exact}
\end{equation}
Summing over the affected connections gives the exact change of $v_{G_r}$. Clause and switch updates may therefore be applied in either order if the second update uses the value already updated by the first. The next transition must be evaluated only after all affected inputs have settled. Supplementary Algorithm~\ref{alg:solver} applies these updates within one iteration of the emulation, and Supplementary Section~\ref{supp:onoff} maps them to ON--OFF \textit{clause} neurons.

\FloatBarrier
\section{Event-driven operations: per-iteration ratio and a synthetic family}\label{supp:event}

\begin{figure}[!tbp]
\centering
\includegraphics[width=\linewidth]{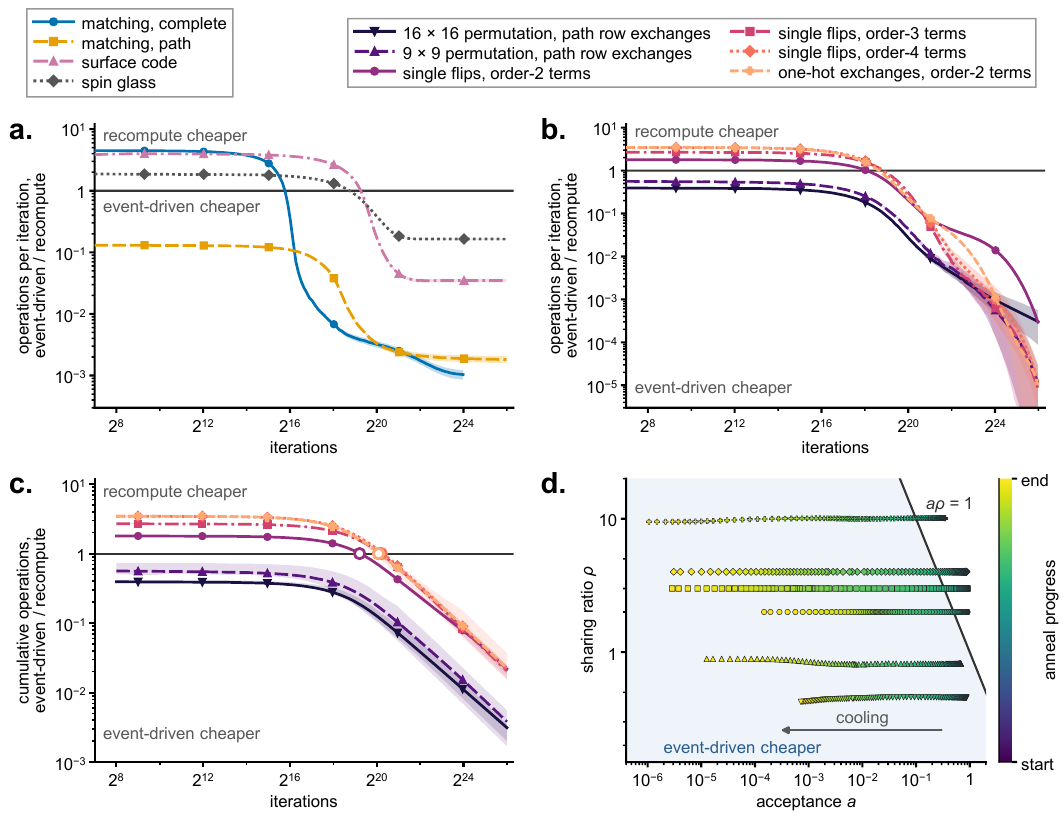}
\caption{\textbf{Event-driven updates: operations per iteration and a synthetic family.}
All panels use slow anneals as in Fig.~\ref{fig:event}. One operation is one local update of one input (event-driven) or one connection read (recomputation). \textbf{a},
Operations per iteration (event-driven / recomputation) per window (1/8 octave) for the four workloads of Fig.~\ref{fig:event}. Workloads: matching (complete
library, $L=30$), matching (path library, $L=50$), surface code ($d=11$, $p_\mathrm{err}=0.12$) and 2D $\pm J$ spin glass (single flips). Lines:
totals over replicas and instances. In every window, the ratio equals $a\rho$. Bands: 95\% bootstrap confidence intervals (replicas
resampled within each instance). \textbf{b}, The same for synthetic problems with Gaussian random coefficients (4 random instances $\times$ 64
replicas, $2^{26}$ iterations). A term of order $\ell$ multiplies $\ell$ spins. Single flips: 24 spins with terms of order $\ell=2$, 3 or 4. One-hot exchanges: 40 spins in groups of 5
with exactly one spin up per group, and all exchanges inside each group, with $\ell=2$. Permutations: $L=9$ or 16 items assigned to $L$ positions,
changed only by exchanging adjacent rows ($L-1$ moves), with $\ell=2$. \textbf{c}, Cumulative operations for the synthetic family. Bands (interquartile range of the per-replica ratio) and
open circles (break-even point) as in Fig.~\ref{fig:event}a. \textbf{d}, Acceptance $a$ versus sharing ratio $\rho$ per window for the synthetic family (as in Fig.~\ref{fig:event}b).
Marker shape: setting. Color: progress through the anneal. Line: $a\rho=1$. Shaded region: event-driven updating requires fewer operations.}\label{fig:supp-event}
\end{figure}

Each workload of Fig.~\ref{fig:event} was run with one slow anneal (Methods Eq.~\eqref{eq:fn-schedule}) from the calibrated $\tau_\mathrm{start}$ to $\tau_\mathrm{start}/1{,}000$, without stopping at the optimum, so that the acceptance passes through its whole range. Graph matching with the complete library uses $L=30$ and the 5 test instances with the lowest probability of reaching $0.98\,\Psi_\mathrm{ref}$ within $2^{22}$ iterations under the automatically calibrated schedule (128 replicas, $2^{24}$ iterations). Graph matching with the path library uses $L=50$ and the first 3 test instances (128 replicas, $2^{26}$ iterations). The surface code uses $d=11$, depolarizing $p_\mathrm{err}=0.12$ and all $d$ logical operators per type (512 shots, $2^{26}$ iterations). The spin glass is the two-dimensional $\pm J$ model\citeSI{Edwards1975} on a $16\times16$ torus, with single spin flips and couplings $\pm1$ of equal probability (4 coupling sets, 64 replicas, $2^{26}$ iterations). Operations were accumulated in windows of 1/8 octave, eight per doubling of the iteration count. The event-driven count omits the one read per iteration of the maintained input of the proposed \textit{move} neuron. Counting it would change the ratio of main text Eq.~\eqref{eq:arho} from $a\rho$ to $a\rho+1/f_\mathrm{in}$. The break-even point is the iteration after which the cumulative ratio stays below 1. If event-driven updating needs more operations early in the run, the break-even point follows the iteration after which the per-window ratio stays below 1, once the operations saved have offset the earlier excess.

With $f_\mathrm{in}$ and $f_\mathrm{out}$ the means within each window, main text Eq.~\eqref{eq:arho} holds in every window to within $4\times10^{-16}$ relative error, the level of rounding error in double-precision arithmetic (Fig.~\ref{fig:supp-event}a).

The synthetic family isolates how the move library sets $\rho$ (Fig.~\ref{fig:supp-event}b). For single spin flips with terms of order $\ell$, recomputing the input of a proposed flip of spin $i$ reads the $\deg(i)$ weighted terms that contain spin $i$, so $f_\mathrm{in}=\deg(i)$. An accepted flip toggles these terms, and each toggled term updates the inputs of the \textit{move} neurons of all $\ell$ of its spins (each spin is flipped by its own neuron). Hence $f_\mathrm{out}=\ell\deg(i)$ and $\rho=\ell$ exactly, measured as 2, 3 and 4. For exchanges within one-hot groups, an accepted spin transition changes two spins whose terms are read by many moves, and $\rho$ is 9.5 to 10.1. For permutations changed by adjacent-row exchanges, a proposed exchange reads the terms of all $2L$ spins in its two rows. An accepted transition, by contrast, changes only four spins, and few moves read their terms. Here $\rho$ is below 1 and falls as $L$ grows, from 0.81--0.89 at $L=9$ to 0.42--0.46 at $L=16$. As for the workloads of Fig.~\ref{fig:event}a, the cumulative operations of event-driven updating fall below those of recomputation early in the run, and from the start for the permutations (Fig.~\ref{fig:supp-event}c). At the end of the run, they are 0.3 to 2.2\% of those of recomputation. Within each setting, $\rho$ varies by at most 11\% over the anneal, whereas the acceptance falls by 3.1 to 5.5 orders of magnitude (Fig.~\ref{fig:supp-event}d). With Gaussian coefficients, no transition leaves the energy unchanged, so the acceptance has no positive floor, unlike in the workloads of Fig.~\ref{fig:event}b. The condition $a\rho<1$ therefore follows from the move library, which sets $\rho$, and the temperature, which sets $a$.

\FloatBarrier
\section{Proposal symmetry and state-dependent proposal normalization}\label{supp:barriers}

A run at a fixed temperature on a connected library, with the Metropolis acceptance of Methods Section~\ref{meth:benchmarks}, samples the Boltzmann distribution asymptotically when each spin transition is proposed with the same probability as its reverse. This section shows that our proposal rule meets this condition, how redrawing the move after a self-loop can violate it, and which correction an asymmetric proposal requires. Let $Q(\muvec,\nuvec)$ be the probability of proposing a distinct feasible state $\nuvec$ from $\muvec$, summed over all moves that take $\muvec$ to $\nuvec$. All experiments use reversible moves and select each move with a probability that does not depend on the current state. If $G_r(\muvec)=\nuvec$, then $G_r(\nuvec)=\muvec$, and $G_r$ is selected with the same probability from both states. Summing over all moves that connect the two states gives $Q(\muvec,\nuvec)=Q(\nuvec,\muvec)$. The symmetry of $\alpha_{k,\muvec\leftrightarrow\nuvec}$ (Methods Eq.~\eqref{eq:alpha-definition}) does not by itself establish proposal symmetry, because it states that a spin transition and its reverse change the same clauses, not that they are proposed equally often.

For a symmetric proposal, Metropolis acceptance gives
\begin{equation}
\mathrm{e}^{-E(\muvec)/\tau}Q(\muvec,\nuvec)\min\{1,\mathrm{e}^{-[E(\nuvec)-E(\muvec)]/\tau}\}=\mathrm{e}^{-E(\nuvec)/\tau}Q(\nuvec,\muvec)\min\{1,\mathrm{e}^{-[E(\muvec)-E(\nuvec)]/\tau}\},
\label{eq:supp-balance}
\end{equation}
which is detailed balance with the Boltzmann distribution restricted to the feasible states, $\pi(\svec)\propto\mathrm{e}^{-E(\svec)/\tau}$ for $\svec\in\Omega$ and $\pi(\svec)=0$ otherwise. The equality holds because $Q$ is symmetric and $\mathrm{e}^{-E(\muvec)/\tau}\min\{1,\mathrm{e}^{-[E(\nuvec)-E(\muvec)]/\tau}\}=\min\{\mathrm{e}^{-E(\muvec)/\tau},\mathrm{e}^{-E(\nuvec)/\tau}\}$ does not change when $\muvec$ and $\nuvec$ are exchanged. Once the distribution of the state equals $\pi$, detailed balance makes each spin transition $\muvec\to\nuvec$ as frequent as its reverse, so $\pi$ is the stationary distribution of the fixed-temperature dynamics. Such dynamics are called reversible with respect to $\pi$, a property distinct from the reversibility of a move. The critical depth and the Catoni difficulty of Supplementary Section~\ref{supp:shortcut-theory} are defined for this dynamics: Metropolis acceptance with symmetric proposals of reversible moves.

Batching retains this property when the batch is drawn at every iteration with probability proportional to its size. Each move is then proposed with probability $(R_b/R)(1/R_b)=1/R$, as in the fully loaded library. In the cyclic batching of Fig.~\ref{fig:library}g, the loaded moves change with the iteration, so the process has no single proposal probability $Q$. At every iteration, however, each loaded move is still selected with a probability that does not depend on the current state, and its reverse spin transitions belong to the same move. The proposal of each iteration is therefore symmetric, and the update of each iteration satisfies detailed balance with the same Boltzmann distribution $\pi$. An update that leaves $\pi$ unchanged, followed by another that does so, also leaves $\pi$ unchanged. At a fixed temperature, cyclic batching therefore keeps $\pi$ as a stationary distribution, provided that the moves of all batches together connect the feasible states. Only the reversibility of the whole process is lost, because the batches follow a fixed order.

Counting a self-loop as an iteration, rather than redrawing the move, preserves the selection probabilities of the moves. Suppose that $R_{\neq}(\muvec)$ of the $R$ moves map the current state $\muvec$ to a distinct state. With one uniform draw, every move has probability $1/R$, and a move with $G_r(\muvec)=\muvec$ produces a self-loop. If instead the move is redrawn until one maps $\muvec$ to a distinct state, each such move is selected with probability $1/R_{\neq}(\muvec)$. When $R_{\neq}(\muvec)\neq R_{\neq}(\nuvec)$, a spin transition $\muvec\to\nuvec$ and its reverse in the same reversible move can then be proposed with the different probabilities $1/R_{\neq}(\muvec)$ and $1/R_{\neq}(\nuvec)$. Redrawing thus introduces a state-dependent normalization that can violate proposal symmetry. A special case is the complete library for an exact-cardinality constraint, with $N_\uparrow$ of $N$ spins up, which contains the exchange of every pair of spins. An exchange changes the state only when its two spins have opposite signs, and every feasible state has $N_\uparrow(N-N_\uparrow)$ such pairs. Hence $R_{\neq}(\muvec)$ is the same for all states, and redrawing leaves the proposal symmetric. Sparse exchange graphs, mixed move libraries and sets of shortcut moves, however, need not have this property.

If the proposal is asymmetric but both directions are present, the Metropolis rule alone no longer satisfies Eq.~\eqref{eq:supp-balance}: a transition proposed more often than its reverse would also be made more often than its reverse. Exact fixed-temperature Boltzmann sampling then uses the Metropolis--Hastings acceptance, which accepts a transition proposed more often than its reverse correspondingly less often,
\begin{equation}
A_{\mathrm{MH}}(\muvec,\nuvec)=\min\left\{1,\exp\!\left[-\frac{E(\nuvec)-E(\muvec)}{\tau}\right]\frac{Q(\nuvec,\muvec)}{Q(\muvec,\nuvec)}\right\}.
\label{eq:supp-mh}
\end{equation}
With this acceptance, both sides of Eq.~\eqref{eq:supp-balance} equal $\min\{\mathrm{e}^{-E(\muvec)/\tau}Q(\muvec,\nuvec),\mathrm{e}^{-E(\nuvec)/\tau}Q(\nuvec,\muvec)\}$, so detailed balance holds again. Let $\Phi=Q(\nuvec,\muvec)/Q(\muvec,\nuvec)$. For an uphill transition, with $\Delta E=E(\nuvec)-E(\muvec)>0$ and acceptance below one, the acceptance is then $\Phi\,\mathrm{e}^{-\Delta E/\tau}=\mathrm{e}^{-(\Delta E-\tau\ln \Phi)/\tau}$, the Metropolis acceptance of the shifted energy change $\Delta E-\tau\ln \Phi$. If $\Phi$ remains bounded away from zero and infinity independently of temperature, $\tau\ln \Phi\rightarrow0$ as $\tau\rightarrow0$. At low temperature the time needed to climb an energy barrier of height $\Delta E$ grows as $\mathrm{e}^{\Delta E/\tau}$, and a bounded $\Phi$ changes this time only by a bounded factor. The Hastings factor $\Phi$ therefore does not change the leading low-temperature exponent, the barrier height $\Delta E$, although it can still alter finite-temperature transition probabilities and finite-time trajectories.

If one proposal direction is absent, or if the ratio $\Phi$ grows without bound or approaches zero with problem size or temperature, this bounded-ratio argument no longer applies. All experiments therefore use symmetric proposals. The Metropolis--Hastings correction is required whenever an asymmetric proposal is used for a fixed-temperature equilibrium claim.

\section{ON--OFF neuromorphic realization}\label{supp:onoff}

A spiking neuron signals only by emitting events and has no negative activity. Every clause output $T_k\in\{-1,+1\}$ can therefore be held by a complementary pair of ON and OFF neurons, with activities
\begin{equation}
T_k^+=\frac{1+T_k}{2},\qquad T_k^-=\frac{1-T_k}{2},\qquad T_k=T_k^+-T_k^-,\qquad T_k^++T_k^-=1.
\label{eq:supp-onoff-state}
\end{equation}
The clause toggle $\chi_k\in\{0,1\}$ of Methods Eq.~\eqref{eq:decoder-sigma-async} generates at most one of two complementary events,
\begin{equation}
e_k^+=T_k^-\chi_k,\qquad e_k^-=T_k^+\chi_k,
\label{eq:supp-onoff-events}
\end{equation}
so that
\begin{equation}
T_k^+\leftarrow T_k^++e_k^+-e_k^-,\qquad T_k^-\leftarrow T_k^- -e_k^+ +e_k^-.
\label{eq:supp-onoff-update}
\end{equation}
In an integrate-and-fire realization, each cell emits a spike when its accumulated input crosses a threshold, and is then reset and cannot fire again for a short refractory period. The two cells of a pair can be cross-coupled, so that a firing cell resets the other and holds it in its refractory period. Only the cell of the new sign can then fire, and a toggle produces exactly one event, as Eq.~\eqref{eq:supp-onoff-events} requires. NeuroSA realizes its neurons with such complementary ON--OFF pairs\citeSI{Chen2025}, and N-HOIM uses the same realization for its \textit{clause} neurons\citeSI{Ahsan2026}. Here, ON--OFF pairs hold the output of every stored clause, objective and auxiliary clauses alike.

With Eq.~\eqref{eq:supp-onoff-state}, the input of \textit{move} neuron $G_r$ (Eq.~\eqref{eq:maintained-input}) becomes
\begin{equation}
v_{G_r}=\sum_kJ_k\alpha_{k,G_r}(T_k^+-T_k^-).
\label{eq:supp-onoff-field}
\end{equation}
Because the \textit{move} neuron maintains this sum (Supplementary Section~\ref{supp:local-updates}), only its changes are transmitted. By Eq.~\eqref{eq:supp-onoff-update}, a toggle changes $T_k^+-T_k^-$ by $2(e_k^+-e_k^-)$, so the local update of Eq.~\eqref{eq:event-clause-local} becomes
\begin{equation}
\delta v_{G_r}^{(T,k)}=2J_k\alpha_{k,G_r}^{\mathrm{current}}(e_k^+-e_k^-).
\label{eq:supp-onoff-clause-pulse}
\end{equation}
The local update of Eq.~\eqref{eq:event-gate-local} becomes
\begin{equation}
\delta v_{G_r}^{(\alpha,k)}=J_k(T_k^+-T_k^-)\left(\alpha_{k,G_r}^{\new}-\alpha_{k,G_r}^{\old}\right).
\label{eq:supp-onoff-gate-pulse}
\end{equation}

Each \textit{move} neuron fires according to Methods Eq.~\eqref{eq:firing}, with the stochastic threshold of Methods Eq.~\eqref{eq:threshold}. Its firing accepts the spin transition of its move from the current state with the Metropolis probability (Methods Section~\ref{meth:benchmarks}). When several \textit{move} neurons can fire at about the same time, an arbitration rule lets at most one spin transition be accepted before the network settles (Methods Eq.~\eqref{eq:single-commit}). Its form depends on the implementation, and the digital emulation of Methods Section~\ref{meth:benchmarks} is one realization. An accepted-transition spike returns through the decoder connections to the ON--OFF \textit{clause} neurons. Once the clause and switch updates settle, the neuron inputs again satisfy Eq.~\eqref{eq:maintained-input}.

\FloatBarrier
\makeatletter\setcounter{NAT@ctr}{0}\makeatother
\bibliographystyleSI{sn-mathphys-num-pub}
\bibliographySI{references}

\end{document}